\documentclass[12pt]{article}
\usepackage{a4,epsfig}

\newcommand{\gev}{\ensuremath{\mathrm{\, GeV}}}
\newcommand{\gevc}{\ensuremath{\mathrm{\, GeV}/c}}
\newcommand{\gevcc}{\ensuremath{\mathrm{\, GeV}/c^2}}
\newcommand{\cm}{\ensuremath{\mathrm{\, cm}}}
\newcommand{\m}{\ensuremath{\mathrm{\, m}}}
\newcommand{\invpb}{\ensuremath{\mathrm{\, pb^{-1}}}}

\newcommand{\nbar}{\ensuremath{\mathrm{\bar{N}_{95}}}}
\newcommand{\alephcoll}{{\tt ALEPH}}
\newcommand{\sq}{\ensuremath{\tilde{\mathrm q}}}
\newcommand{\snu}{\ensuremath{\tilde{\nu}}}
\newcommand{\slep}{\ensuremath{\tilde{\ell}}}

\newcommand{\stau}{\ensuremath{\tilde{\tau}}}

\newcommand{\stq}{\ensuremath{\mathrm{\tilde{t}}}}
\newcommand{\ee}{\ensuremath{\mathrm{e^+e^-}}}
\newcommand{\tautau}{\ensuremath{\tau^+\tau^-}}
\newcommand{\ww}{\ensuremath{\mathrm{W^+W^-}}}

\newcommand{\qq}{\ensuremath{\mathrm{q\bar{q}}}}

\newcommand{\Pl}{\ensuremath{\mathrm{l}}}
\newcommand{\Pf}{\ensuremath{\mathrm{f}}}
\newcommand{\PW}{\ensuremath{\mathrm{W}}}
\newcommand{\PZ}{\ensuremath{\mathrm{Z}}}
\newcommand{\Pe}{\ensuremath{\mathrm{e}}}
\newcommand{\Pq}{\ensuremath{\mathrm{q}}}
\newcommand{\Pc}{\ensuremath{\mathrm{c}}}
\newcommand{\llnu}{\ensuremath{\mathrm{ll\nu}}}
\newcommand{\nch}{\ensuremath{N_{\mathrm{ch}}}}
\newcommand{\mvis}{\ensuremath{M_{\mathrm{vis}}}}
\newcommand{\nlep}{\ensuremath{N_{\mathrm{lep}}}}
\newcommand{\elep}{\ensuremath{E_{\mathrm{lep}}}}
\newcommand{\enlep}{\ensuremath{E_{\mathrm{non lep}}}}
\newcommand{\ehad}{\ensuremath{E_{\mathrm{had}}}}
\newcommand{\evis}{\ensuremath{E_{\mathrm{vis}}}}
\newcommand{\ptmiss}{\ensuremath{p^{\mathrm{miss}}_{\perp}}}
\newcommand{\onehalf}{\raisebox{0.1ex}{${\frac{1}{2}}$}}
\newcommand{\onequarter}{\raisebox{0.1ex}{${\frac{1}{4}}$}}
\newcommand{\newc}{\newcommand}
\newc{\emiss}{{\not \!\! E}}
\newc{\R}{$R$}
\newc{\charginom}{M_{\tilde \chi}^{+}}
\newc{\mue}{\mu_{\tilde{e}_{iL}}}
\newc{\mud}{\mu_{\tilde{d}_{jL}}}
\newc{\beq}{\begin{equation}}
\newc{\eeq}{\end{equation}}
\newc{\barr}{\begin{eqnarray}}
\newc{\earr}{\end{eqnarray}}
\newc{\ra}{\rightarrow}
\newc{\lam}{\lambda}
\newc{\eps}{\epsilon}
\def\tevc{{\rm \, Te\kern-0.125em V/c^2}}
\newc{\eq}[1]{(\ref{eq:#1})}
\newc{\eqs}[2]{(\ref{eq:#1},\ref{eq:#2})}
\newc{\etal}{{\it et al.}\ }
\newc{\Hbar}{{\bar H}}
\newc{\Ubar}{{\bar U}}
\newc{\Dbar}{{\bar D}}
\newc{\Ebar}{{\bar E}}
\newc{\eg}{{\it e.g.}\ }
\newc{\ie}{{\it i.e.}\ }
\newc{\nonum}{\nonumber}
\newc{\lab}[1]{\label{eq:#1}}
\newc{\lle}[3]{L_{#1}L_{#2}\Ebar_{#3}}
\newc{\lqd}[3]{L_{#1}Q_{#2}\Dbar_{#3}}
\newc{\udd}[3]{\Ubar_{#1}\Dbar_{#2}\Dbar_{#3}}
\newc{\dpr}[2]{({#1}\cdot{#2})}
\newc{\rpv}{{\not R_p}}
\newc{\rpvm}{{\not \! R_p}}
\newc{\rp}{$R_p$}
\newc{\gsim}{ \,  \scriptstyle{\stackrel{>}{\sim}}\displaystyle \, }
\newc{\lsim}{ \,  \scriptstyle{\stackrel{<}{\sim}}\displaystyle \, }

\font\ninerm=cmr9

\begin{document}
\begin{titlepage}

\title{
Search for Supersymmetry with a dominant R-Parity\\
violating $LL{\bar E}$ Coupling in ${\mathrm e}^+{\mathrm e}^-$ Collisions\\
at centre-of-mass energies of 130~GeV to 172~GeV
\vspace{1cm}}
\date{}
\author{The ALEPH Collaboration$^*)$}

\maketitle

\begin{picture}(160,1)
\put(10,110){\rm EUROPEAN LABORATORY FOR PARTICLE PHYSICS (CERN)}
\put(115,94){\parbox[t]{45mm}{CERN PPE/97-151}}
\put(115,88){\parbox[t]{45mm}{December 1, 1997}}
\end{picture}

\vspace{0cm}
\begin{abstract}
\vspace{.5cm}
A search for pair-production of supersymmetric particles under the 
assumption that R-parity is violated via a dominant $LL{\bar E}$ coupling 
has been performed using the
data collected by \alephcoll{} at centre-of-mass energies of 130--172\gev. 
The observed candidate events in the data are in agreement with the
Standard Model expectation. This is translated into lower limits on the mass
of charginos, neutralinos, sleptons, sneutrinos and squarks.  
For instance, charginos with masses less than 73\gevcc{} and neutralinos
with masses less than 23\gevcc{} are excluded at 95\% confidence level for
any generation structure of the $LL{\bar E}$ coupling, and for
 neutralino, slepton or sneutrino LSPs.
\end{abstract}
\vfill
\centerline{\em Submitted to Zeitschrift f\"ur Physik C}
\vskip .5cm
\noindent
--------------------------------------------\hfil\break
{\ninerm $^*)$ See next pages for the list of authors}

\end{titlepage}

\newpage
\pagestyle{empty}
\newpage
\small
%
%
\newlength{\saveparskip}
\newlength{\savetextheight}
\newlength{\savetopmargin}
\newlength{\savetextwidth}
\newlength{\saveoddsidemargin}
\newlength{\savetopsep}
\setlength{\saveparskip}{\parskip}
\setlength{\savetextheight}{\textheight}
\setlength{\savetopmargin}{\topmargin}
\setlength{\savetextwidth}{\textwidth}
\setlength{\saveoddsidemargin}{\oddsidemargin}
\setlength{\savetopsep}{\topsep}
%
%
\setlength{\parskip}{0.0cm}
\setlength{\textheight}{25.0cm}
\setlength{\topmargin}{-1.5cm}
\setlength{\textwidth}{16 cm}
\setlength{\oddsidemargin}{-0.0cm}
\setlength{\topsep}{1mm}
\pretolerance=10000
\centerline{\large\bf The ALEPH Collaboration}
\footnotesize
\vspace{0.5cm}
{\raggedbottom
\begin{sloppypar}
\samepage\noindent
R.~Barate,
D.~Buskulic,
D.~Decamp,
P.~Ghez,
C.~Goy,
J.-P.~Lees,
A.~Lucotte,
\mbox{M.-N.~Minard},
\mbox{J.-Y.~Nief},
B.~Pietrzyk
\nopagebreak
\begin{center}
\parbox{15.5cm}{\sl\samepage
Laboratoire de Physique des Particules (LAPP), IN$^{2}$P$^{3}$-CNRS,
74019 Annecy-le-Vieux Cedex, France}
\end{center}\end{sloppypar}
\vspace{2mm}
\begin{sloppypar}
\noindent
G.~Boix,
M.P.~Casado,
M.~Chmeissani,
J.M.~Crespo,
M.~Delfino, 
E.~Fernandez,
M.~Fernandez-Bosman,
Ll.~Garrido,$^{15}$
E.~Graug\`{e}s,
A.~Juste,
M.~Martinez,
G.~Merino,
R.~Miquel,
Ll.M.~Mir,
P.~Morawitz,
I.C.~Park,
A.~Pascual,
J.A.~Perlas,
I.~Riu,
F.~Sanchez
\nopagebreak
\begin{center}
\parbox{15.5cm}{\sl\samepage
Institut de F\'{i}sica d'Altes Energies, Universitat Aut\`{o}noma
de Barcelona, 08193 Bellaterra (Barcelona), Spain$^{7}$}
\end{center}\end{sloppypar}
\vspace{2mm}
\begin{sloppypar}
\noindent
A.~Colaleo,
D.~Creanza,
M.~de~Palma,
G.~Gelao,
G.~Iaselli,
G.~Maggi,
M.~Maggi,
N.~Marinelli,
S.~Nuzzo,
A.~Ranieri,
G.~Raso,
F.~Ruggieri,
G.~Selvaggi,
L.~Silvestris,
P.~Tempesta,
A.~Tricomi,$^{3}$
G.~Zito
\nopagebreak
\begin{center}
\parbox{15.5cm}{\sl\samepage
Dipartimento di Fisica, INFN Sezione di Bari, 70126
Bari, Italy}
\end{center}\end{sloppypar}
\vspace{2mm}
\begin{sloppypar}
\noindent
X.~Huang,
J.~Lin,
Q. Ouyang,
T.~Wang,
Y.~Xie,
R.~Xu,
S.~Xue,
J.~Zhang,
L.~Zhang,
W.~Zhao
\nopagebreak
\begin{center}
\parbox{15.5cm}{\sl\samepage
Institute of High-Energy Physics, Academia Sinica, Beijing, The People's
Republic of China$^{8}$}
\end{center}\end{sloppypar}
\vspace{2mm}
\begin{sloppypar}
\noindent
D.~Abbaneo,
R.~Alemany,
U.~Becker,
\mbox{P.~Bright-Thomas},
D.~Casper,
M.~Cattaneo,
F.~Cerutti,
V.~Ciulli,
G.~Dissertori,
H.~Drevermann,
R.W.~Forty,
M.~Frank,
F.~Gianotti,
R.~Hagelberg,
J.B.~Hansen,
J.~Harvey,
P.~Janot,
B.~Jost,
I.~Lehraus,
P.~Mato,
A.~Minten,
L.~Moneta,
A.~Pacheco,
\mbox{J.-F.~Pusztaszeri},$^{20}$
F.~Ranjard,
L.~Rolandi,
D.~Rousseau,
D.~Schlatter,
M.~Schmitt,
O.~Schneider,
W.~Tejessy,
F.~Teubert,
I.R.~Tomalin,
M.~Vreeswijk,
H.~Wachsmuth,
A.~Wagner$^{1}$
\nopagebreak
\begin{center}
\parbox{15.5cm}{\sl\samepage
European Laboratory for Particle Physics (CERN), 1211 Geneva 23,
Switzerland}
\end{center}\end{sloppypar}
\vspace{2mm}
\begin{sloppypar}
\noindent
Z.~Ajaltouni,
F.~Badaud
G.~Chazelle,
O.~Deschamps,
A.~Falvard,
C.~Ferdi,
P.~Gay,
C.~Guicheney,
P.~Henrard,
J.~Jousset,
B.~Michel,
S.~Monteil,
\mbox{J-C.~Montret},
D.~Pallin,
P.~Perret,
F.~Podlyski,
J.~Proriol,
P.~Rosnet
\nopagebreak
\begin{center}
\parbox{15.5cm}{\sl\samepage
Laboratoire de Physique Corpusculaire, Universit\'e Blaise Pascal,
IN$^{2}$P$^{3}$-CNRS, Clermont-Ferrand, 63177 Aubi\`{e}re, France}
\end{center}\end{sloppypar}
\vspace{2mm}
\begin{sloppypar}
\noindent
T.~Fearnley,
J.D.~Hansen,
J.R.~Hansen,
P.H.~Hansen,
B.S.~Nilsson,
B.~Rensch,
A.~W\"a\"an\"anen
\begin{center}
\parbox{15.5cm}{\sl\samepage
Niels Bohr Institute, 2100 Copenhagen, Denmark$^{9}$}
\end{center}\end{sloppypar}
\vspace{2mm}
\begin{sloppypar}
\noindent
G.~Daskalakis,
A.~Kyriakis,
C.~Markou,
E.~Simopoulou,
A.~Vayaki
\nopagebreak
\begin{center}
\parbox{15.5cm}{\sl\samepage
Nuclear Research Center Demokritos (NRCD), Athens, Greece}
\end{center}\end{sloppypar}
\vspace{2mm}
\begin{sloppypar}
\noindent
A.~Blondel,
\mbox{J.-C.~Brient},
F.~Machefert,
A.~Roug\'{e},
M.~Rumpf,
A.~Valassi,$^{6}$
H.~Videau
\nopagebreak
\begin{center}
\parbox{15.5cm}{\sl\samepage
Laboratoire de Physique Nucl\'eaire et des Hautes Energies, Ecole
Polytechnique, IN$^{2}$P$^{3}$-CNRS, 91128 Palaiseau Cedex, France}
\end{center}\end{sloppypar}
\vspace{2mm}
\begin{sloppypar}
\noindent
T.~Boccali,
E.~Focardi,
G.~Parrini,
K.~Zachariadou
\nopagebreak
\begin{center}
\parbox{15.5cm}{\sl\samepage
Dipartimento di Fisica, Universit\`a di Firenze, INFN Sezione di Firenze,
50125 Firenze, Italy}
\end{center}\end{sloppypar}
\vspace{2mm}
\begin{sloppypar}
\noindent
R.~Cavanaugh,
M.~Corden,
C.~Georgiopoulos,
T.~Huehn,
D.E.~Jaffe
\nopagebreak
\begin{center}
\parbox{15.5cm}{\sl\samepage
Supercomputer Computations Research Institute,
Florida State University,
Tallahassee, FL 32306-4052, USA $^{13,14}$}
\end{center}\end{sloppypar}
\vspace{2mm}
\begin{sloppypar}
\noindent
A.~Antonelli,
G.~Bencivenni,
G.~Bologna,$^{4}$
F.~Bossi,
P.~Campana,
G.~Capon,
V.~Chiarella,
G.~Felici,
P.~Laurelli,
G.~Mannocchi,$^{5}$
F.~Murtas,
G.P.~Murtas,
L.~Passalacqua,
M.~Pepe-Altarelli
\nopagebreak
\begin{center}
\parbox{15.5cm}{\sl\samepage
Laboratori Nazionali dell'INFN (LNF-INFN), 00044 Frascati, Italy}
\end{center}\end{sloppypar}
\vspace{2mm}
\begin{sloppypar}
\noindent
L.~Curtis,
S.J.~Dorris,
A.W.~Halley,
J.G.~Lynch,
P.~Negus,
V.~O'Shea,
C.~Raine,
J.M.~Scarr,
K.~Smith,
P.~Teixeira-Dias,
A.S.~Thompson,
E.~Thomson,
F.~Thomson
\nopagebreak
\begin{center}
\parbox{15.5cm}{\sl\samepage
Department of Physics and Astronomy, University of Glasgow, Glasgow G12
8QQ,United Kingdom$^{10}$}
\end{center}\end{sloppypar}
\vspace{2mm}
\begin{sloppypar}
\noindent
O.~Buchm\"uller,
S.~Dhamotharan,
C.~Geweniger,
G.~Graefe,
P.~Hanke,
G.~Hansper,
V.~Hepp,
E.E.~Kluge,
A.~Putzer,
J.~Sommer,
K.~Tittel,
S.~Werner,
M.~Wunsch
\begin{center}
\parbox{15.5cm}{\sl\samepage
Institut f\"ur Hochenergiephysik, Universit\"at Heidelberg, 69120
Heidelberg, Fed.\ Rep.\ of Germany$^{16}$}
\end{center}\end{sloppypar}
\vspace{2mm}
\begin{sloppypar}
\noindent
R.~Beuselinck,
D.M.~Binnie,
W.~Cameron,
P.J.~Dornan,
M.~Girone,
S.~Goodsir,
E.B.~Martin,
A.~Moutoussi,
J.~Nash,
J.K.~Sedgbeer,
P.~Spagnolo,
M.D.~Williams
\nopagebreak
\begin{center}
\parbox{15.5cm}{\sl\samepage
Department of Physics, Imperial College, London SW7 2BZ,
United Kingdom$^{10}$}
\end{center}\end{sloppypar}
\vspace{2mm}
\begin{sloppypar}
\noindent
V.M.~Ghete,
P.~Girtler,
E.~Kneringer,
D.~Kuhn,
G.~Rudolph
\nopagebreak
\begin{center}
\parbox{15.5cm}{\sl\samepage
Institut f\"ur Experimentalphysik, Universit\"at Innsbruck, 6020
Innsbruck, Austria$^{18}$}
\end{center}\end{sloppypar}
\vspace{2mm}
\begin{sloppypar}
\noindent
A.P.~Betteridge,
C.K.~Bowdery,
P.G.~Buck,
P.~Colrain,
G.~Crawford,
A.J.~Finch,
F.~Foster,
G.~Hughes,
R.W.L.~Jones,
E.P.~Whelan,
M.I.~Williams
\nopagebreak
\begin{center}
\parbox{15.5cm}{\sl\samepage
Department of Physics, University of Lancaster, Lancaster LA1 4YB,
United Kingdom$^{10}$}
\end{center}\end{sloppypar}
\vspace{2mm}
\begin{sloppypar}
\noindent
I.~Giehl,
C.~Hoffmann,
K.~Jakobs,
K.~Kleinknecht,
G.~Quast,
B.~Renk,
E.~Rohne,
H.-G.~Sander,
P.~van~Gemmeren,
C.~Zeitnitz
\nopagebreak
\begin{center}
\parbox{15.5cm}{\sl\samepage
Institut f\"ur Physik, Universit\"at Mainz, 55099 Mainz, Fed.\ Rep.\
of Germany$^{16}$}
\end{center}\end{sloppypar}
\vspace{2mm}
\begin{sloppypar}
\noindent
J.J.~Aubert,
C.~Benchouk,
A.~Bonissent,
G.~Bujosa,
J.~Carr,
P.~Coyle,
C.~Diaconu,
A.~Ealet,
D.~Fouchez,
O.~Leroy,
F.~Motsch,
P.~Payre,
M.~Talby,
A.~Sadouki,
M.~Thulasidas,
A.~Tilquin,
K.~Trabelsi
\nopagebreak
\begin{center}
\parbox{15.5cm}{\sl\samepage
Centre de Physique des Particules, Facult\'e des Sciences de Luminy,
IN$^{2}$P$^{3}$-CNRS, 13288 Marseille, France}
\end{center}\end{sloppypar}
\vspace{2mm}
\begin{sloppypar}
\noindent
M.~Aleppo, 
M.~Antonelli,
F.~Ragusa
\nopagebreak
\begin{center}
\parbox{15.5cm}{\sl\samepage
Dipartimento di Fisica, Universit\`a di Milano e INFN Sezione di
Milano, 20133 Milano, Italy.}
\end{center}\end{sloppypar}
\vspace{2mm}
\begin{sloppypar}
\noindent
R.~Berlich,
W.~Blum,
V.~B\"uscher,
H.~Dietl,
G.~Ganis,
C.~Gotzhein,
H.~Kroha,
G.~L\"utjens,
G.~Lutz,
C.~Mannert,
W.~M\"anner,
\mbox{H.-G.~Moser},
R.~Richter,
\mbox{A.~Rosado-Schlosser},
S.~Schael,
R.~Settles,
H.~Seywerd,
H.~Stenzel,
W.~Wiedenmann,
G.~Wolf
\nopagebreak
\begin{center}
\parbox{15.5cm}{\sl\samepage
Max-Planck-Institut f\"ur Physik, Werner-Heisenberg-Institut,
80805 M\"unchen, Fed.\ Rep.\ of Germany\footnotemark[16]}
\end{center}\end{sloppypar}
\vspace{2mm}
\begin{sloppypar}
\noindent
J.~Boucrot,
O.~Callot,$^{12}$
S.~Chen,
M.~Davier,
L.~Duflot,
\mbox{J.-F.~Grivaz},
Ph.~Heusse,
A.~H\"ocker,
A.~Jacholkowska,
M.M.~Kado,
D.W.~Kim,$^{2}$
F.~Le~Diberder,
J.~Lefran\c{c}ois,
\mbox{A.-M.~Lutz},
\mbox{M.-H.~Schune},
L.~Serin,
E.~Tournefier,
\mbox{J.-J.~Veillet},
I.~Videau,
D.~Zerwas
\nopagebreak
\begin{center}
\parbox{15.5cm}{\sl\samepage
Laboratoire de l'Acc\'el\'erateur Lin\'eaire, Universit\'e de Paris-Sud,
IN$^{2}$P$^{3}$-CNRS, 91405 Orsay Cedex, France}
\end{center}\end{sloppypar}
\vspace{2mm}
\begin{sloppypar}
\noindent
\samepage
P.~Azzurri,
G.~Bagliesi,$^{12}$
S.~Bettarini,
C.~Bozzi,
G.~Calderini,
R.~Dell'Orso,
R.~Fantechi,
I.~Ferrante,
A.~Giassi,
A.~Gregorio,
F.~Ligabue,
A.~Lusiani,
P.S.~Marrocchesi,
A.~Messineo,
F.~Palla,
G.~Rizzo,
G.~Sanguinetti,
A.~Sciab\`a,
G.~Sguazzoni,
J.~Steinberger,
R.~Tenchini,
C.~Vannini,
A.~Venturi,
P.G.~Verdini
\samepage
\begin{center}
\parbox{15.5cm}{\sl\samepage
Dipartimento di Fisica dell'Universit\`a, INFN Sezione di Pisa,
e Scuola Normale Superiore, 56010 Pisa, Italy}
\end{center}\end{sloppypar}
\vspace{2mm}
\begin{sloppypar}
\noindent
G.A.~Blair,
L.M.~Bryant,
J.T.~Chambers,
J.~Coles,
M.G.~Green,
T.~Medcalf,
P.~Perrodo,
J.A.~Strong,
J.H.~von~Wimmersperg-Toeller
\nopagebreak
\begin{center}
\parbox{15.5cm}{\sl\samepage
Department of Physics, Royal Holloway \& Bedford New College,
University of London, Surrey TW20 OEX, United Kingdom$^{10}$}
\end{center}\end{sloppypar}
\vspace{2mm}
\begin{sloppypar}
\noindent
D.R.~Botterill,
R.W.~Clifft,
T.R.~Edgecock,
S.~Haywood,
P.~Maley,
P.R.~Norton,
J.C.~Thompson,
A.E.~Wright
\nopagebreak
\begin{center}
\parbox{15.5cm}{\sl\samepage
Particle Physics Dept., Rutherford Appleton Laboratory,
Chilton, Didcot, Oxon OX11 OQX, United Kingdom$^{10}$}
\end{center}\end{sloppypar}
\vspace{2mm}
\pagebreak
\begin{sloppypar}
\noindent
\mbox{B.~Bloch-Devaux},
P.~Colas,
B.~Fabbro,
G.~Fa\"if,
E.~Lan\c{c}on,
\mbox{M.-C.~Lemaire},
E.~Locci,
P.~Perez,
H.~Przysiezniak,
J.~Rander,
\mbox{J.-F.~Renardy},
A.~Rosowsky,
A.~Roussarie,
A.~Trabelsi,
B.~Vallage
\nopagebreak
\begin{center}
\parbox{15.5cm}{\sl\samepage
CEA, DAPNIA/Service de Physique des Particules,
CE-Saclay, 91191 Gif-sur-Yvette Cedex, France$^{17}$}
\end{center}\end{sloppypar}
\nopagebreak
\vspace{2mm}
\begin{sloppypar}
\noindent
S.N.~Black,
J.H.~Dann,
H.Y.~Kim,
N.~Konstantinidis,
A.M.~Litke,
M.A. McNeil,
G.~Taylor
\nopagebreak
\begin{center}
\parbox{15.5cm}{\sl\samepage
Institute for Particle Physics, University of California at
Santa Cruz, Santa Cruz, CA 95064, USA$^{19}$}
\end{center}\end{sloppypar}
\vspace{2mm}
\begin{sloppypar}
\noindent
C.N.~Booth,
C.A.J.~Brew,
S.~Cartwright,
F.~Combley,
M.S.~Kelly,
M.~Lehto,
J.~Reeve,
L.F.~Thompson
\nopagebreak
\begin{center}
\parbox{15.5cm}{\sl\samepage
Department of Physics, University of Sheffield, Sheffield S3 7RH,
United Kingdom$^{10}$}
\end{center}\end{sloppypar}
\vspace{2mm}
\begin{sloppypar}
\noindent
K.~Affholderbach,
A.~B\"ohrer,
S.~Brandt,
G.~Cowan,
J.~Foss,
C.~Grupen,
L.~Smolik,
F.~Stephan 
\nopagebreak
\begin{center}
\parbox{15.5cm}{\sl\samepage
Fachbereich Physik, Universit\"at Siegen, 57068 Siegen,
 Fed.\ Rep.\ of Germany$^{16}$}
\end{center}\end{sloppypar}
\vspace{2mm}
\begin{sloppypar}
\noindent
M.~Apollonio,
L.~Bosisio,
R.~Della~Marina,
G.~Giannini,
B.~Gobbo,
G.~Musolino
\nopagebreak
\begin{center}
\parbox{15.5cm}{\sl\samepage
Dipartimento di Fisica, Universit\`a di Trieste e INFN Sezione di Trieste,
34127 Trieste, Italy}
\end{center}\end{sloppypar}
\vspace{2mm}
\begin{sloppypar}
\noindent
J.~Putz,
J.~Rothberg,
S.~Wasserbaech,
R.W.~Williams
\nopagebreak
\begin{center}
\parbox{15.5cm}{\sl\samepage
Experimental Elementary Particle Physics, University of Washington, WA 98195
Seattle, U.S.A.}
\end{center}\end{sloppypar}
\vspace{2mm}
\begin{sloppypar}
\noindent
S.R.~Armstrong,
E.~Charles,
P.~Elmer,
D.P.S.~Ferguson,
Y.~Gao,
S.~Gonz\'{a}lez,
T.C.~Greening,
O.J.~Hayes,
H.~Hu,
S.~Jin,
P.A.~McNamara III,
J.M.~Nachtman,$^{21}$
J.~Nielsen,
W.~Orejudos,
Y.B.~Pan,
Y.~Saadi,
I.J.~Scott,
J.~Walsh,
Sau~Lan~Wu,
X.~Wu,
J.M.~Yamartino,
G.~Zobernig
\nopagebreak
\begin{center}
\parbox{15.5cm}{\sl\samepage
Department of Physics, University of Wisconsin, Madison, WI 53706,
USA$^{11}$}
\end{center}\end{sloppypar}
}
\footnotetext[1]{Now at Schweizerischer Bankverein, Basel, Switzerland.}
\footnotetext[2]{Permanent address: Kangnung National University, Kangnung,
Korea.}
\footnotetext[3]{Also at Dipartimento di Fisica, INFN Sezione di Catania,
Catania, Italy.}
\footnotetext[4]{Also Istituto di Fisica Generale, Universit\`{a} di
Torino, Torino, Italy.}
\footnotetext[5]{Also Istituto di Cosmo-Geofisica del C.N.R., Torino,
Italy.}
\footnotetext[6]{Supported by the Commission of the European Communities,
contract ERBCHBICT941234.}
\footnotetext[7]{Supported by CICYT, Spain.}
\footnotetext[8]{Supported by the National Science Foundation of China.}
\footnotetext[9]{Supported by the Danish Natural Science Research Council.}
\footnotetext[10]{Supported by the UK Particle Physics and Astronomy Research
Council.}
\footnotetext[11]{Supported by the US Department of Energy, grant
DE-FG0295-ER40896.}
\footnotetext[12]{Also at CERN, 1211 Geneva 23,Switzerland.}
\footnotetext[13]{Supported by the US Department of Energy, contract
DE-FG05-92ER40742.}
\footnotetext[14]{Supported by the US Department of Energy, contract
DE-FC05-85ER250000.}
\footnotetext[15]{Permanent address: Universitat de Barcelona, 08208 Barcelona,
Spain.}
\footnotetext[16]{Supported by the Bundesministerium f\"ur Bildung,
Wissenschaft, Forschung und Technologie, Fed. Rep. of Germany.}
\footnotetext[17]{Supported by the Direction des Sciences de la
Mati\`ere, C.E.A.}
\footnotetext[18]{Supported by Fonds zur F\"orderung der wissenschaftlichen
Forschung, Austria.}
\footnotetext[19]{Supported by the US Department of Energy,
grant DE-FG03-92ER40689.}
\footnotetext[20]{Now at School of Operations Research and Industrial
Engireering, Cornell University, Ithaca, NY 14853-3801, U.S.A.}
\footnotetext[21]{Now at University of California at Los Angeles (UCLA),
Los Angeles, CA 90024, U.S.A.}
%
%
\setlength{\parskip}{\saveparskip}
\setlength{\textheight}{\savetextheight}
\setlength{\topmargin}{\savetopmargin}
\setlength{\textwidth}{\savetextwidth}
\setlength{\oddsidemargin}{\saveoddsidemargin}
\setlength{\topsep}{\savetopsep}
\normalsize
\newpage
\pagestyle{plain}
\setcounter{page}{1}

\section{Introduction}
The minimal supersymmetric extension of the Standard Model (SM) requires 
that the SM particle content is doubled and an extra Higgs $SU(2)_L$ doublet is added. 
The most general interactions of these particles  invariant under 
the  $SU(3)_c\times SU(2)_L\times U(1)_Y$ gauge
symmetry are those of the Minimal Supersymmetric Standard Model (MSSM) \cite{MSSM} 
plus the additional  superpotential terms \cite{rpsuper}
\begin{equation}
W_{\rpv} = \lam_{ijk}\lle{i}{j}{k}+\lam'_{ijk}\lqd{i}{j}{k}+\lam''_{ijk}\udd{i}{j}{k}.
\label{eqrpv}
\end{equation}
Here $L$ ($Q$) are the lepton (quark) doublet superfields, and
$\Dbar,\Ubar$ ($\Ebar$) are the 
down-like and up-like quark (lepton) singlet superfields, respectively; $\lambda, \lambda', \lambda''$ are Yukawa couplings, and $i,j,k=1,2,3$ are generation indices.
The simultaneous presence of the last two terms leads to rapid proton decay, and
the solution of this problem in the MSSM is to exclude all terms in
Eq.(\ref{eqrpv}) by imposing conservation of R-parity  ($R_p=-1^{3B+L+2S}$)\footnote{Here $B$
  denotes the baryon number, $L$ the lepton number and $S$ the spin of a
  field.}, a discrete multiplicative  quantum number \cite{fayet}. This solution is not unique, and a number of models \cite{rpv.models} predict only a subset of the terms in (\ref{eqrpv}), thus protecting the proton from decay. These alternative solutions are denoted ``R-parity violation''.

R-parity violation has two major consequences for collider searches. Firstly, 
the Lightest Supersymmetric Particle (LSP) is not stable and  decays to SM
particles. Consequently the  signatures are very different from the 
classic missing energy signatures of R-parity conserving models.
And secondly, supersymmetric particles (sparticles) can be produced singly via the $LL{\bar
  E}$ coupling at LEP,  either in  s-channel resonance \cite{resonant.1,resonant.2}, 
or in $\gamma e$ collisions \cite{egamma}, a possibility which is not addressed here. 
This paper focuses on the pair-production of sparticles, 
 which subsequently decay violating R-parity. 
Two simplifying assumptions are made throughout the analysis:
\begin{itemize}
\item{Only one term in  Eq.(\ref{eqrpv}) is non-zero. 
 The analysis presented here is restricted to signals from the $LL{\bar
  E}$ couplings. When the results are translated into limits, 
  it is also assumed that only one of the possible nine
 $\lambda_{ijk}$ couplings\footnote{The $\lambda_{ijk}$ coupling is antisymmetric
 in the $i$ and $j$ indices,  $j>i$ is taken here.} is non-zero. 
 The derived limits correspond to the most conservative choice of the coupling.}
\item{The lifetime of the LSP is negligible, i.e.\ the mean path of flight is less than $1$cm.}
\end{itemize} 
The second assumption restricts the sensitivity of this analysis in  $\lambda$, 
which is however probed well below existing upper-limits from low energy
constraints. No assumption on the nature of the LSP is made.

The reported search results use data collected by the \alephcoll{} detector in
1995-1996 at centre-of-mass energies from 130 to 172\gev{}. 
The total data sample used in the analysis corresponds to
 an integrated recorded luminosity of 27.5\invpb.
 The results complement the previously reported \alephcoll{} searches for
 R-parity violating Supersymmetry (SUSY) at LEP~1 energies \cite{lep1.jfg}, and
 the searches for charginos and neutralinos at energies up to 136\gev{}
 \cite{lep15.jfg}.

The outline of this paper is as follows: after reviewing the phenomenology of R-parity
violating SUSY models and existing limits in Sections~\ref{pheno} and \ref{ex.lims}, a brief description of the \alephcoll{}
detector is given in Section~\ref{aleph.detector}. The data and Monte Carlo (MC)
samples and the search analyses are
described in Sections~\ref{dataandmc} and ~\ref{searches}, and the results and their interpretation
within the MSSM  are discussed in
Section~\ref{results}. Finally conclusions are drawn in Section~\ref{conclusions}. 

\section{Phenomenology}\label{pheno}
Within minimal Supersymmetry all SM fermions have scalar SUSY partners: the
sleptons, sneutrinos and squarks. The SUSY equivalent of the gauge and Higgs
bosons are the charginos and neutralinos, which are the mass eigenstates of the
(${\tilde W^+}, {\tilde H^+}$) and (${\tilde \gamma},{\tilde Z},{\tilde
  H^0_1},{\tilde H^0_2}$) fields, respectively, with obvious notation.
 The lightest SUSY particle takes
a special role in R-parity conserving models: it must be stable
\cite{MSSM}. Cosmological arguments \cite{cosmo} then require it to be neutral,
and the only possible LSP candidates are the neutralino, the sneutrino and the
gravitino. 

If R-parity is violated, the LSP can decay to SM particles,  and the
above cosmological arguments do not apply. 
The LSP candidates relevant to this analysis are the neutralino, the chargino, the
sleptons and the sneutrinos.  Squark LSPs are not considered, since they cannot
  decay directly via the purely leptonic $LL{\bar E}$ operator, and would
  instead have to undergo a 4-body decay,  thus acquire a substantial lifetime
  and fall outside the assumption of negligible lifetime. 
  It is also assumed that gravitinos are heavy enough to effectively decouple. 
 Gluinos, which cannot be the LSP if the gaugino masses are universal at the
 GUT scale \cite{MSSM}, are assumed to be heavy enough to play no role for the phenomenology
 at LEP.

\begin{figure}[t]
\begin{center}
\makebox[\textwidth]{
\epsfig{figure=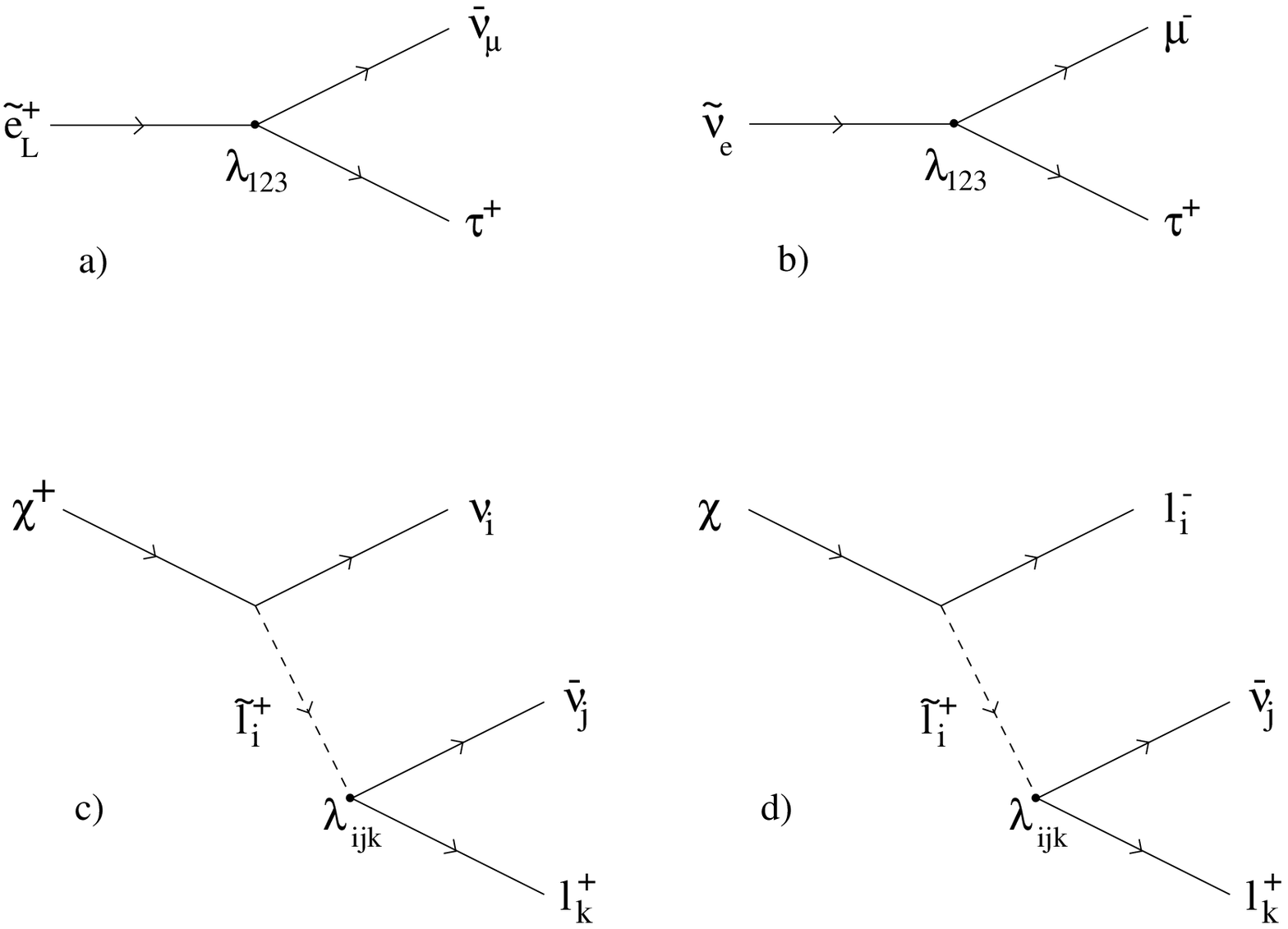,width=.8\textwidth}}
\caption[.]{\em\label{rpv.decays.feyn}{Examples of  R-parity violating 
decays of (a) left-handed selectrons and (b) electron-sneutrinos 
via the coupling $\lambda_{123}$, and  
c) charginos and d) neutralinos via  slepton exchange.}}
\end{center}
\end{figure}

The production cross sections do not depend on the size of the 
R-parity violating Yukawa coupling $\lambda$, since the pair-production
of sparticles only involves gauge couplings\footnote{Ignoring 
 t-channel processes in which the R-parity violating coupling appears twice.}. 
 The sparticle decay modes 
are classified according to their topologies:
all decays proceeding via the lightest neutralino are throughout
referred to as the ``indirect'' decay modes. The final states produced
by the other decays, the ``direct'' decay modes, consist of two
or three leptons as summarised in Table~\ref{rpv.decays}. 
 Fig.~\ref{rpv.decays.feyn}a and b show examples of direct selectron and
 electron-sneutrino decays,
 Fig.~\ref{rpv.decays.feyn}c and d show 
 examples of a direct chargino decay and a neutralino decay via slepton exchange. Note that
  the classification into direct decay modes is made on the basis of the
  topology of the decay, and it is therefore immaterial whether the exchanged
  slepton (or sneutrino) in the chargino or neutralino decays  
  is real or virtual.
 
\begin{table}
\centering
\begin{tabular}{|c|c|}
\hline 
Sparticle  &  Decay Mode ($\lambda_{ijk}$)\\
\hline 
$\chi^+$ & $\nu_i \nu_j l^+_k$ , $l^+_i l^+_j l^-_k$ , $l^+_i \nu_j \nu_k$ , $\nu_i l^+_j \nu_k$ \\
$\chi$ &  ${\bar \nu}_i l^+_j l^-_k$ , ${\bar \nu}_j l^+_i l^-_k$ , $\nu_i
l^-_j l^+_k$ , $\nu_j l^-_i l^+_k$\\
${\tilde l}^-_{iL}$ & $\nu_j l^-_k$ \\
 ${\tilde l}^-_{kR}$ & $\nu_i l^-_j$ , $\nu_j l^-_i$ \\
 ${\tilde \nu}_{i}$ & $l^-_j l^+_k$ \\
\hline
\end{tabular}
\caption[.]{\em  Direct R-parity violating decay modes for a non-zero coupling
  $\lambda_{ijk}$. 
  Here
  $i,j,k$ are generation indices. For example,  the electron
  sneutrino can decay via the coupling  $\lambda_{123}$ to ${\tilde \nu}_{e} \ra \mu^- \tau^+$.}
\label{rpv.decays}
\end{table}

The branching ratios  of the direct to indirect decay modes  explicitly depend  on the
{\em a priori} unknown size of the Yukawa coupling $\lambda$, the masses and couplings
of the decaying sparticle and the lighter SUSY states, 
and the nature of the LSP \cite{rpvneut}.  For example, charginos 
dominantly decay directly if the sleptons and sneutrinos are lighter than the
 lightest neutralino\footnote{In some particular cases, a subset of the direct
   decays of the gauginos are not possible with a single non-zero
   coupling~$\lambda_{ijk}$ since  
  gauginos can decay to sleptons (or sneutrinos) of {\it all three} flavours. 
  In these instances at least two couplings must be non-zero,
  although one of the couplings may be much smaller
  than the other.}, independent of the size of the coupling
 $\lambda$. If the masses of the sleptons or
  sneutrinos lie between the mass of the chargino and the lightest neutralino, the direct decays
 of charginos can   dominate for large values of the R-parity violating
 coupling and  if the neutralino couples higgsino-like. 
 In another example the direct decays of  right-handed 
 sleptons can dominate even when the neutralino is the LSP provided the
 R-parity violating coupling is large and the neutralino couples higgsino-like. 
 In order to be as model independent as possible,   all topologies arising from both
 classes of decays  are considered in the subsequent analyses.

Following the above terminology, the lightest neutralino can decay 
{\em directly} to two  leptons\footnote{In the following the term ``lepton''
  shall denote ``charged lepton''.}
 and a neutrino,
either via 2-body decays to lighter sleptons or sneutrinos, or via a 3-body
decay. The flavours of the decay products of the neutralino depend on the flavour structure
of the Yukawa coupling $\lambda_{ijk}$.
 Heavier neutralinos can also decay {\em indirectly} to the lightest neutralino:
 $\chi' \ra \Pf\bar{\Pf} \chi$.  

The chargino can decay {\em indirectly} to the neutralino:
$\chi^+ \ra \Pf\bar{\Pf'} \chi$. The chargino can also decay {\em directly} 
to SM particles: $\chi^+ \ra l^+ l^- l^+$ or $\chi^+ \ra \nu
 \nu  l^+$. This typically happens when the sleptons/sneutrinos are
 lighter than the chargino, or when the chargino is the LSP.
 Throughout this paper the gauge unification condition \cite{MSSM}
\beq
{M_1 = \frac{5}{3} \tan^2\theta_W M_2} \label{gauge.uni}
\eeq
is assumed. Under this assumption the chargino cannot be the LSP if $M_{\chi^+}>45\gevcc$ --
 the LEP~1 chargino mass limit \cite{lep1.jfg}--, but it is noted that the
 search analyses cover chargino LSP topologies.

Sleptons and sneutrinos can decay {\em indirectly} to the lightest neutralino:
${\tilde l} \ra l \chi$ and ${\tilde \nu} \ra \nu \chi$. If the chargino is
lighter than the sleptons or sneutrinos, the decays ${\tilde l} \ra \nu \chi^+$
and ${\tilde \nu} \ra l^- \chi^+$ are viable decay modes. 
In the following the
decays to charginos  are not considered, since the chargino mass limit derived in
Section~\ref{chargino.limit} is beyond the slepton and sneutrino masses of 
interest to this analysis.
If the sleptons or sneutrinos are the LSPs, pairs of sleptons can decay {\em
 directly} to acoplanar leptons,
and pairs of sneutrinos to four-lepton final states. 

Stops and sbottoms are the most likely candidates for the lightest
scalar quark states because of the potential for large mixing angles between
the left and right handed states, and because of the large  Yukawa couplings 
of the third generation quarks.  They can  decay {\em indirectly} to the lightest
neutralino: e.g. ${\tilde t} \ra c \chi$,  $\tilde b
\ra b \chi$. For the decays to the chargino similar remarks apply as for the
sleptons and sneutrinos. The squarks cannot decay {\em directly}
to SM particles at tree level via the purely leptonic $LL{\bar E}$ coupling.

\section{Existing Limits and the LSP Decay Length}\label{ex.lims}
The  lower limits  on sparticle masses from  precision measurements 
of the Z-width and direct searches
at LEP~1 \cite{lep1.jfg} are: $M_{\chi^+},M_{\tilde l},M_{\tilde \nu}>M_Z/2$,
 and $M_{\tilde t},M_{\tilde b}>M_Z/2$ for negligible
mixing between left-right squark states, and $M_{\tilde t} > 41\gevcc{}$
in the most conservative mixing scenario ($\phi_{mix}^{\tilde t}=56^{\circ}$).
Furthermore, from the \alephcoll{} searches at $\sqrt{s}=130$--136~GeV
(LEP~1.5) $M_{\chi^+}>65\gevcc$ \cite{lep15.jfg}, assuming the lightest
neutralino to be the LSP.

\begin{figure}[t]
\begin{center}
\makebox[\textwidth]{
\epsfig{figure=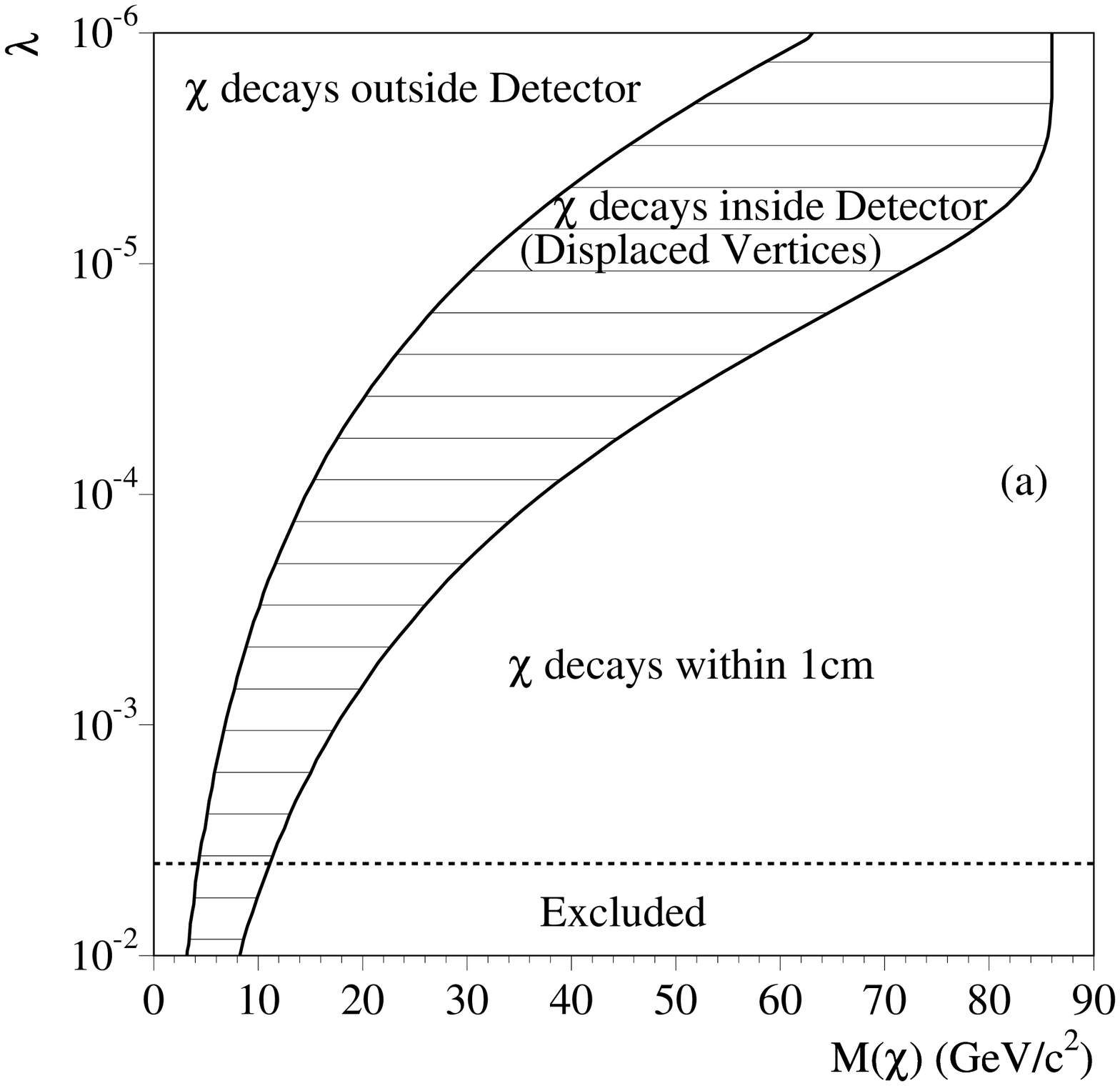,width=0.5\textwidth}\hfill
\epsfig{figure=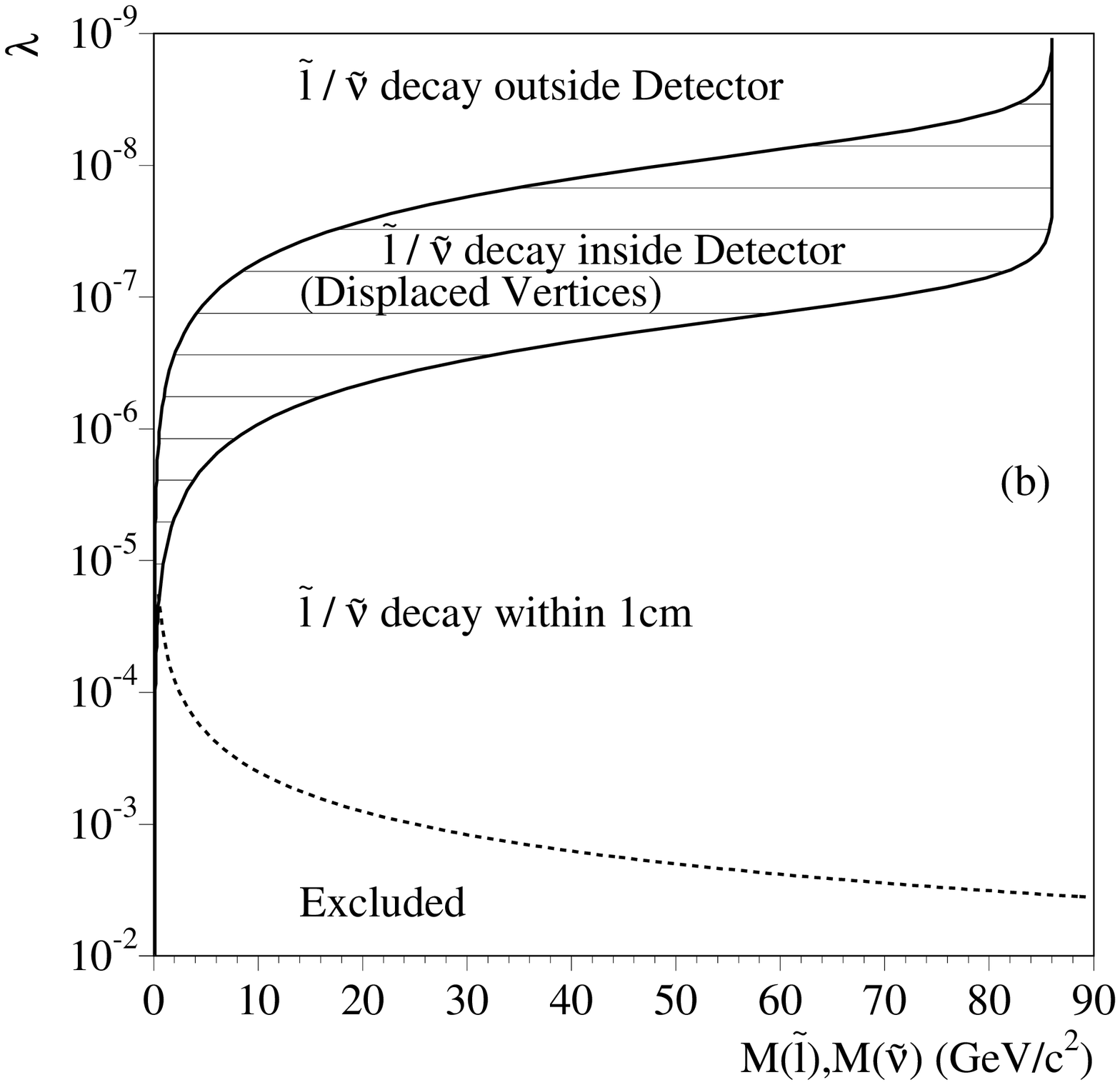,width=0.5\textwidth}}
\caption[.]{\em\label{decay.length}{Regions in the $(\tilde M, \lambda)$-plane
    where  pair-produced LSPs at $\sqrt{s}=172\gev$ have a
    mean decay length of  $X<1\cm$, $1\cm <X< 3\m$ (displaced vertices), and
    $3\m<X$ (LSP decays outside detector) for a) neutralinos (with $M_{\tilde
    f}=100\gevcc$) and b) sleptons and
    sneutrinos. The dashed lines show the
   low energy limit on $\lambda_{133}$ from Eq.(\ref{lam133limit}).}}
\end{center}
\end{figure}

In addition to the above SUSY mass limits,
 upper-bounds on the size of the coupling $\lambda$  from low energy constraints
 exist \cite{resonant.1}. The most stringent limit requires \cite{neutrino.majorana}:
\beq
\lambda_{133} < 0.004 \left( \frac{M_{\tilde l}} {100 \gevcc} \right) \label{lam133limit}
\eeq
The coupling strength determines the mean decay length $X$ of the {\em direct}
 decays of the LSP,  which is given by \cite{dawson}
\barr
X_{\chi} (\mathrm{cm}) &=&   0.3 \lambda^{-2} \left(\frac {M_{\tilde f}} {100
 \gevcc}\right)^4 \left(\frac {\gevcc} {M_{\chi}}\right)^5 (\beta \gamma),  \nonumber \\
X_{\tilde l} , X_{\tilde \nu} (\mathrm{cm}) &=&  10^{-12} \lambda^{-2} \left({\frac{\gevcc
 }{M_{{\tilde l},{\tilde \nu}}}}\right) (\beta \gamma),  \label{eq.decay.length}
\earr
for neutralino and slepton/sneutrino decays, respectively, where the Lorentz factor
$\beta \gamma = p/M$. Fig.~\ref{decay.length} shows 
 regions  where the LSP decays within $X<1\cm$,  
 i.e.\ the region applicable to  this analysis,  together with the limit from
 Eq.(\ref{lam133limit}).  Also indicated in Fig.~\ref{decay.length} are the regions 
 where the LSP  decays within the detector but with a mean decay length
  exceeding $1\cm$, therefore producing displaced
 vertices, and regions where the LSP decays outside the detector. In the latter
 case the signatures are identical to the R-parity conserving signals if the LSP
 is neutral and weakly interacting (neutralino, sneutrino), or they resemble heavy 
 stable charged particle signatures if the LSP is charged  (slepton, chargino).

 The assumption of negligible
 lifetime  restricts the sensitivity of this analysis to neutralino masses exceeding 
 $M_{\chi}\gsim$ 10\gevcc. Close to the kinematic limit, gauginos can be
 probed down to $\lambda \gsim 10^{-5}$ for $M_{\tilde f} = 100 \gevcc$, 
 and sleptons and sneutrinos down to $\lambda \gsim 10^{-7}$.

\section{\label{aleph.detector}The ALEPH Detector}
The \alephcoll{} detector is described in detail in
Ref.~\cite{bib:detectorpaper}. An account of the performance of the
detector and a description of the standard analysis algorithms can be
found in Ref.~\cite{bib:performancepaper}. Here, only a brief
description of the detector components and the algorithms relevant for
this analysis is given.

The trajectories of charged particles are measured with
a silicon
vertex detector, a cylindrical drift chamber, and a large time
projection chamber (TPC). They are immersed in a 1.5~T axial field provided
by a superconducting solenoidal coil.
The electromagnetic calorimeter (ECAL), placed between the TPC and the coil,
is a highly segmented sampling calorimeter which is used to identify electrons
and photons and to measure their energy. 
The luminosity monitors extend the calorimetric coverage
down to 24~mrad from the beam axis.
An additional shielding against beam related background installed
before the 1996 running reduces the acceptance by 10~mrad.
The hadron calorimeter (HCAL) consists of the iron return yoke of the magnet
instrumented with streamer tubes. It provides a measurement of hadronic energy
and, together with the external muon chambers, muon identification.

The calorimetry and tracking information are combined
in an energy flow algorithm, classifying a set of
energy flow ``particles'' as photons, neutral
hadrons and charged particles. Hereafter, charged particle tracks
reconstructed with at least four hits in the TPC,
and originating from within a cylinder of length~20~cm and radius~2~cm
coaxial with the beam and centred at the nominal collision point,
will be referred to as {\it good tracks}.

Lepton identification is described in~\cite{bib:performancepaper}.
Electrons are identified
using the transverse and longitudinal shower shapes in ECAL.
Muons are separated from hadrons by their characteristic penetrating
pattern in HCAL and the presence of hits in the muon chambers.

\section{Data and Monte Carlo Samples}\label{dataandmc}
This analysis uses data collected by \alephcoll{} in 1996 at centre-of-mass
energies of 161.3~GeV (11.1\invpb), 170.3~GeV (1.1\invpb) and
172.3~GeV (9.6\invpb). In the search for sfermions the sensitivity is 
increased  by including also the LEP~1.5
data recorded in 1995 at $\sqrt{s}=130$--136~GeV (5.7\invpb).

For the purpose of designing selections and evaluating efficiencies,
samples of signal events for all accessible final states have been
generated using {\tt SUSYGEN}~\cite{susygen} for a wide range of signal
masses. 
A subset of these has been processed through the
full \alephcoll{} detector
simulation and reconstruction programs, whereas efficiencies for
intermediate points have been interpolated using a fast, simplified simulation.

For the stop, the decays via loop diagrams to a charm quark and the lightest
neutralino result in a lifetime  larger than the typical
hadronisation time scale. The scalar bottom can also develop a 
substantial lifetime in certain regions of parameter space. This has been 
taken into account by modifying the {\tt SUSYGEN} MC program 
to allow stops and sbottoms to hadronise prior to their decays according 
to the spectator model \cite{laurent}.

Samples of all major backgrounds have been generated and passed through the
full simulation, corresponding to at least 20 times the collected
luminosity in the data. Events from $\gamma\gamma\to$hadrons, $\ee\to\qq{}$
and four-fermion events
from $\PW\Pe\nu$, $\PZ\gamma^*$ and $\PZ\Pe\Pe$
were produced with {\tt PYTHIA}~\cite{pythia}, with an invariant mass cut
for the resonance of $0.2\gevcc$ for $\PZ\gamma^*$ and $\PW\Pe\nu$,
and $2\gevcc$ for $\PZ\Pe\Pe$.
Pairs of W bosons were generated with {\tt KORALW}~\cite{koralw}.
 Pair production of leptons
was simulated with {\tt UNIBAB}~\cite{unibab} (electrons) and
{\tt KORALZ}~\cite{koralz}
(muons and taus), and the process $\gamma\gamma\to$leptons with
{\tt PHOT02}~\cite{phot02}.

\section{\label{searches}Selection Criteria}
The topologies expected from sparticle pair production
decaying via a dominant $LL{\bar E}$
coupling share the signature of leptons in the final state.
They can consist of as little as two acoplanar leptons in the simplest case,
or they may consist of as many as six leptons plus four neutrinos in the most
complicated case. In addition to the purely leptonic topologies,
the cascade decays of squarks or heavier gauginos into lighter gaugino
states may produce multi-jet and multi-lepton final states. 

In the following sections the selections
of the various topologies are described in turn.
A brief summary of all selections, the
expected number of background events from SM processes, and the number of
candidates selected in the data is shown in Table~\ref{tops}.
\begin{table}
\centering
\begin{tabular}{|l|l|c|c|}
\hline
Selection & signal process & Background   & Data \\
\hline
Six Leptons & $\chi^+\chi^-\to \mathrm{llllll}$ & 0.02 & 0 \\
Six Leptons plus $\emiss$ &
  $\chi^+\chi^-\to\Pl\nu\Pl\nu\chi\chi\to\Pl\nu\Pl\nu\llnu\llnu$ & 0.12 & 0 \\
& $\chi^+\chi^-\to\Pl\nu\chi\Pl\Pl\Pl\to\Pl\nu\llnu\Pl\Pl\Pl$ & & \\
& $\chi'\chi\to\Pl\Pl\chi\chi\to\Pl\Pl\llnu\llnu$  & & \\
& $\slep\slep\to\Pl\chi\Pl\chi\to\Pl\llnu\Pl\llnu$ & & \\
Four  Leptons & $\snu\snu\to\mathrm{llll}$ & 0.90 & 0 \\
Four Leptons plus $\emiss$ & $\chi\chi\to\llnu\llnu$ & 0.47 & 1 \\
& $\chi'\chi\to\nu\nu\chi\chi\to\nu\nu\llnu\llnu$ & &\\
& $\snu\snu\to\nu\chi\nu\chi\to\nu\llnu\nu\llnu$ & &\\
& $\snu\snu\to\nu\chi\Pl\Pl\to\nu\llnu\Pl\Pl$ & &\\
& $\slep\slep\to\Pl\chi\Pl\nu\to\Pl\llnu\Pl\nu$ & &\\
& $\chi^+\chi^-\to\Pl\nu\chi\Pl\nu\nu\to\Pl\nu\llnu\Pl\nu\nu$ & & \\
& $\chi^+\chi^-\to\Pl\nu\nu\Pl\Pl\Pl$ & &\\
Acoplanar Leptons & $\slep\slep\to\Pl\nu\Pl\nu$& 12$^{(*)}$& 15 \\
& $\chi^+\chi^-\to\Pl\nu\nu\Pl\nu\nu$ & &\\
Leptons and Hadrons & $\chi^+\chi^-\to\Pq\Pq\Pq\Pq\chi\chi\to\Pq\Pq\Pq\Pq\llnu\llnu$ & 1.43 & 1 \\
& $\chi^+\chi^-\to\Pq\Pq\Pl\nu\chi\chi\to\Pq\Pq\Pl\nu\llnu\llnu$ & & \\
& $\chi^+\chi^-\to\Pq\Pq\chi\Pl\Pl\Pl\to\Pq\Pq\llnu\Pl\Pl\Pl$ & & \\
& $\chi^+\chi^-\to\Pq\Pq\chi\Pl\nu\nu\to\Pq\Pq\llnu\Pl\nu\nu$ & & \\
& $\chi'\chi\to\Pq\Pq\chi\chi\to\Pq\Pq\llnu\llnu$ & &\\
& $\sq\sq\to\Pq\chi\Pq\chi\to\Pq\llnu\Pq\llnu$ & &\\
\hline
\end{tabular}
\caption[.]{\em The selections, the signal processes giving rise to the above
topologies, the number of background events expected, and the number of
candidate events selected in the data ($\sqrt{s}=130-172\gev$).
The value marked (*) contains 10.3 events of irreducible background from
WW production.}
\label{tops}
\end{table}
The positions of the most important cuts of all selections have been
chosen such that the expected upper limit 
 (~\nbar)  without the presence of a signal is minimised\cite{n95}.  
This minimum was determined using the Monte Carlo
for background and signal, focussing on signal masses close to the
high end of the sensitivity region.

\subsection{Six Leptons}
Six lepton topologies are expected from the production of pairs of charginos,
which  decay via sneutrinos into three leptons each.
To select this topology the analysis 
requires at least five, but no more than nine good tracks, of which
at least four should be identified as
leptons (i.e.\ electrons or muons). To ensure that the tracks are well
separated, the event is clustered into four (and three) jets using the
Durham algorithm, and a minimum Durham scale~$y_4$ of 0.002 (and
$y_3$ of 0.01) is required between all the jets. After this, a total
background of 0.02 events is expected, predominantly
coming from $\ee\to\PZ\ee$.

\subsection{Six Leptons plus Missing Energy}
This topology is expected from the indirect decays of charginos, neutralinos
and sleptons.
The selection requires a visible mass of at least 25\gevcc{} and
at least five, but no more than eleven good tracks, with at least
two of them identified as leptons.
Fig.~\ref{datamc}a shows the distribution of the number of identified
leptons~\nlep{} for data, background Monte Carlo and events from
$\chi^+\chi^-\to\Pl\nu\Pl\nu\llnu\llnu$ at an intermediate stage of
the selection.
In addition the amount of neutral hadronic energy is limited to
$6\%\sqrt{s}$ and $17\%$ of the total energy of all
good tracks.
Since missing energy is expected
for the signal, the events should have a visible mass of less than
$85\%\sqrt{s}$ and a minimum missing transverse momentum of $2\%\sqrt{s}$.
The remaining background from \qq{} and \tautau{} is reduced by requiring $y_4$
to be at least 0.004. The total background after all cuts amounts to
0.12 events expected in the data, mainly
consisting of events from \qq, $\PZ\gamma^*$ and $\PZ\ee$.

\subsection{Four Leptons}
A final state of four leptons is expected from the direct decays of pairs
of sneutrinos.
For the purpose of defining selections, the possible lepton flavour
combinations ($\Pl_i\Pl_k\Pl_i\Pl_k$ or $\Pl_j\Pl_k\Pl_j\Pl_k$) can be
divided into three classes according
to the number of taus: final states with no taus, two taus or four taus. 
For all cases a common preselection is applied, requiring a visible
mass of at least 30\gevcc{} and four, five or six good tracks in the event.
To reject background from \tautau, events are clustered into jets, which should
be well separated ($y_4>4\times 10^{-4}$ and
$y_3>0.007$) and contain at least one good track.
The discriminating power of $y_4$ is illustrated in Fig.~\ref{datamc}b,
comparing the distribution for data, background Monte Carlo and events
from direct sneutrino decays.

For a signal with four taus, the remaining background is reduced further
by requiring that no energy be reconstructed in a cone of $12^{\circ}$
around the beam axis.
This cut introduces an inefficiency due to beam related background
and electronic noise, which
was measured to be 0.5\% (4\%, 2\%) at centre-of-mass energy of
130--136 (161, 172)~GeV,
using events triggered at random beam crossings.
In addition, the amount of neutral hadronic energy~\ehad{} should be less
than 30\% of the visible energy.

The requirements on \ehad{} and energies at low angles can be dropped
for signal final states with two taus
(no taus) by introducing new requirements on the lepton content of the event:
there should be no conversions reconstructed and two muons or
electrons identified (at least three leptons identified),
with a total leptonic energy~\elep{} fulfilling $\ehad < 30\%\elep$
($<15\%\elep$). For two taus, a missing transverse momentum of at least
$2\%\sqrt{s}$ can be required to suppress the remaining background, whereas
for no taus, events should have less than 25\gevc{} of missing momentum
along the beam axis.

For the case of two or more non-zero Yukawa couplings, final states with an
odd number of taus are accessible in sneutrino pair production.
Since the four tau selection contains
no cut on leptonic energy, such final states are selected
at least as efficiently as four tau final states.

The total background expected by the inclusive combination of all three
subselections amounts to 0.90 events.
Most of this background consists of events from $\PZ\gamma^*$ and $\PZ\ee$.

\subsection{Four Leptons plus Missing Energy}
A final state with four leptons of arbitrary flavour and missing energy
can be produced in decays of charginos, neutralinos, sleptons and
sneutrinos (Table~\ref{tops}).
It is selected
using criteria similar to the ones defined for the four-lepton final
states: events should have four, five or six good tracks, of which at
least one should correspond to an identified electron or muon.
A total visible mass of at least 16\gevcc{} and a missing transverse
momentum of more than 5\gevc{} is required. The total neutral hadronic
energy in the event should be less than the total leptonic energy.
The remaining background from \qq{} and \tautau{} is reduced further by
requiring $y_4$ to be greater than $6\times 10^{-4}$.
In addition, events are clustered
into jets using the JADE algorithm and a $y_{\mathrm{cut}}$
of $m_{\tau}^2/s$ to form tau-like jets, at least four of which are
required to contain good tracks. After these cuts, a background
of 0.47 events is expected in the total data sample, mainly consisting
of four-fermion events.

\subsection{Acoplanar Leptons}
Final states with two leptons and missing energy are expected from direct
decays of sleptons and charginos. Depending on the process and the generation
structure of the $LL{\bar E}$ operator, the charged leptons
can be of equal flavour
(e.g.\ left-handed sleptons) or of arbitrary flavour (charginos).
Selections for the topology of two acoplanar leptons have
already been developed for the search for sleptons under
the assumption that R-parity is
conserved: for $\sqrt{s}=130$--136~GeV, the selection described in
\cite{rpc.lep15} is used, whereas for $\sqrt{s}=161$--172~GeV the
analysis published in \cite{slepton.paper} is extended to allow for
mixed lepton flavours.

For $\Pe\mu$ final states, the requirement for two identified leptons
of the same flavour is replaced by the requirement for one electron
and one muon. For $\Pe\tau$ ($\mu\tau$), the leading lepton should
be an electron (muon) with momentum less than 75\gevc{}.
In case there is a second lepton identified, its momentum should be
less than 30\gevc{} at $\sqrt{s}=161$~GeV (25\gevc{} at
$\sqrt{s}=172$~GeV).

All these subselections have irreducible background from leptonic WW events,
which is particularly large when the flavour structure of the signal process
requires to use inclusive combinations of the subselections.
Therefore subtracting this background using the method suggested
in \cite{pdg.subtract} increases the sensitivity of the analysis.

\subsection{Leptons and Hadrons}
Final states with leptons and hadrons are expected from charginos,
neutralinos and squarks decaying to the lightest neutralino.
Depending on the masses of the supersymmetric particles and on the
lepton flavour composition in the neutralino decays, signal events
populate different regions in track multiplicity~\nch,
visible mass~\mvis{} and leptonic energy~\elep. As the properties
of background events change as a function of these variables,
three different subselections have been developed to select the full range
of signal events at a small background level (Table~\ref{hlcuts}).

\begin{figure}
\begin{center}
\makebox[\textwidth]{
\epsfig{figure=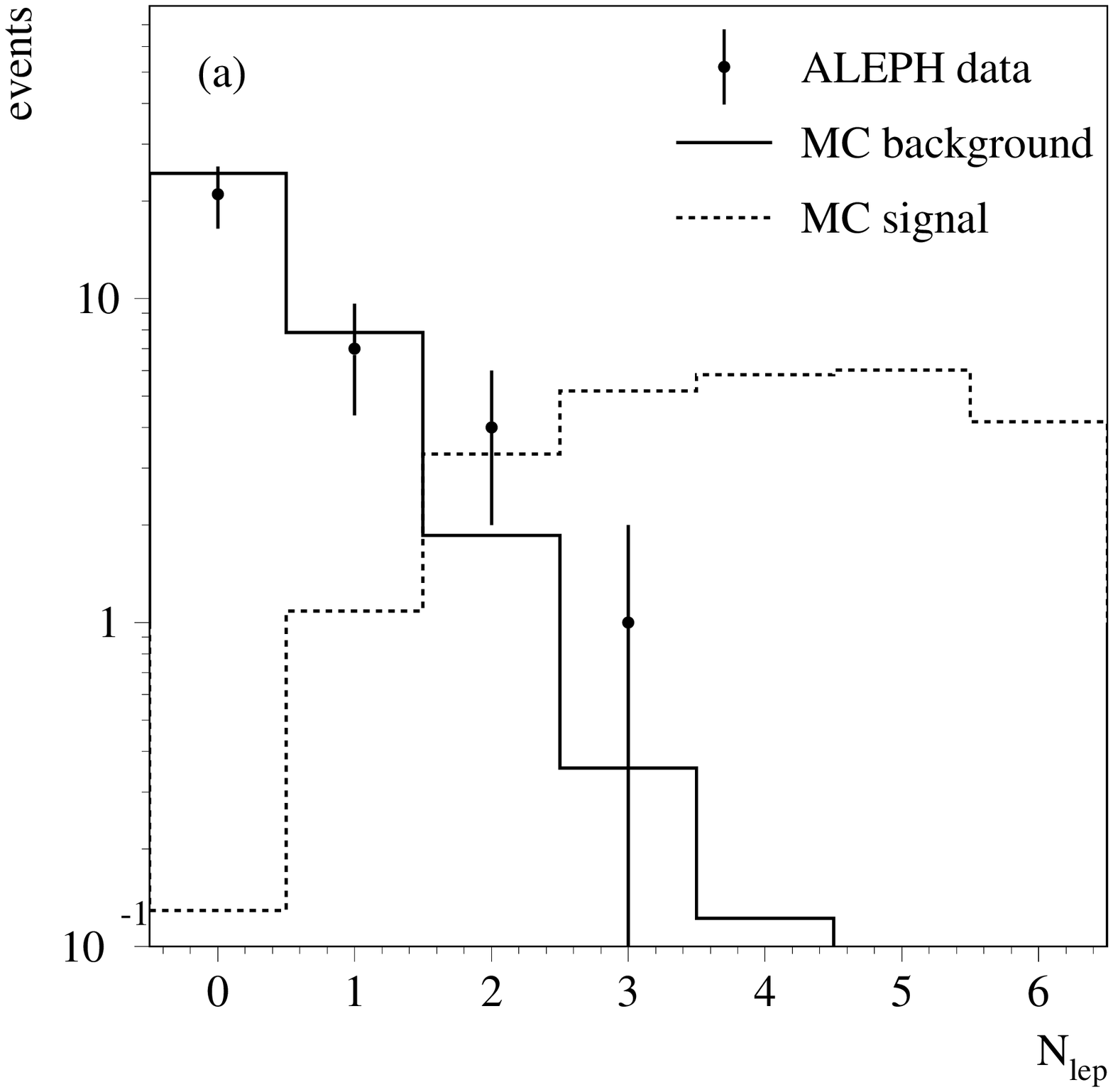,width=0.5\textwidth}\hfill
\epsfig{figure=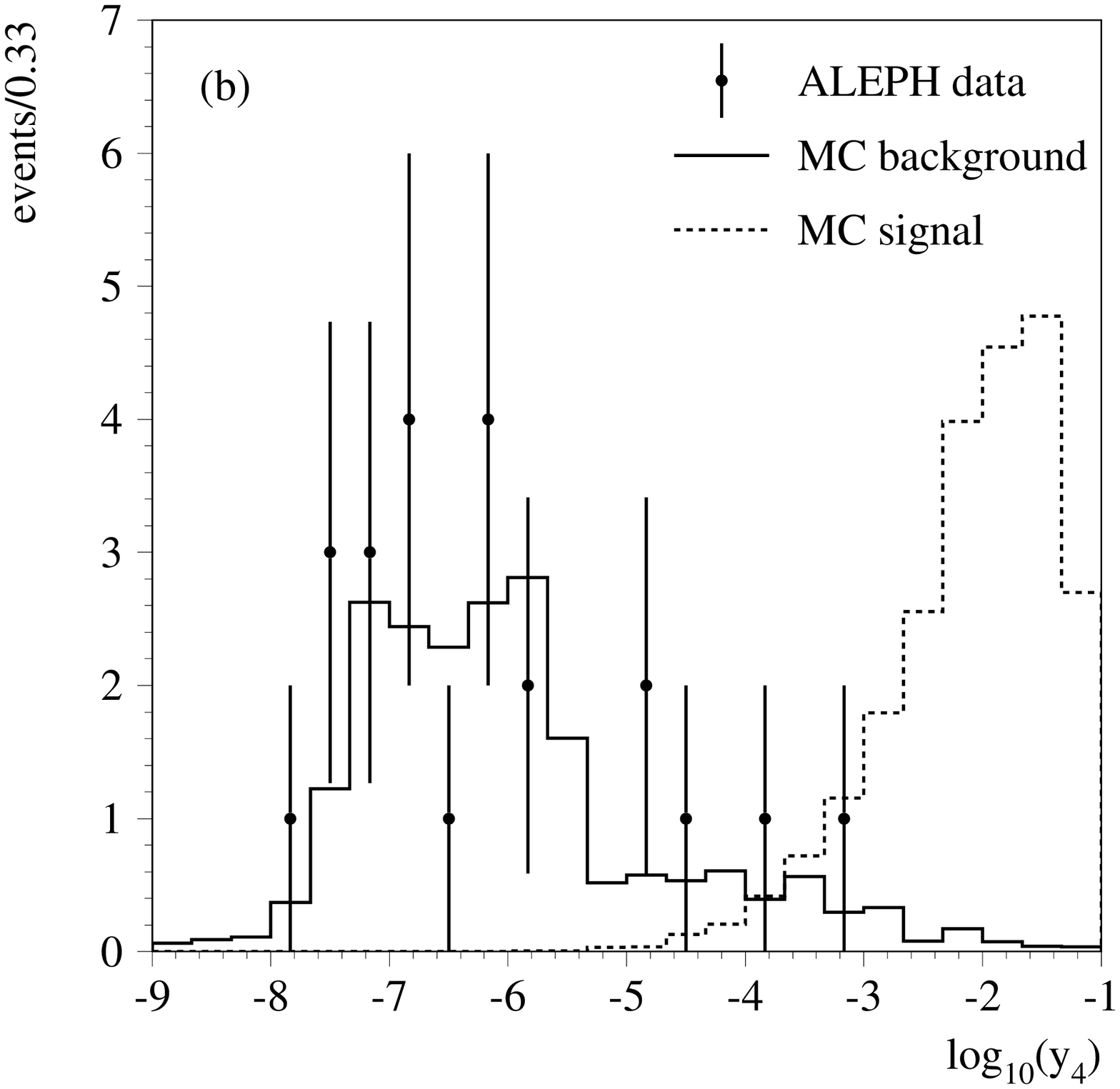,width=0.5\textwidth}}
\makebox[\textwidth]{
\epsfig{figure=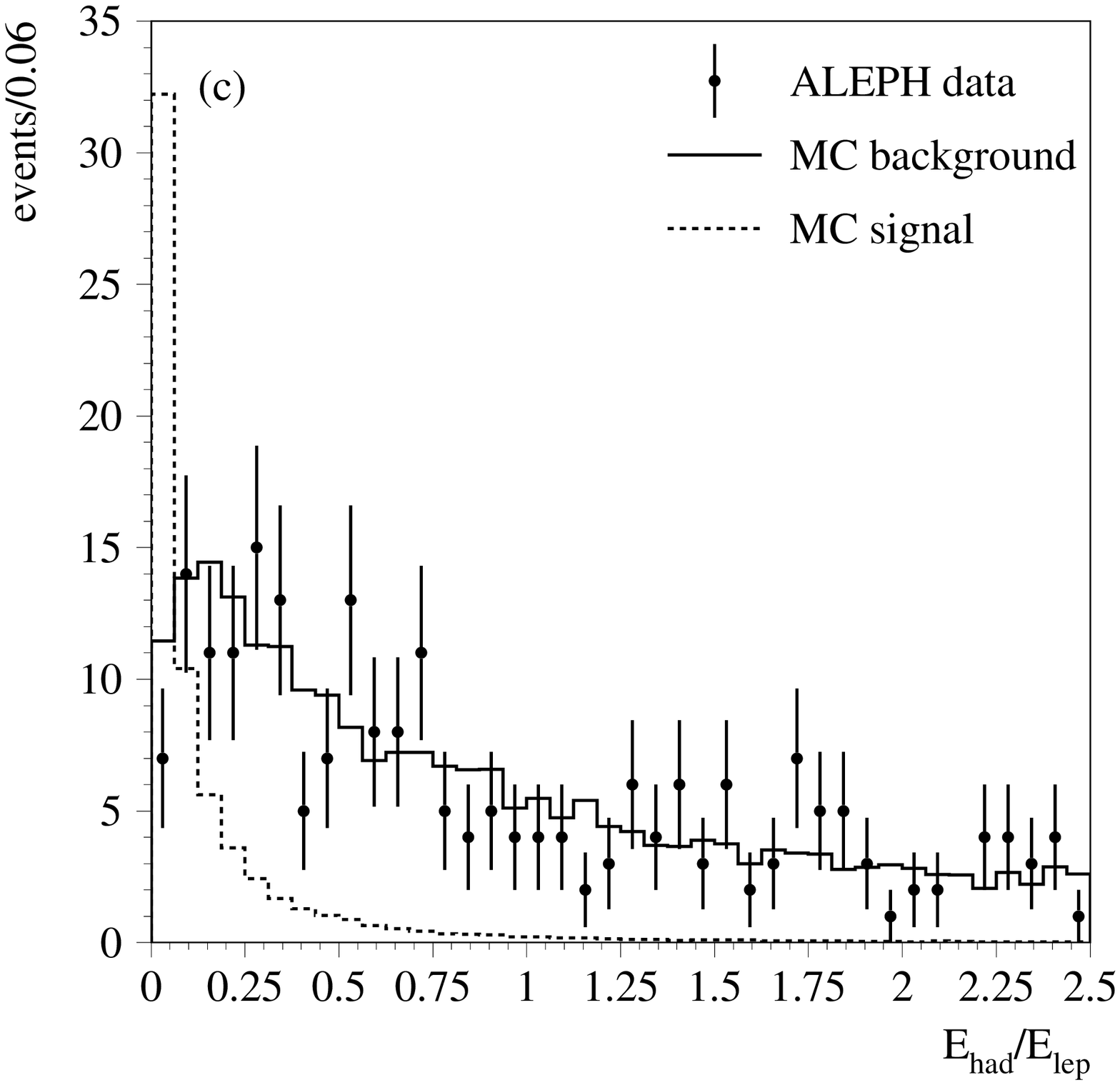,width=0.5\textwidth}\hfill
\epsfig{figure=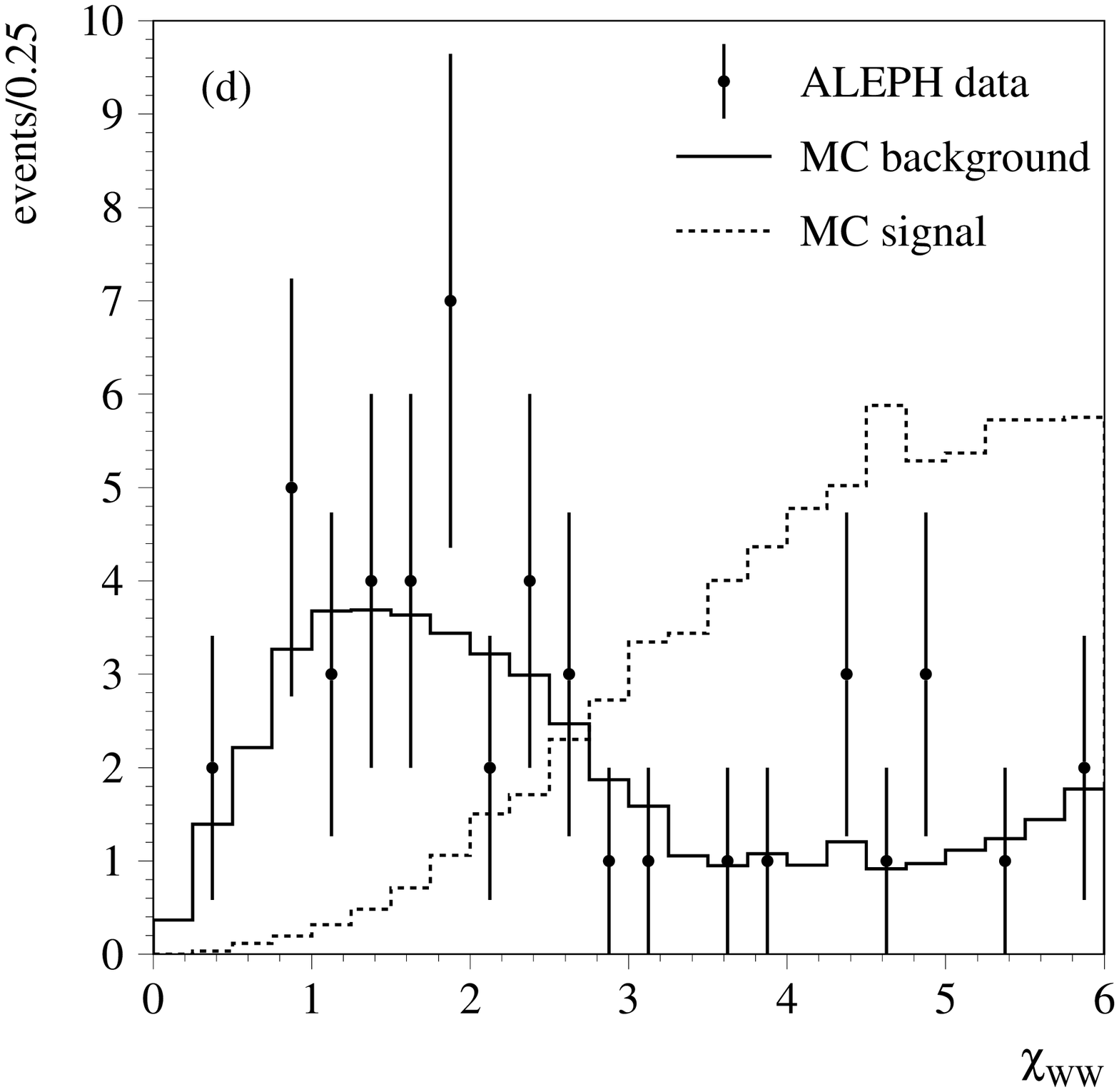,width=0.5\textwidth}}
\caption{\label{datamc}\em The distributions of 
a) number of identified leptons~$N_{lep}$ as used in the ``Six Lepton
plus Missing Energy'' selection
b) $y_4$ as used in the ``Four Lepton'' selection and
c) \ehad/\elep{} and d) $\chi_{WW}$ as used in the ``Leptons and Hadrons''
selection.
The data (dots) at $\sqrt{s}=$161--172~GeV are compared to the
background Monte Carlo (full histograms).
The dashed histograms show typical signal distributions
in arbitrary normalisation: a) $\chi^+\chi^-\to\Pl\nu\Pl\nu\chi\chi$
for $\lambda_{122}$ and $\lambda_{133}$, b) $\snu\snu\to\Pl\Pl\Pl\Pl$ for
all couplings, c) and d) $\chi^+\chi^-\to\Pl\nu\Pq\Pq\chi\chi$ or
$\Pq\Pq\Pq\Pq\chi\chi$ for $\lambda_{122}$ and $\lambda_{133}$.
Only a subset of the cuts is applied to
preserve sufficient statistics.}
\end{center}
\end{figure}

All three subselections are based on the central requirement of
large leptonic energy, supplemented with cuts on the amount of
neutral hadronic energy~\ehad{} (Fig.~\ref{datamc}c) and
non-leptonic energy~\enlep{}.
Due to the presence of at least two neutrinos, signal events are expected
to contain some missing momentum. This is used to suppress the background
by requiring a minimum missing transverse momentum~\ptmiss.
Background from hadronic events with energetic initial state
radiation photons is reduced by removing events with large missing
momentum~$p^{\mathrm{miss}}_{\mathrm z}$ along the beam axis
(for photons escaping
at small polar angles) or by requiring the charged
multiplicity~$\nch^{\mathrm{jet}}$ in all jets
found with $y_{\mathrm{cut}}=0.005$ to be at least one
(for photons in the detector).
Most of the remaining background is then rejected by selecting
spherical events using $y_3, y_4, y_5$
and the event thrust. At this stage background at $\sqrt{s}=161$--172~GeV
dominantly comes from
$\ww\to l\nu\qq$. The kinematic properties of these events can be exploited
to suppress the background by defining
\begin{center}
$\chi^2_{WW} = (\frac{m_{qq}-m_W}{10\gevcc})^2 +
               (\frac{m_{l\nu}-m_W}{10\gevcc})^2 +
               (\frac{p_l-43\gevc}{\Delta p_l})^2$.
\end{center}
Here $m_{qq}$ is the hadronic mass, i.e.\ the mass of the event after
removing the leading
lepton, $m_{l\nu}$ is the mass of the leading lepton and the missing momentum,
and $p_l$ is the momentum of the leading lepton. The spread~$\Delta p_l$ of
lepton momenta from WW is approximated by 5\gevc{} at $\sqrt{s}=161$~GeV
and 5.8\gevc{} at $\sqrt{s}=172$~GeV. As can be seen in Fig.~\ref{datamc}d,
WW events are likely to occur
at small $\chi^2_{WW}$, and can therefore be rejected by requiring a minimum
$\chi^2_{WW}$ for events to be selected.

Subselection~I is designed to select final states with large leptonic
energy and at least two jets, this way covering most of the parameter
space. For charginos decaying to $\Pl\nu\Pq\Pq\chi\chi$ with a small
mass difference between the chargino and the lightest neutralino, the
efficiency is increased with subselection~II, concentrating on events
with small multiplicity and large leptonic energy fraction.
For small masses of the lightest neutralino, signal events tend to have a
smaller  leptonic energy fraction such that additional cuts on the event shape
are needed to suppress the background (subselection~III). 
Final states with hadrons and leptons as expected from chargino, neutralino
and squark decays are efficiently selected  by using the inclusive combination
of all three subselections. The background amounts
to 1.43 events expected in the total data sample.
This background mainly consists of events from $\qq (\gamma)$
and W pair production.

\begin{table}[t]
\begin{center}
\begin{tabular}{|c|c|c|} \hline
 subselection~I & subselection~II & subselection~III\\
\hline \hline
$\nch{} \geq 5$ & $15 \geq \nch{} \geq 5$ & $\nch{} \geq 11$\\
$25\gevcc{} < \mvis{}$ & $20\gevcc{} < \mvis{} < 75\%\sqrt{s}$ & 
  $55\%\sqrt{s} < \mvis{} < 80\%\sqrt{s}$\\
\hline
$\ptmiss>3.5\%\sqrt{s}$ & $\ptmiss>2.5\%\sqrt{s}$ & 
  $\ptmiss>5\%\sqrt{s}$ \\
$|p^{\mathrm{miss}}_{\mathrm z}| < 27\gevc{}$ &  & 
   $\nch^{\mathrm{jet}} \geq 1$\\
\hline
 & $y_3 > 0.009$ & $y_3>0.025$\\
 & $y_4 > 0.0026$ & $y_4>0.012$\\
$y_5 > 0.006$ &  & $y_5 > 0.004$ \\
 & & $\mathrm{Thrust} < 0.85$\\
\hline
$\nlep \geq 1$ & $\nlep \geq 1$ & $\nlep \geq 1$ \\
$\enlep < 54\%\sqrt{s}$ & $\enlep < 70\%\sqrt{s}$ & \\
$\ehad < 28\%\evis$ & $\ehad < 22\%\elep$ & $\elep > 20\% \ehad$\\
\hline
\multicolumn{3}{|c|}
{$\chi_{\mathrm{WW}}>3.3$ (for $\sqrt{s}=161$~GeV),
$\chi_{\mathrm{WW}}>3.5$ (for $\sqrt{s}=172$~GeV)}\\
\hline
\end{tabular}
\caption{\label{hlcuts}\em The complete list of cuts as defined for the leptons and
hadrons selection.}
\end{center}
\end{table}

\section{\label{results}Results}
In the data recorded at $\sqrt{s}=$130--172~GeV, corresponding to an
integrated luminosity of 27.5~\invpb, a total of 17 events is selected.
This is in agreement with the expectation from Standard Model backgrounds
of 14.5 events.
Out of these, 15 events are selected by the acoplanar lepton selection, with a
subtractable background from $\ww\to\Pl\nu\Pl\nu$ of 10.3 events.
All of these events show clear characteristics of WW-events, and
are split up into the different lepton flavours as shown
in Table~\ref{wwcands}. The highest number of candidates is observed in the  
$\mu \tau$ channel (seven candidates with a total expected
background of 2.6), which also shares two candidates with the $e \mu$ channel.
The probability for seeing such an upwards fluctuation in any of the
six channels is $\sim 10\%$.

The other two events are selected by the ``four leptons plus missing energy''
selection and the ``leptons and hadrons'' selection, respectively.
The former is consistent with coming from
$\ee\to\PZ\gamma^*\to\ee\tau^+\tau^-$, whereas the latter can be interpreted
as $\ww\to\Pe\nu\qq$.
\begin{table}
\centering
\begin{tabular}{|c|cccccc|}
\hline 
Topology       & $ee$ & $\mu\mu$ & $\tau\tau$ & $e\mu$ & $e\tau$ & $\mu\tau$ \\
\hline 
WW background  & 1.6  & 1.7      & 1.2   &      3.6 &    2.3     & 2.3  \\ 

Selected in Data &1&1&1&5&3&7\\
\hline 
\end{tabular}
\caption[.]{\em The 15 candidate events
  selected in the data by the acoplanar lepton
  selection, listed according to the topology in 
  which they are selected, and the WW background
  expectation. Some of the background and candidate events are in common
  to several selections.}
\label{wwcands}
\end{table}

In the following sections, the absence of any significant excess of events in the data
with respect to the Standard Model expectation is used to set limits
on the production of charginos and neutralinos, sleptons, sneutrinos and
squarks. The systematic error on the efficiencies is of the order of 3\%,
dominated by the statistical uncertainty due to limited Monte Carlo statistics,
with small additional contributions from lepton identification and
energy flow reconstruction. It is taken into account by conservatively
reducing the selection efficiency by one standard deviation.

\subsection{\label{chargino.limit}Charginos and Neutralinos}
Charginos and heavier neutralinos can decay either {\em indirectly} via the lightest
neutralino, or {\em directly} via (possibly virtual) sleptons or sneutrinos.
The corresponding branching fractions of the direct and indirect decays, as well as
the branching fractions of the direct decays into different leptonic final
states (c.f. Table~\ref{rpv.decays})  in general depend
on the field content and masses of the charginos and neutralinos, the
sfermion mass spectrum and the Yukawa coupling~$\lambda$. Furthermore,
because of possible mixing in the third generation sfermion sector, staus,
stops and sbottoms can be substantially lighter than their first or second
generation partners. The effect of light staus 
 is to increase the tau branching ratio in the indirect decays (e.g. $\chi^+ \ra
 \tau \nu \chi$) with respect to the other indirect decay modes,
 whereas light stops and sbottoms increase the hadronic
 branching ratios of the indirect decays. Light staus can
 also affect the BRs of the direct decay modes, increasing the BRs to $e, \mu$
 or $\tau$ final states depending on the generation structure of the R-parity
 violating couplings. 

 To constrain a model with such a large number of unknown parameters,
 limits were set that are independent of the various branching ratios.
 For this purpose, the signal topologies are classified into three distinct
 cases: the {\it direct topologies} (when both charginos decay directly),
 the {\it indirect topologies} (when both charginos decay indirectly), and
 the {\it mixed topologies} (when one chargino decays directly, one
 indirectly).
 Secondly, the branching ratios of the various decays involved in both
 indirect and direct decays are varied freely, and the limit is set
 using the most conservative choice.

 Limits have been evaluated in the framework of the MSSM, where the
 masses of the gauginos can be calculated from the three parameters
 $M_2,\mu$ and $\tan{\beta}$. The cross sections of 
 neutralinos (charginos) receive a positive (negative) contribution due to
 t-channel selectron  (electron-sneutrino) exchange, respectively, and thus 
 depend also on $m_{\tilde l}$ and $m_{\tilde \nu}$. A common 
 slepton and sneutrino mass~$m_0$ at the GUT scale was assumed, 
 which according to the renormalisation group equations \cite{guts} links
 the slepton and sneutrino masses at the electroweak scale
 by\footnote{Ignoring effects from the R-parity violating couplings.}
\begin{eqnarray}\label{eq:gut}
m^2_{\tilde{l}_R} &=&
  m^2_0 + 0.22\,M^2_{2} - 
  \sin^2\theta_W M^2_{\mathrm{Z}} \cos 2\beta \nonumber \\
m^2_{\tilde{l}_L} &=&
  m^2_0 + 0.75\,M^2_{2} -
  \onehalf (1-2\sin^2\theta_W) M^2_{\mathrm{Z}} \cos 2\beta \nonumber \\
m^2_{\tilde{\nu}} &=&
  m^2_0 + 0.75\,M^2_{2} +
  \onehalf M^2_{\mathrm{Z}} \cos 2\beta.
\end{eqnarray}

 In summary, the limits derived in this approach are 
 independent of the branching ratios of the gauginos, and only depend on the 
 four parameters $M_2, \mu, \tan{\beta}, m_0$, which determine the masses and
 the cross sections of the charginos and neutralinos. Therefore
 the limits are by construction valid for any size or
 generation structure of the R-parity violating coupling $\lambda$, they
 apply for neutralino, slepton or sneutrino LSPs alike, and are independent
 of mixing between the third generation sfermions.
 It should be noted that the branching ratios which set the limit may not
 correspond to a physically viable model in certain cases (i.e.\ in specific
 points in parameter space $M_2, \mu, \tan{\beta}, m_0$), and hence the real
 limit within a specific model may be even stronger than the conservative
 and more general limit presented in this section.

As discussed in Section~\ref{ex.lims}, the lightest neutralino can have
a decay length of more than 1\cm{} when $m_{\chi}\lsim 10\gevcc$ for couplings 
which are not already excluded by low energy constraints.
Since long-lived sparticles are not considered
in this analysis, regions in parameter space with $m_{\chi}< 10\gevcc$
are ignored in the following.
 Limits on the charginos and neutralinos are derived in
 Sections~\ref{dom.ind} and \ref{dom.dir} for the two extreme
 cases of 100\% indirect and 100\%  direct topologies, respectively,
 and the intermediate case of mixed topologies is investigated in
 Section~\ref{dom.mix}. 
Due to the large cross section for pair production of charginos, the data
recorded at $\sqrt{s}=$130--136~GeV do not improve the sensitivity
of the analysis, and therefore have not been included here. 

\subsubsection{Dominance of indirect decays \label{dom.ind}}
In this scenario all charginos and neutralinos are assumed to decay
to the lightest neutralino, which then decays violating R-parity
into two charged leptons and a neutrino.
The indirect topologies generally correspond to the cases where the
sleptons and sneutrinos are heavier than the charginos and
the neutralinos. When the sleptons or sneutrinos are lighter than the
charginos (or the heavier neutralinos) and heavier than the
lightest neutralino, the indirect decays will also  
dominate provided that the neutralino couples gaugino-like and/or the
coupling $\lambda$ is small.

For charginos the
``Leptons and Hadrons'' selection is combined with the
``Six Leptons plus Missing Energy'' selection, and for
neutralinos ($\chi \chi$) and $(\chi' \chi$) the inclusive combination of
the ``Leptons and Hadrons'' and the ``Four
and Six Leptons plus Missing Energy'' analyses was used.
Signal efficiencies were determined as
a function of $M_{\chi^+}, M_{\chi'}, M_{\chi}$ and the choice
of generation indices $i,j,k$  of the coupling $\lambda_{ijk}$.
In general, efficiencies scale with the mass of the lightest neutralino,
and become smaller with decreasing neutralino mass. Final states
with a large number of electrons or muons are selected with high efficiencies
compared to processes involving hadronic decays or couplings
allowing the lightest neutralino to decay into taus.
A set of efficiencies for choices of the lepton flavour
corresponding to the smallest efficiencies is shown in Table~\ref{effics}.
\begin{table}
\centering
\begin{tabular}{|l|c|c|c|}
\hline
Signal Process & Topology & Masses (\gevcc) & Efficiency (\%) \\
\hline
$\chi^+\chi^-\to\PW^*\PW^*\tau\tau\nu\tau\tau\nu$ & indirect &
  $m_{\chi^+}=85$, $m_{\chi}=30$ & 40\\
 & & $m_{\chi^+}=85$, $m_{\chi}=70$  & 48\\
$\chi^+\chi^-\to\Pe\tau\tau\Pe\tau\tau$ & direct & $m_{\chi^+}=85$ & 73\\
$\chi^+\chi^-\to\tau\nu\nu\tau\nu\nu$ & direct & $m_{\chi^+}=85$ & 44\\
$\chi^+\chi^-\to\tau\nu\nu\PW^*\Pe\tau\nu$ & mixed &
  $m_{\chi^+}=85$, $m_{\chi}=30$ & 53\\
\hline
$\chi'\chi\to\PZ^*\tau\tau\nu\tau\tau\nu$ & indirect &
 $m_{\chi'}=95$, $m_{\chi}=75$ & 47\\
$\chi\chi\to\tau\tau\nu\tau\tau\nu$ & direct & $m_{\chi}=40$ & 30\\
\hline
$\stau\stau\to \tau\tau\tau\tau\nu\tau\tau\nu$ & indirect &
 $m_{\stau}=50$, $m_{\chi}=30$ & 62\\
 & & $m_{\stau}=50$, $m_{\chi}=10$ & 51\\
$\slep\slep\to\tau\nu\tau\nu$ & direct & $m_{\slep}=50$ &37\\
$\stau\stau\to \tau\nu \tau\tau\tau\nu$ & mixed &
 $m_{\stau}=50$, $m_{\chi}=30$ & 46\\
\hline
$\snu\snu\to\nu\nu\tau\tau\nu\tau\tau\nu$ & indirect & 
 $m_{\snu}=50$, $m_{\chi}=30$ & 41\\
 & & $m_{\snu}=50$, $m_{\chi}=10$ & 12\\
$\snu\snu\to\tau\tau\tau\tau$ & direct & $m_{\snu}=50$ & 42\\
$\snu\snu\to\tau\tau\nu\tau\tau\nu$ & mixed & 
 $m_{\snu}=50$, $m_{\chi}=30$ & 50\\
\hline
$\stq\stq\to\Pc\Pc\tau\tau\nu\tau\tau\nu$ & indirect & 
 $m_{\stq}=50$, $m_{\chi}=30$ & 19\\
\hline
\end{tabular}
\caption[.]{\em Selection efficiencies at $\sqrt{s}=172\gev$ for a
representative set of signal processes,
with a lepton flavour composition in the final state leading
to the smallest efficiencies.}
\label{effics}
\end{table}
The efficiencies used for setting limits were checked to give a
conservative estimate for two-body decay cascades via light sfermions
as well as three-body decays of charginos and neutralinos.

For a given value of $m_0$ and $\tan\beta$, limits are derived in the
$(\mu,M_2)$ plane for the worst case in terms of third generation
mixing angles and
the lepton flavour composition of the final state.
In most points this worst case is identified as ($ijk$)=(133) with
$\chi\to\tau\tau\nu$,
corresponding to a maximum number of taus in the final state,
with small squark masses, leading to a large hadronic branching fraction.
The limits set this way are by construction independent of the choice
of generation indices or third generation mixing angles.

\begin{figure}[t]
\begin{center}
\makebox[\textwidth][l]{
\epsfig{figure=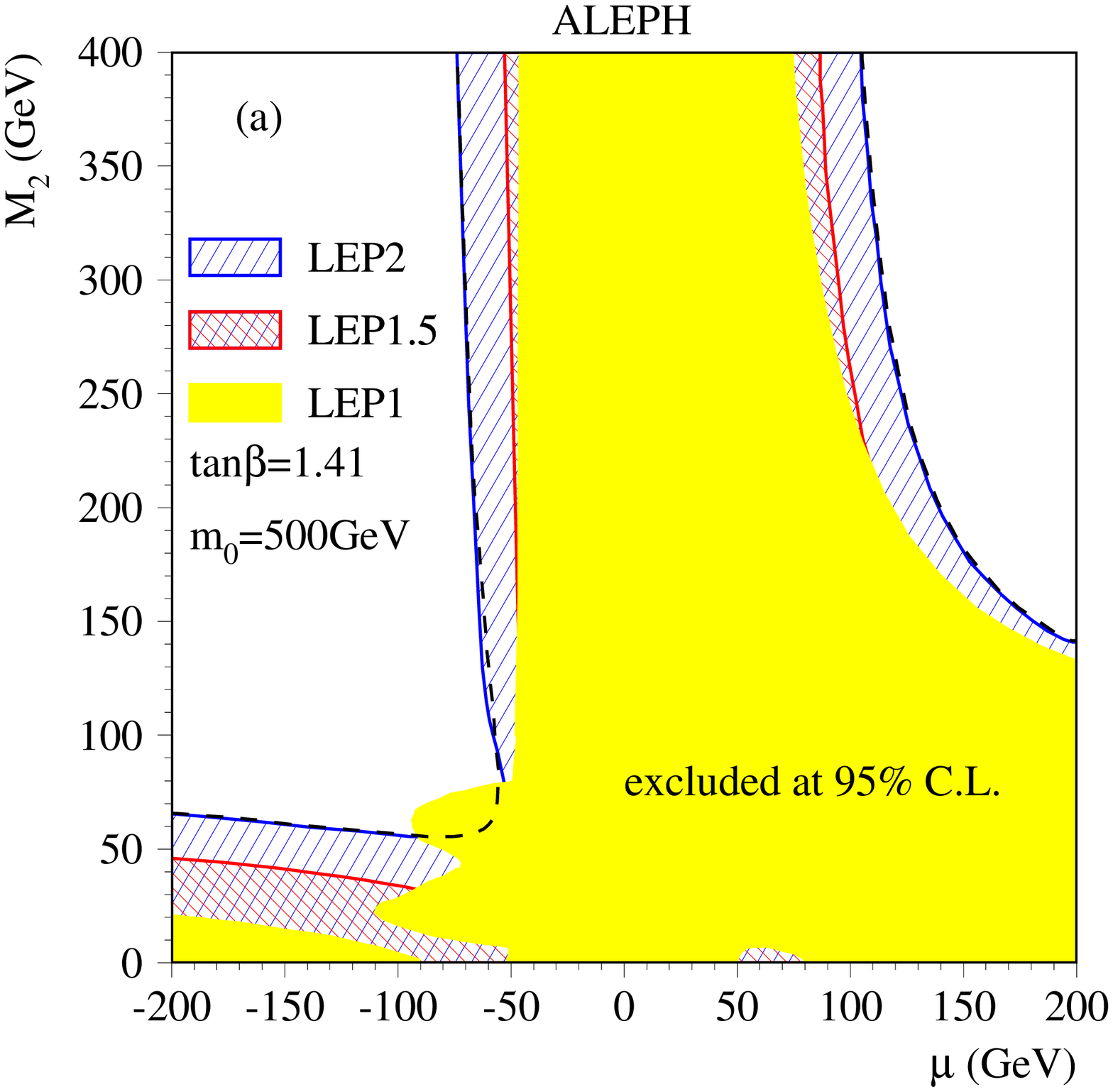,width=0.5\textwidth}
\epsfig{figure=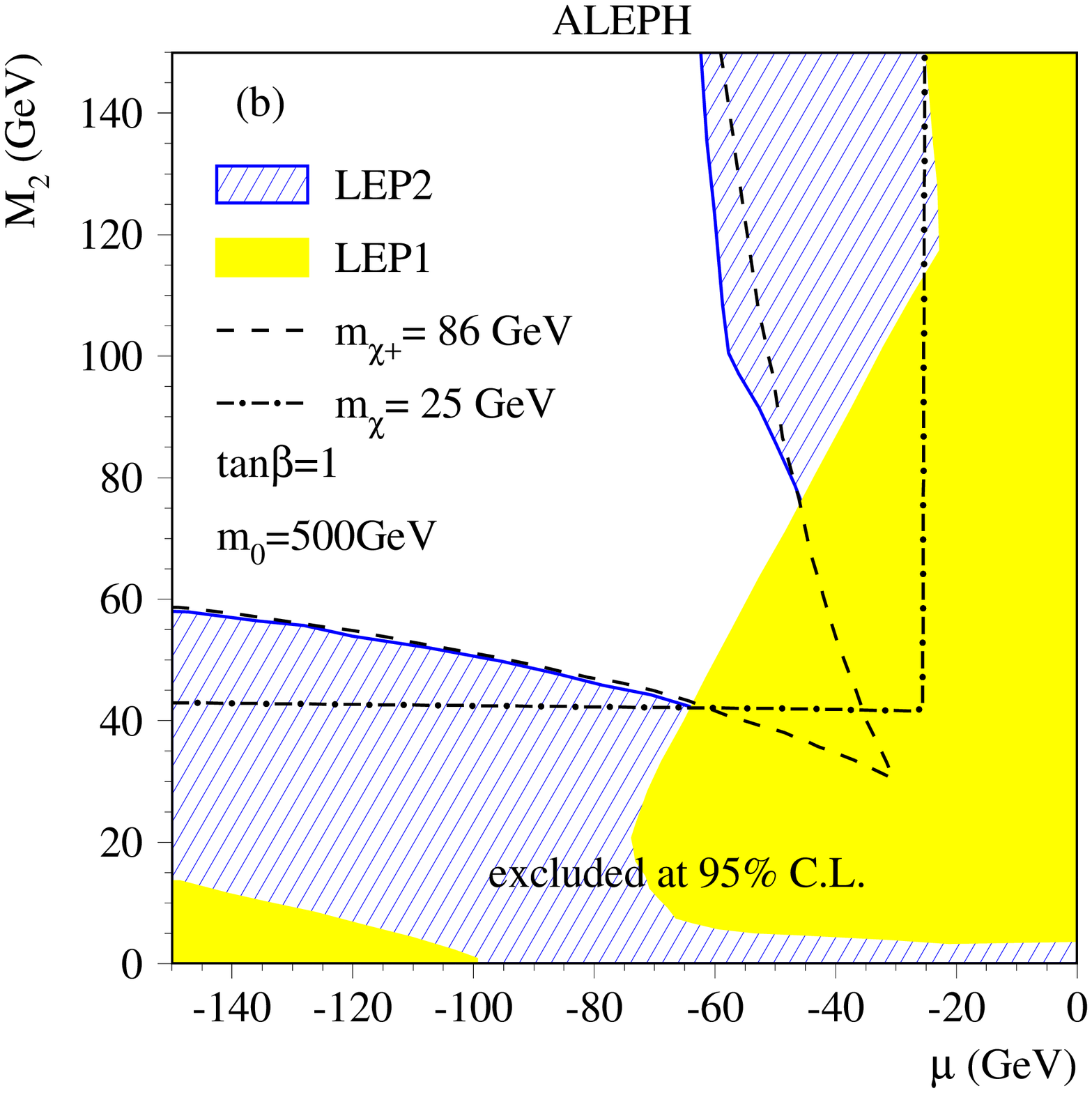,width=0.5\textwidth}}
\caption[.]{\em\label{lle.indirect}{Regions in the $(\mu,M_2)$ plane excluded
 at $95\%$ C.L. at $m_0=500$\gevcc{} and a) $\tan\beta=1.41$ or
b) $\tan\beta=1$, assuming that the {indirect} decays dominate.
The superimposed dashed and dash-dotted lines show the kinematic limit
$M_{\chi^+}=86$\gevcc, and a fixed neutralino mass of
$M_{\chi}=25$\gevcc. The neutralino limit of $M_{\chi}=25$\gevcc{}
is set at $\tan\beta=1$ and $(\mu,M_2) \sim (-60,40)$ by an
interplay of the LEP1 and LEP2 exclusion limits.}}
\end{center}
\end{figure}

\begin{figure}[t]
\begin{center}
\makebox[\textwidth][l]{
\epsfig{figure=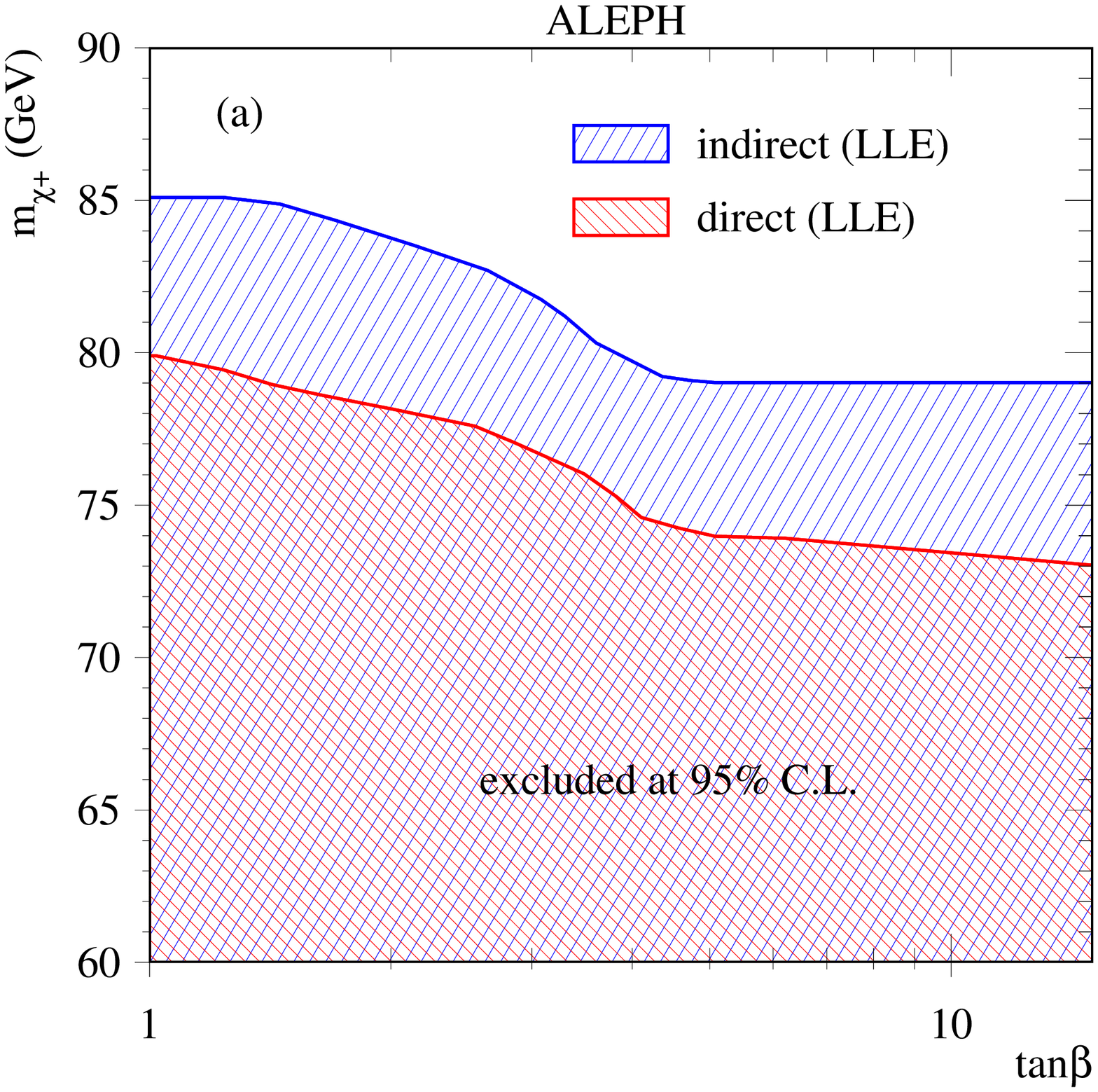,width=0.5\textwidth}\hfill
\epsfig{figure=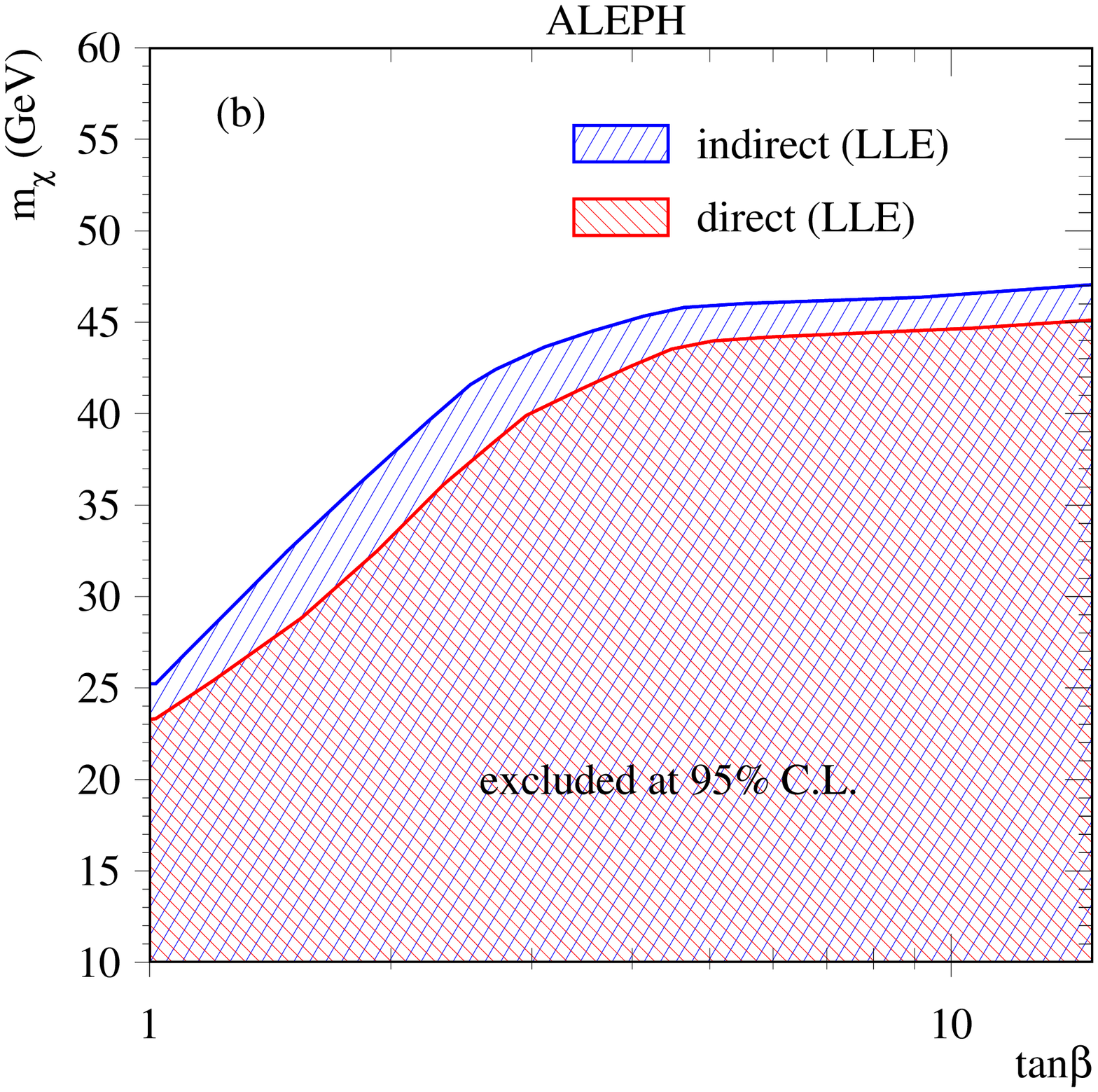,width=0.5\textwidth}}
\caption[.]{\em\label{limit_mc_tanb}{The $95\%$ C.L. limit on a) the
chargino mass and b) the lightest neutralino mass as a function of
$\tan\beta$, assuming the dominance of either {direct} or {indirect}
decay modes. The limits hold for any choice of $\mu, M_2, m_0$ and
the generation indices $i,j,k$ of the coupling $\lambda_{ijk}$.}}
\end{center}
\end{figure}

For each point in $\mu\! -\! M_2\! -\! m_0\! -\! \tan\beta$, the
$\nbar$-prescription is applied to decide which combination of
chargino and neutralino searches gives the best exclusion power and should
therefore be used to set the limit.
Fig.~\ref{lle.indirect}a shows the limits obtained
in the $(\mu,M_2)$ plane for a fixed value of $\tan\beta$ and $m_0$, from
which a lower limit on the chargino and neutralino masses can be derived.
Scanning over $m_0$, these limits are shown as a function of $\tan\beta$ in
Fig.~\ref{limit_mc_tanb}.
Since the worst case limit
is basically set by the purely hadronic decays, the
$\tan\beta$-dependence of the two mass limits
is dictated mainly by the relative change
of the chargino and neutralino mass isolines in the $(\mu,M_2)$ plane
with respect to $\tan\beta$.

For small $m_0$, contributions from t-channel $\snu$-exchange suppress
the pair production of charginos in the gaugino region. However, according
to Eq.~\ref{eq:gut} selectrons are expected to be light in the same region
of parameter space, enhancing the cross section for $\chi\chi$ and 
$\chi'\chi$ production. In contrast to scenarios with conservation of
R-parity, both these processes lead to visible final states, allowing
to exclude these regions up to large chargino masses.

At values of $\tan\beta$ close to one, small neutralino masses are excluded
by an interplay of limits on $\chi\chi^{\prime}$-production from
LEP1~\cite{lep1.jfg}
and the LEP2 chargino and neutralino limits (Fig.~\ref{lle.indirect}b),
in the case of $\tan\beta=1$ still allowing neutralino masses as small
as 25\gevcc.

\subsubsection{Dominance of direct decays\label{dom.dir}}
In this scenario the charginos and the heavier neutralinos are assumed to decay 
directly to SM particles. This generally corresponds to the cases when the 
 sleptons or the  sneutrinos are the LSP. Furthermore, when the sleptons or sneutrinos are
 lighter than the charginos (or the heavier neutralinos) and heavier than the
 lightest neutralino, the direct decays can
 dominate provided that the neutralino couples higgsino-like and $\lambda$ is
 large. 

Charginos can decay either into one charged lepton plus
two neutrinos or into three charged leptons, leading to two-, four- or
six-lepton topologies. The composition of lepton flavours appearing
in these final states depends on the field content of the
chargino, the generation indices and the details of the mass
spectrum. For simplicity, the inclusive combination of all
corresponding selections
is used. All branching fractions and flavour compositions have been scanned
to identify the overall most conservative limit, which in general
is set by charginos decaying dominantly into two taus
via a coupling involving all three lepton flavours. For such couplings,
selections for all possible flavour combinations have to be combined,
leading to the largest possible background and number of candidate events.
If in addition the branching fraction into two taus is large, selection
efficiencies are smallest, resulting in the most conservative limit.

For the scenario considered here, all neutralinos are assumed to decay
to two charged leptons plus a neutrino. Using the ``Four Leptons plus
missing Energy'' selection, efficiencies have been calculated as a
function of the neutralino masses for each possible flavour composition
in the final state. As before, the smallest
efficiency -- corresponding to a maximum number of taus in the final state --
is used to set  limits independent of the choice of generation indices.

\begin{figure}[t]
\begin{center}
\makebox[\textwidth][l]{
\epsfig{figure=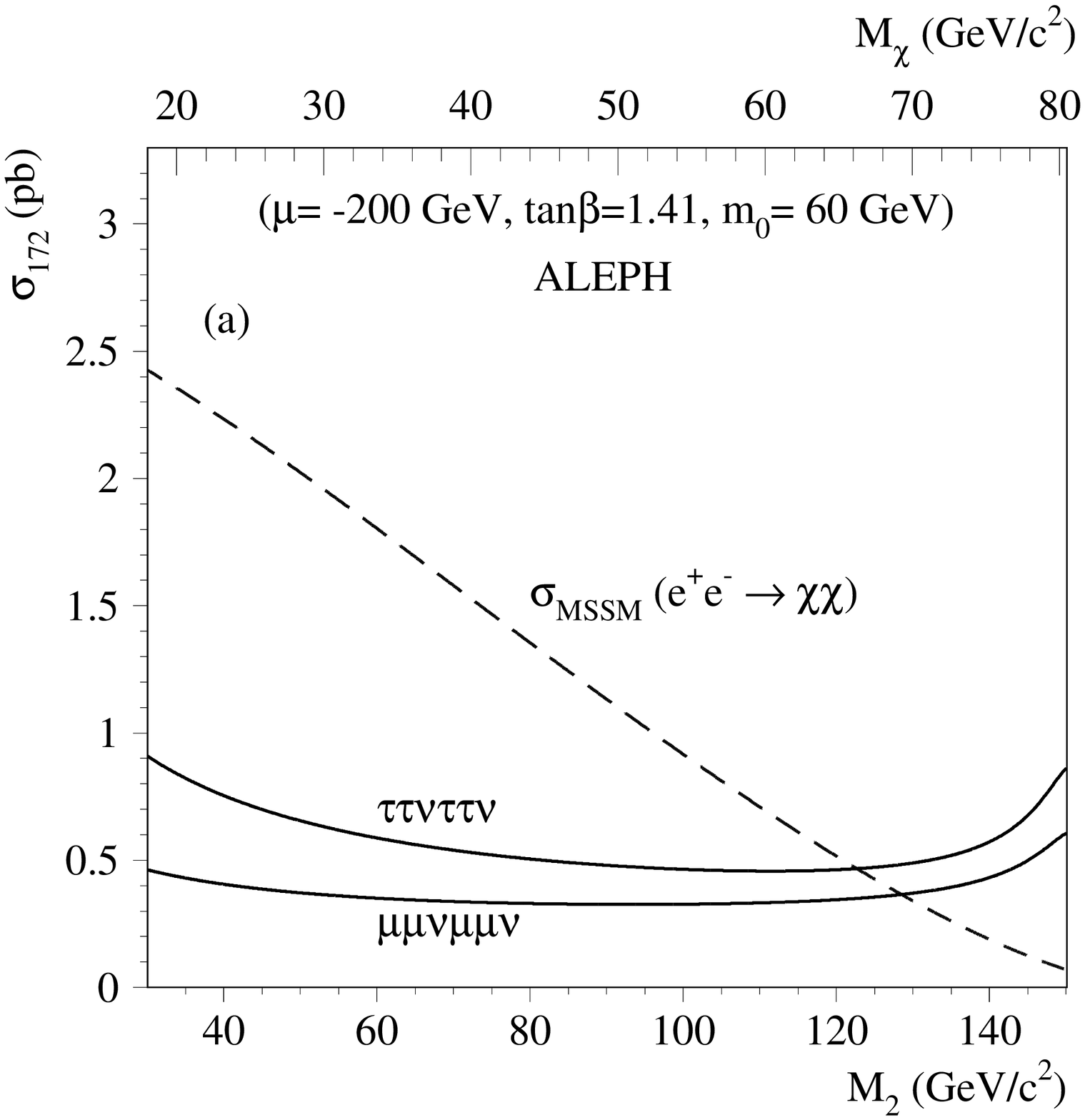,width=0.5\textwidth}\hfill
\epsfig{figure=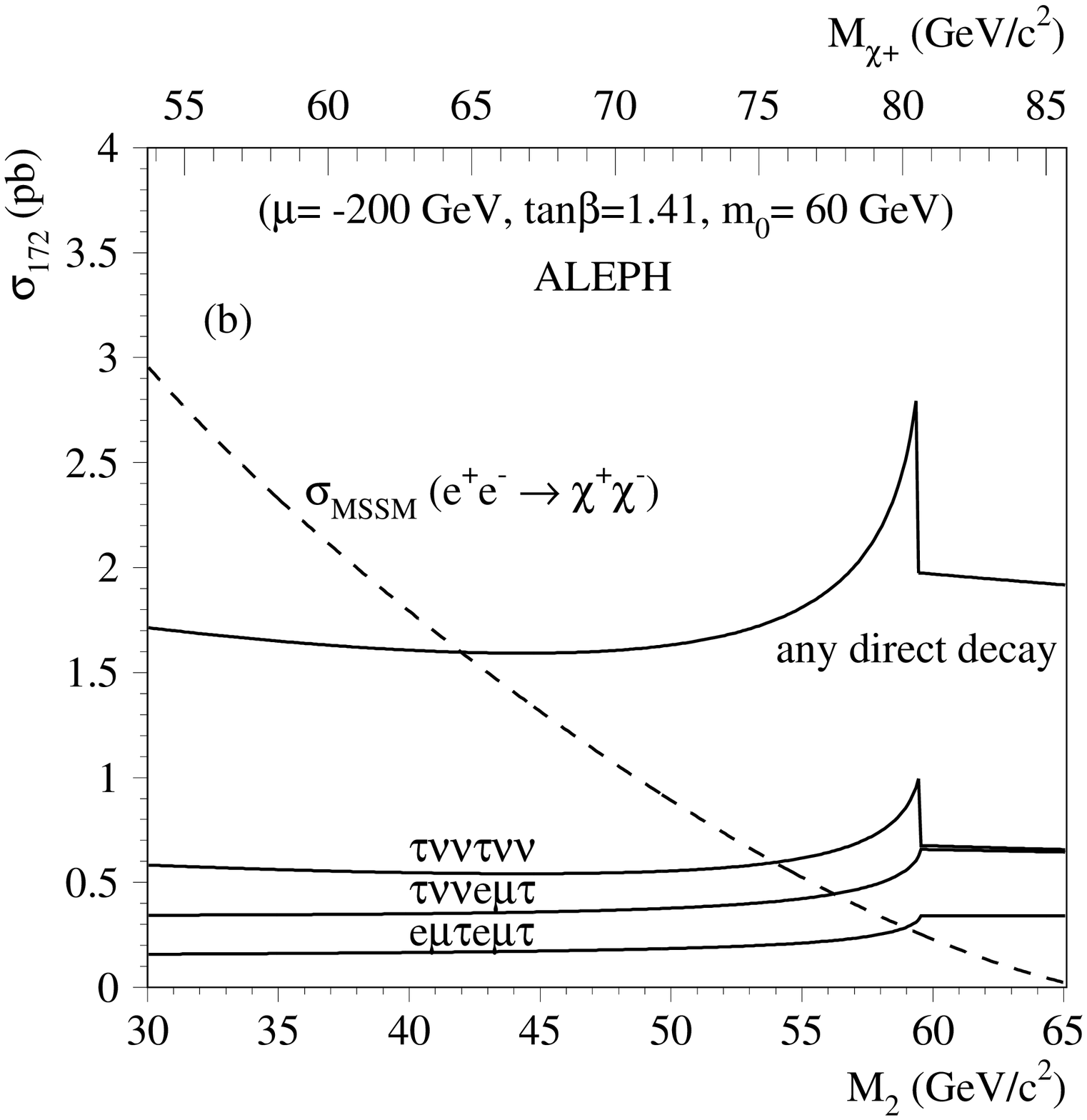,width=0.5\textwidth}}
\caption[.]{\em\label{lle.direct}{Cross sections at $\sqrt{s}=$~172~GeV
excluded at 95\% C.L. for pair production of a) the lightest neutralino
and b) the lightest chargino, for various final states as a function
of $M_2$ at $\tan\beta=1.41$, $m_0=60$\gevcc, $\mu=-200$\gevcc.
In b) candidates selected at $\sqrt{s}=$~161~GeV are restricted to
$m_{\chi^+}<80.5$\gevcc.}}
\end{center}
\end{figure}

In analogy to the procedure described in the previous section, limits from
chargino and neutralino searches are set for each point in
$\mu\! -\! M_2\! -\! \tan\beta\! -\! m_0$ parameter space.
Fig.~\ref{lle.direct} shows an example of the limit
obtained in the gaugino region at $m_0=60$\gevcc.
Due to the destructive
interference of the $s$- and $t$-channel contributions to the
chargino cross section, the limit set by the chargino search
does not reach the kinematic limit at small $m_0$. On the other hand,
the production cross section for $\chi\chi$ is enhanced at small
selectron masses, allowing charginos well beyond the
kinematic limit to be excluded in certain regions of parameter space.

Limits on the masses of the lightest chargino and neutralino as a function
of $\tan\beta$ are obtained by scanning the parameter space
in $\mu\! -\! M_2\! -\! m_0$ (Fig.~\ref{limit_mc_tanb}).
Charginos with masses less than 73\gevcc{} and
neutralinos with masses less than 23\gevcc{} are excluded at 95\% confidence
level for any choice of generation
indices $i,j,k$, and for neutralino, slepton and sneutrino LSPs.

\subsubsection{Mixed Topologies \label{dom.mix}}
For the extreme case of
$\Gamma (\chi^+\to\Pf\Pf\chi)\! =\!\Gamma (\chi^+\to\Pf\Pf\Pf)$, mixed
topologies with one direct and one indirect decay are produced
in 50\% of the events from chargino pair
production. They are selected by the inclusive combination of the
``Leptons and Hadrons'' and the ``Four
and Six Leptons plus Missing Energy'' analyses, with efficiencies similar
to the ones obtained for indirect topologies. As the other half of the
events produced consists of direct and indirect topologies, the selection
is combined with the corresponding analyses described in the previous
sections. The worst case limit is set for $(i,j,k)=$(1,2,3), (2,3,1) or
(1,3,2). In this case all three
lepton flavours are accessible in the direct decays, and therefore the
combination of all acoplanar lepton selections has to be used, leading to
a maximum number of candidates in the data. For the indirect decays, the
lightest neutralino can decay into $\Pe\tau\nu$ with a large branching
fraction, which corresponds to the smallest efficiency for this generation
structure. 
Therefore this case has been used to set limits independent of the
choice for $(i,j,k)$. This limit is at least as constraining
as the limit for direct topologies, the exact position depending on the
size of the coupling~$\lambda_{ijk}$.

\subsection{Sleptons}
A slepton can decay either directly to a lepton and a neutrino, or indirectly to
a lepton and a  neutralino, which subsequently decays to two leptons and a neutrino. 
The decays to charginos are kinematically inaccessible for most of the
slepton mass range considered in this section (see previous section).
The three types of topologies  from the pair-production of sleptons are
classified as the {\it direct topologies} (when both sleptons decay directly), 
the {\it indirect topologies} (when both sleptons decay indirectly), and the
{\it mixed topologies} (when one slepton decays directly, one indirectly). 

\begin{figure}
\begin{center}
\makebox[\textwidth]{
\epsfig{figure=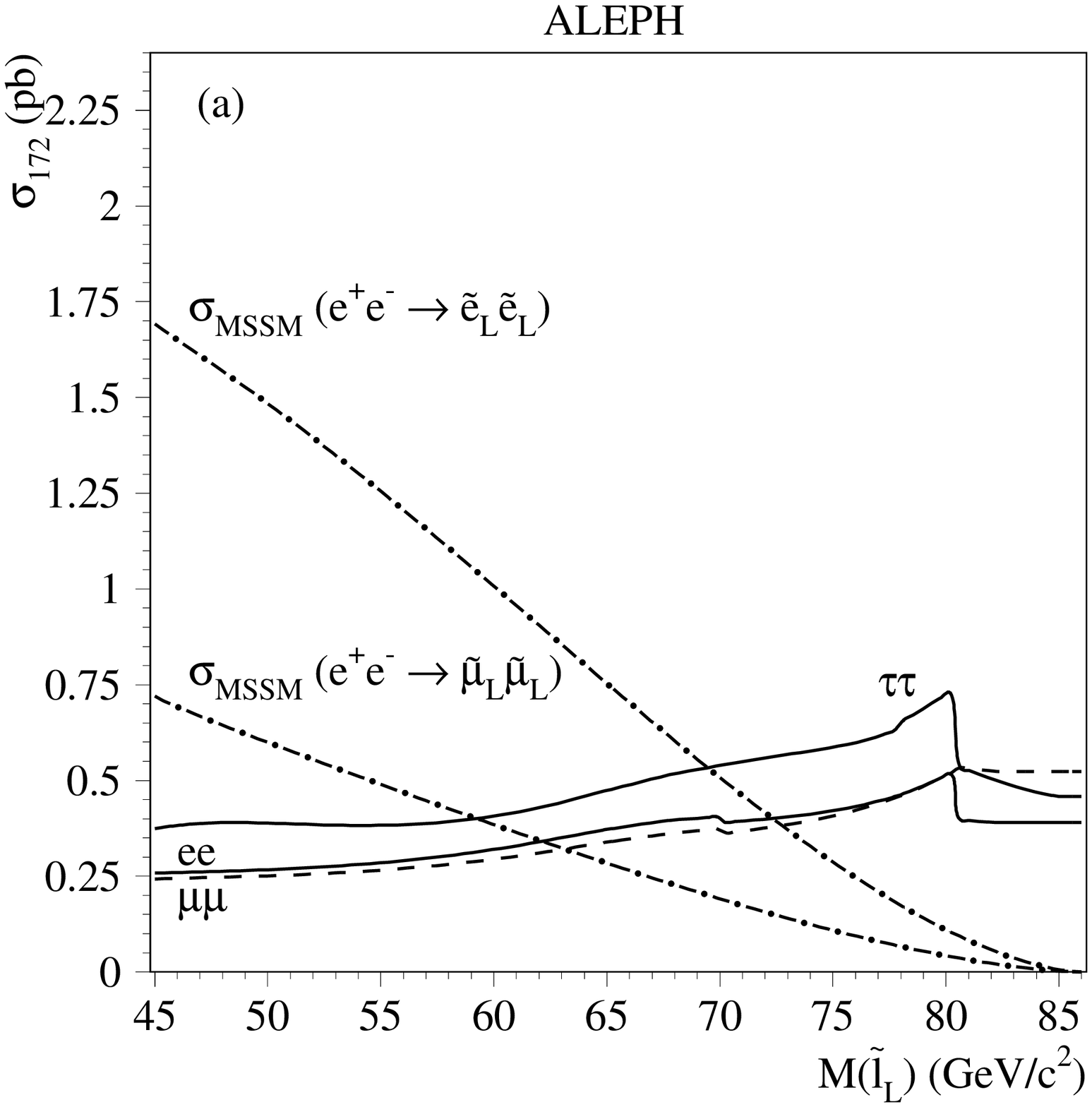,width=0.5\textwidth}\hfill
\epsfig{figure=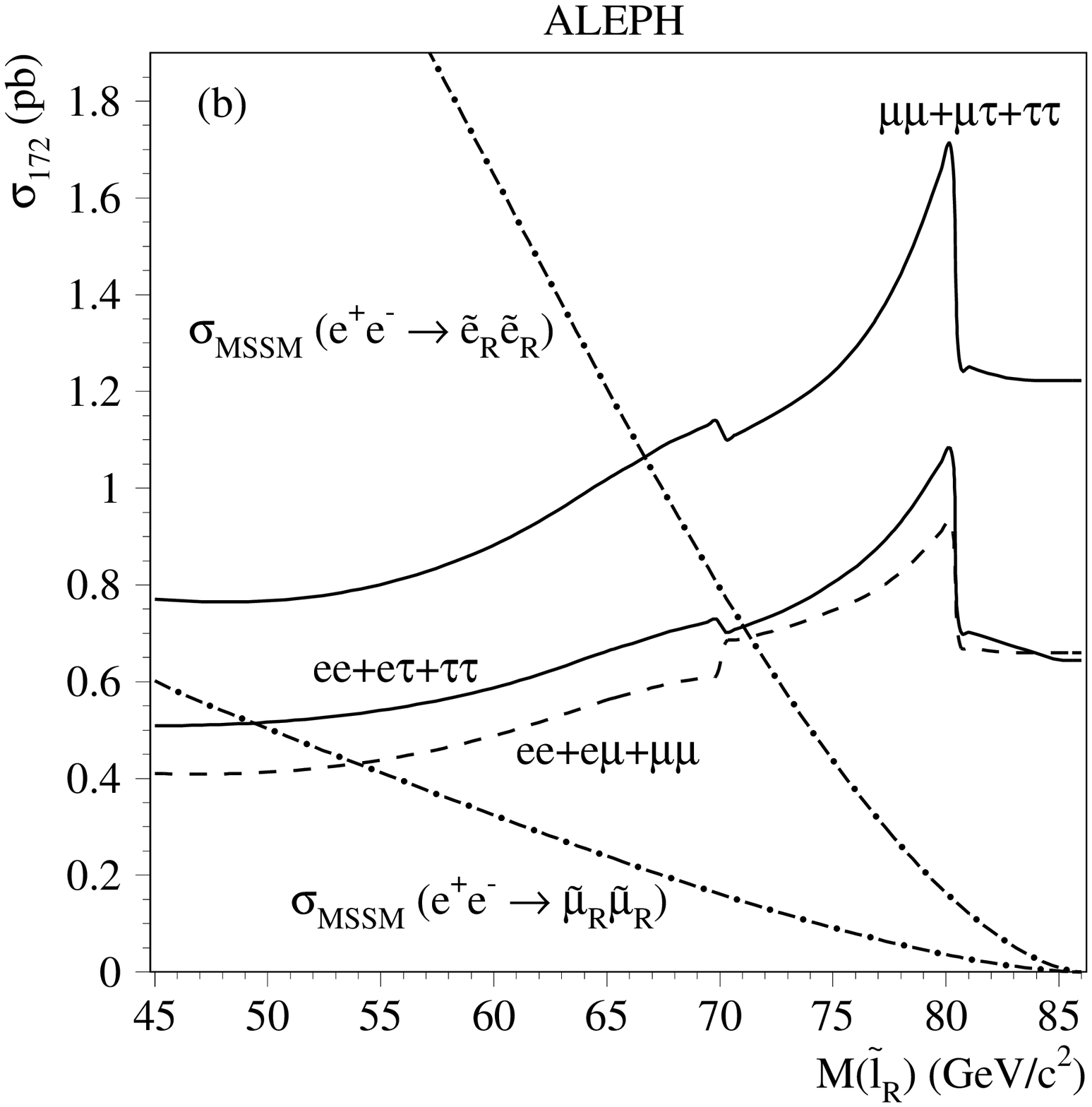,width=0.5\textwidth}}
\makebox[\textwidth]{
\epsfig{figure=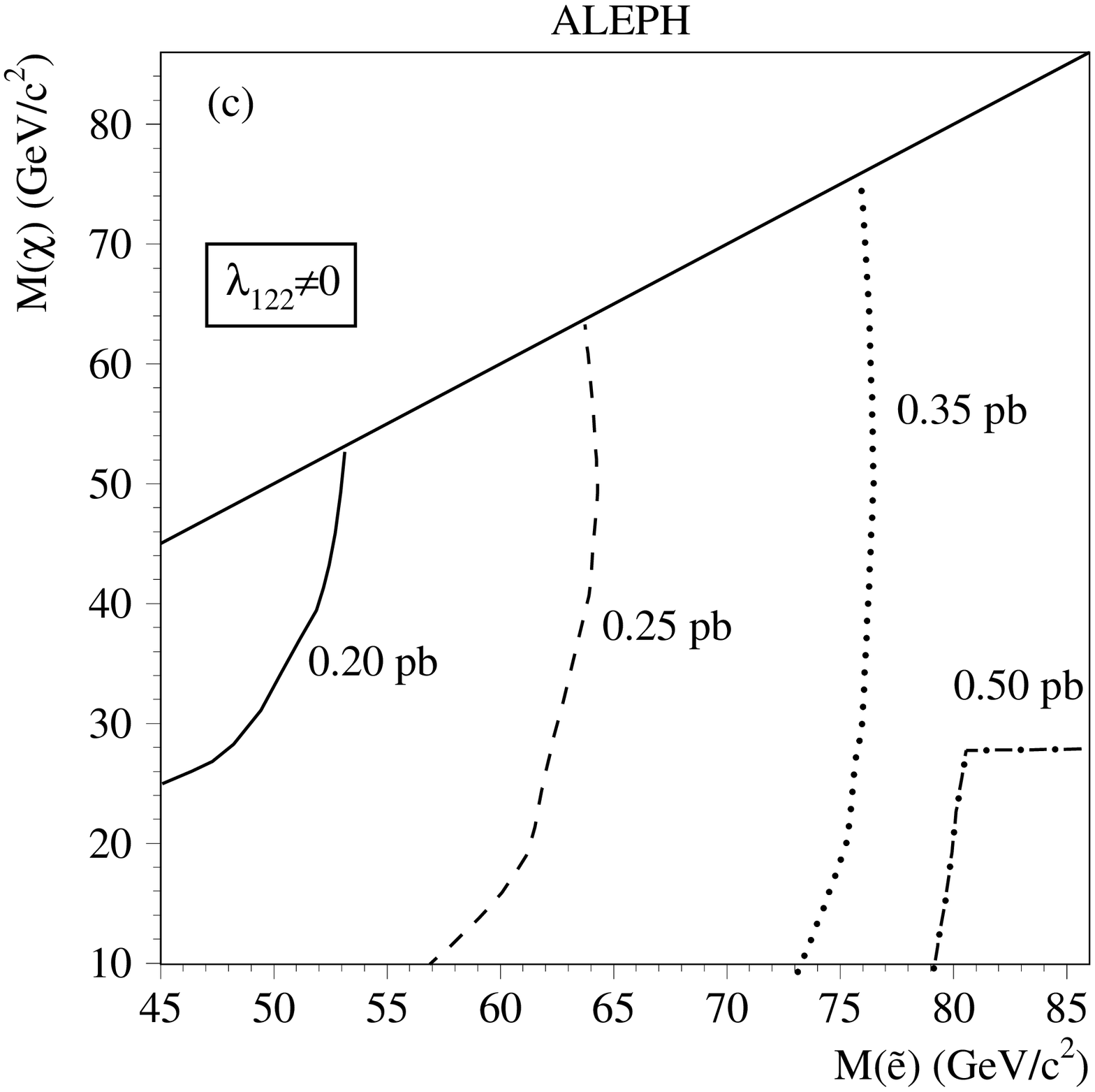,width=0.5\textwidth}\hfill
\epsfig{figure=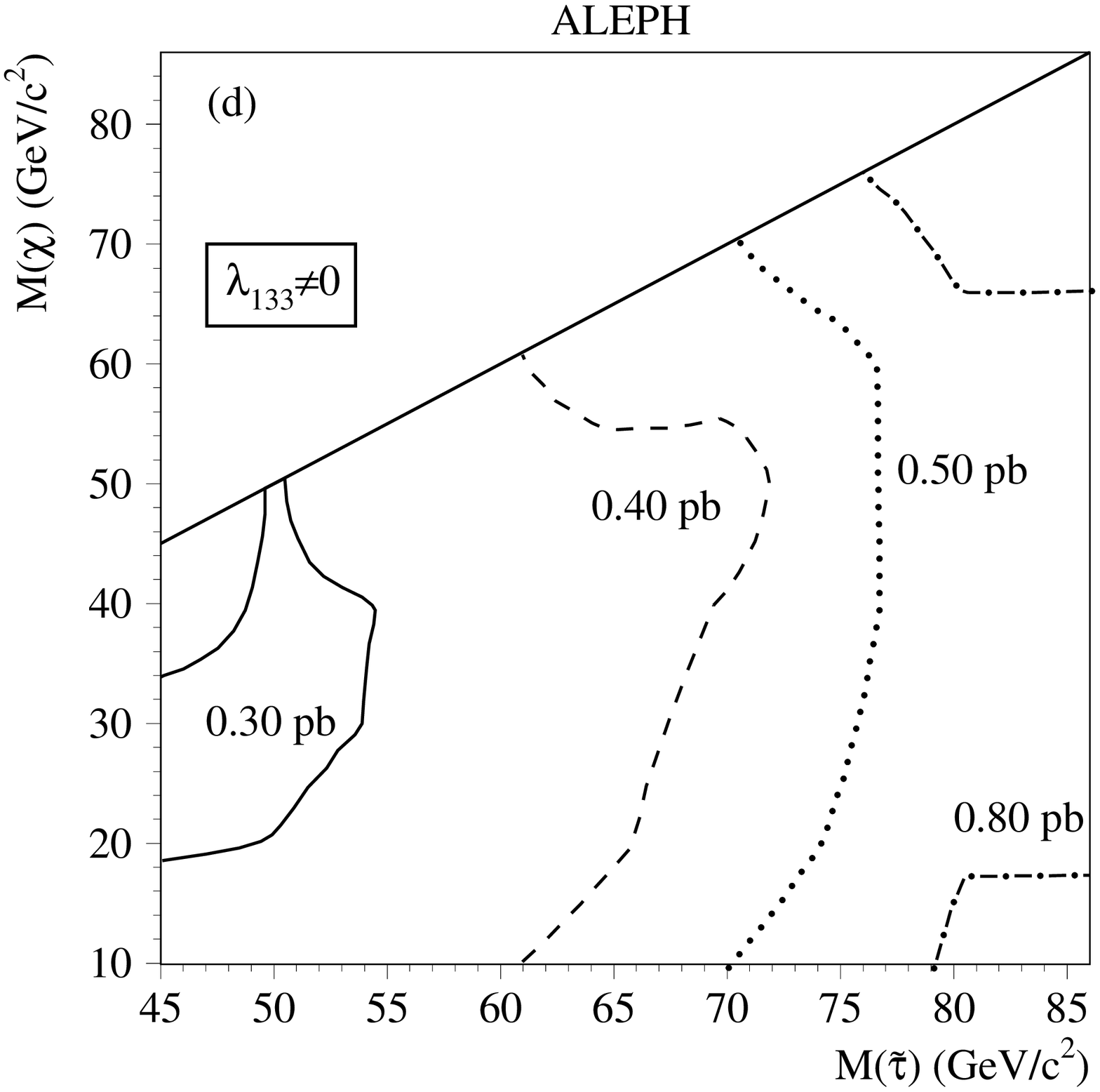,width=0.5\textwidth}}
\caption{\label{excl.xsec.acop}\em The $95\%$ C.L.  slepton exclusion
  cross sections scaled to $\sqrt{s}=172\gev$. For the purpose of these plots
  a $\beta^3/s$ cross section dependence, valid for 
 scalar pair-production in the {\em s}-channel,  was assumed. 
Fig. a) and b) show $\sigma_{\rm excluded}^{172}$ for the direct decays of
left-handed  and right-handed sleptons respectively. Superimposed are
the MSSM cross sections at $\tan{\beta} = 2$ for selectron production
($M_2=50\gevcc,\mu=-200\gevcc$) and smuon production. Fig. c) and d) show
contours of $\sigma_{\rm excluded}^{172}$ for the indirect decays 
in the ($M_{\chi},M_{\tilde l}$) plane for the best-case exclusion ($\tilde e \tilde e$,
$\lambda_{122}$) and for the worst-case exclusion ($\tilde \tau \tilde \tau$,
$\lambda_{133}$).}
\end{center}
\end{figure}

 For the {\em direct topologies} of left-handed sleptons
 the acoplanar lepton selection was used. Individual efficiencies for  the final states 
 $ee$, $\mu\mu$ and $\tau \tau$ are calculated as a function of the
 slepton mass, and for the three energies $\sqrt{s}=133, 161, 172\gev$. The
 various final states correspond to different choices of the generation indices
 $i,j,k$ of the R-parity violating coupling $\lambda_{ijk}$. The efficiencies
 are relatively constant as a function of $M_{\tilde l}$, and typical values 
 are given in Table~\ref{effics}. 
 Subtracting the background from Table~\ref{wwcands} according
 to the prescription given in \cite{pdg.subtract}, the exclusion 
 cross sections scaled to $\sqrt{s}=172\gev$ are shown in Fig.~\ref{excl.xsec.acop}a
 for the three final states.   

 Right-handed sleptons can decay to {\em two} final states in the direct
 topology with a $50\%$ branching
 ratio each for a given choice of the  generation indices $i,j,k$: ${\tilde
 l}_{kR} \ra \nu_i l_j$ and ${\tilde l}_{kR} \ra \nu_j l_i$. For example, for the
 coupling $\lambda_{121}$ 
 pair-produced  right-handed selectrons would produce  the acoplanar topologies
 $\onequarter ee \oplus \onehalf e\mu \oplus \onequarter \mu \mu$ with the given branching ratios.
 The results for admixtures of acoplanar lepton states using
 the above branching ratios are shown in
 Fig.~\ref{excl.xsec.acop}b. The exclusion cross sections for the right-handed
 slepton topologies are larger than the exclusion cross sections for their
 left-handed partners due to the higher background.  

 For the {\em indirect topologies}, which consist of six leptons and two
 neutrinos,   an inclusive combination of 
 the  ``Six Leptons plus $\emiss$'' and the ``Four Leptons plus $\emiss$''
 selection is used, the latter one improving the efficiencies in the region of 
 low and  very high neutralino masses. The efficiencies
 (Table~\ref{effics}) mainly depend on $\Delta M = M_{\tilde l} - M_{\chi}$, and
 are smallest for staus with a $\lambda_{133}$ coupling at large $\Delta M$.
  Including the one candidate event observed
 in the ``Four Leptons plus $\emiss$'' selection (without background subtraction),
 the $95\%$ C.L. exclusion cross sections  scaled to
 $\sqrt{s}=172\gev$ are derived, and are shown in
 Fig.~\ref{excl.xsec.acop}c,d for selectrons and a dominant coupling
 $\lambda_{122}$, and for staus with a coupling $\lambda_{133}$. The two cases 
 correspond to  final states with a  maximum number of electrons or muons (which have
   the largest selection efficiencies),
  and a maximum number of taus (with the smallest selection efficiencies), respectively.  

 Pairs of sleptons can produce up to $50\%$ {\em mixed topologies} 
 in the extreme case when $\Gamma({\tilde l \ra l \nu}) = \Gamma({\tilde l \ra l
 \chi})$. The { mixed topologies} are  selected by the ``Four
 Leptons plus $\emiss$'' selection with similar efficiencies to the {indirect
 topologies}  (Table~\ref{effics}). 

The above results are now interpreted within the MSSM. 
 Limits at the  $95\%$ C.L.  are derived on the masses of the sleptons 
 in the ($M_{\chi},M_{{\tilde l}_R}$) plane, assuming that  only
${\tilde l}_R {\tilde l}_R$ production contributes. 
This assumption is
conservative because of (a) the smaller cross section of right-handed sleptons
compared to left-handed sleptons for pure {\em s}-channel production, and (b)
the larger exclusion cross sections in the direct topologies of right-handed
sleptons compared to the left-handed sleptons. 
The selectron limits are shown in a typical point in the gaugino region
($\mu=-200, \tan{\beta}=2$). 

\begin{figure}
\begin{center}
\makebox[\textwidth]{
\epsfig{figure=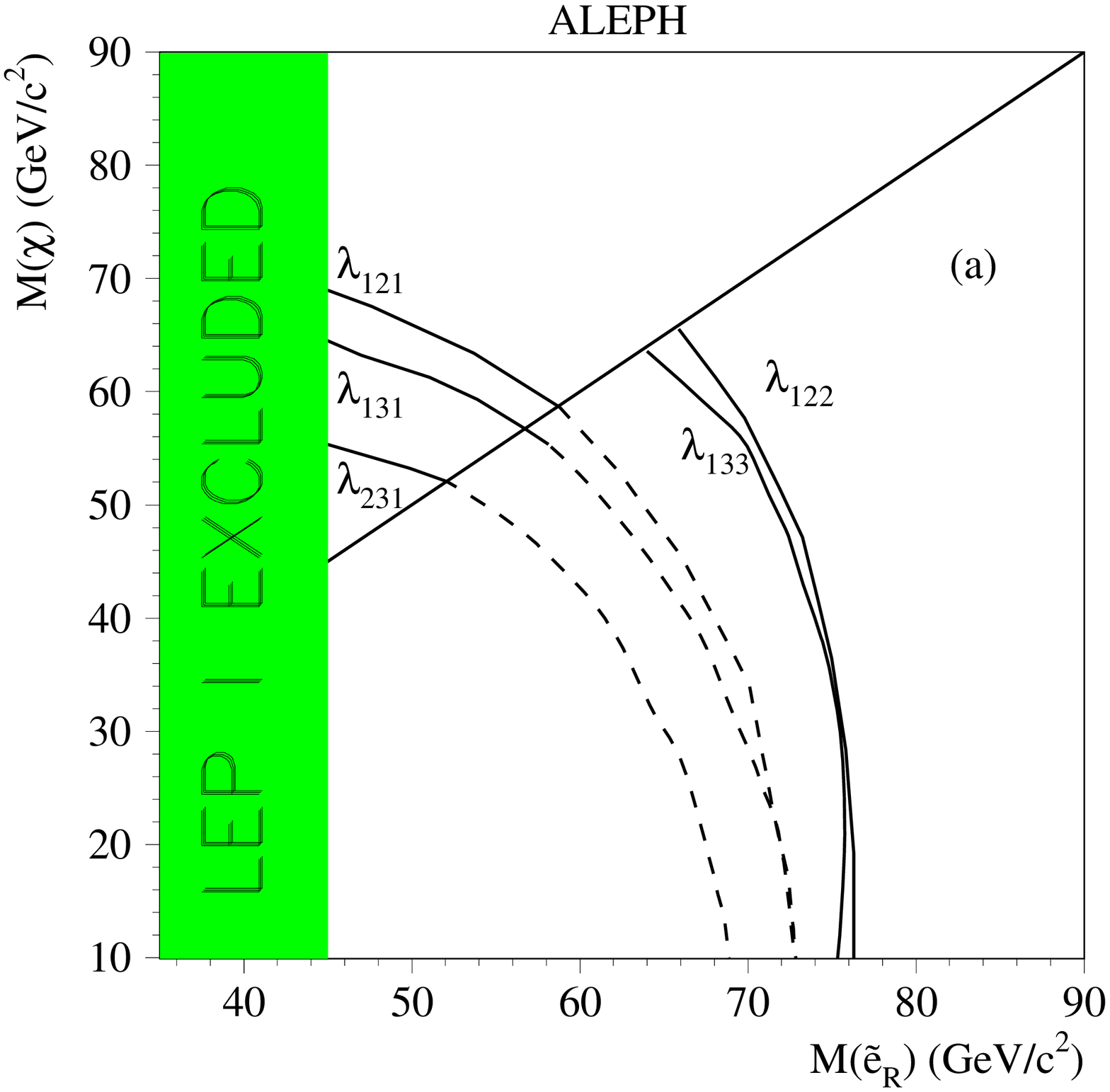,width=0.5\textwidth}\hfill
\epsfig{figure=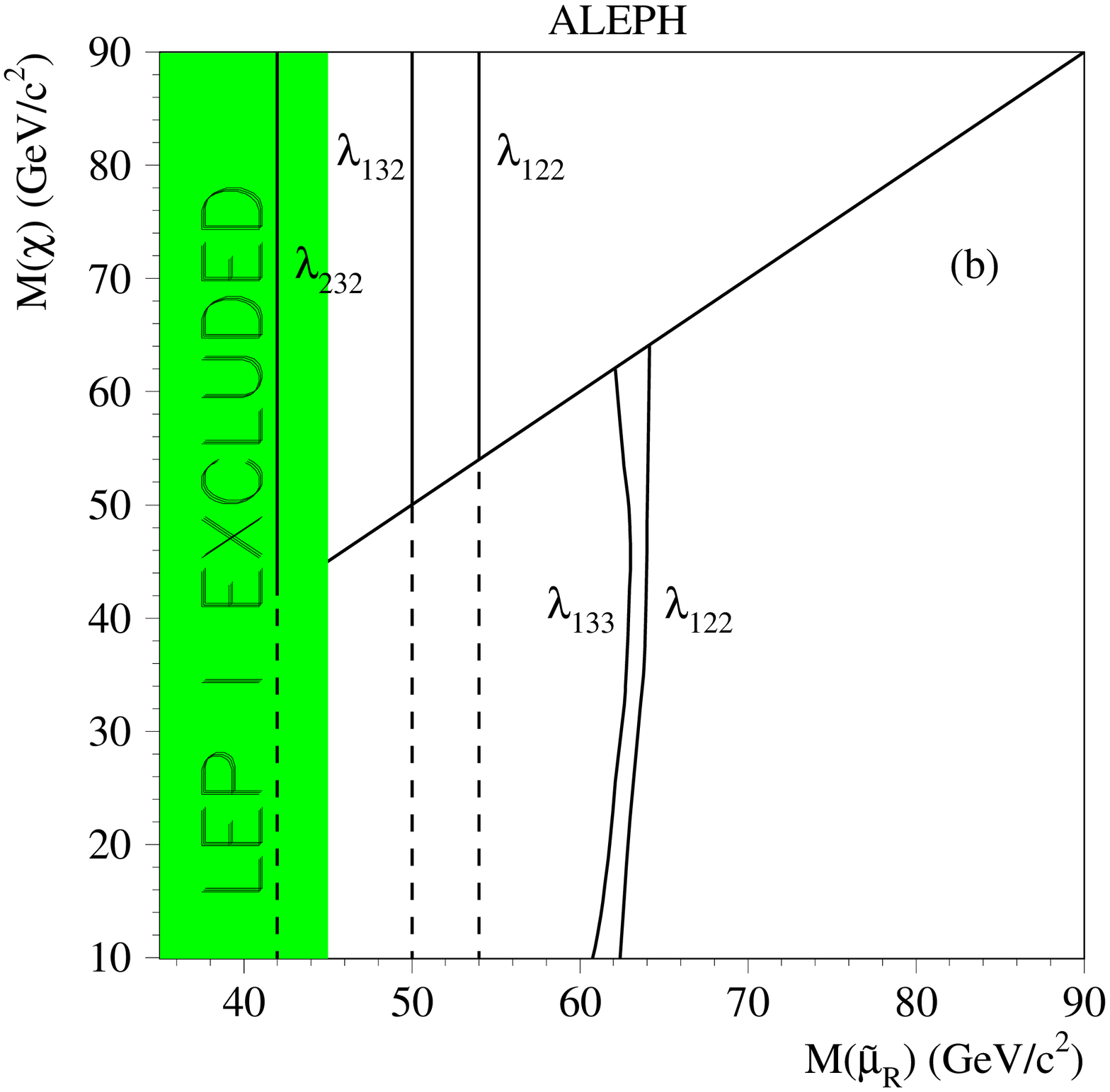,width=0.5\textwidth}}
\epsfig{figure=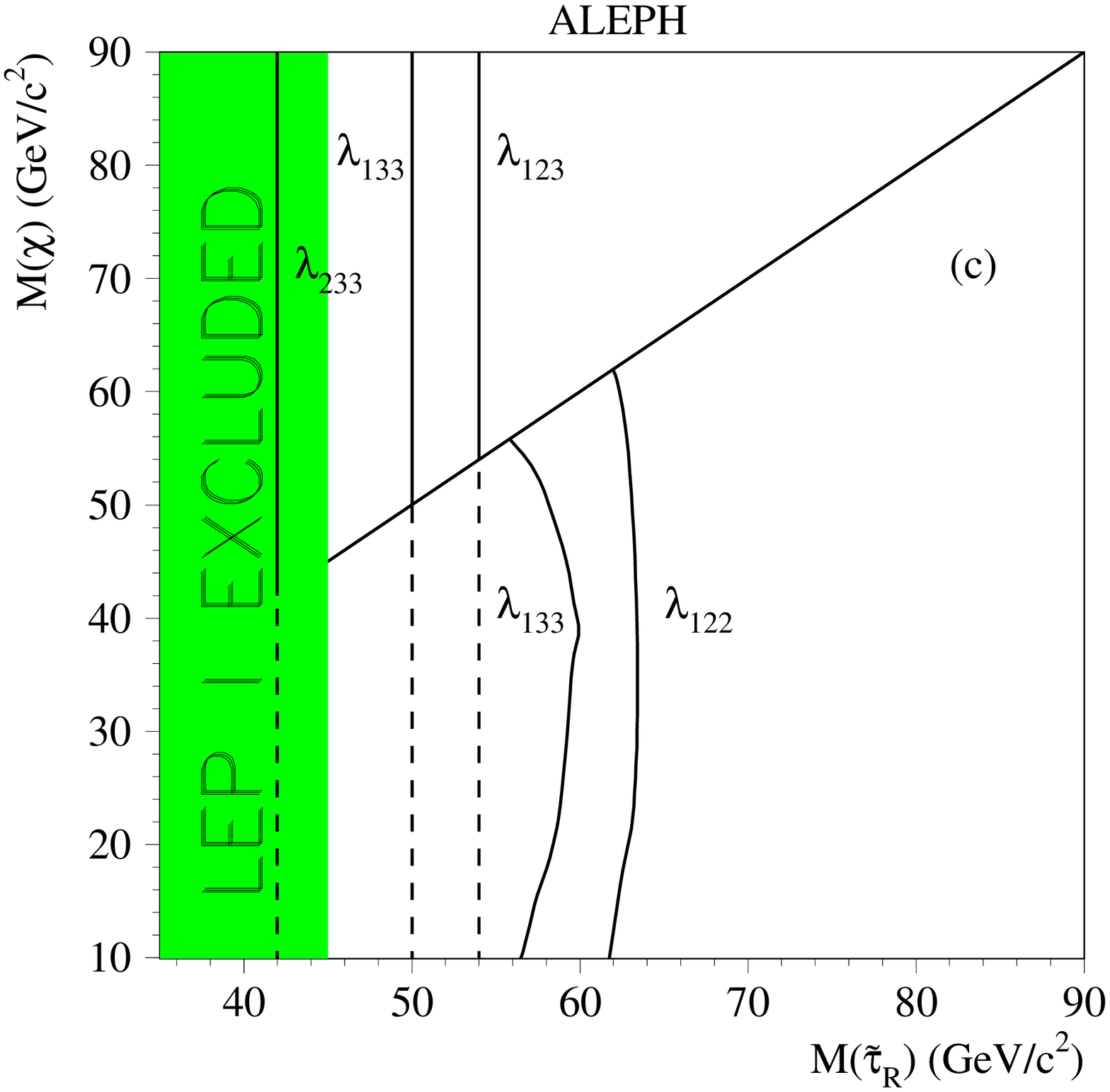,width=0.5\textwidth}
\caption{\label{sleps}\em The $95\%$ C.L. limits in the
  $(m_{\chi},m_{\tilde{l}_R})$ plane at $\tan\beta=2$. 
 Above the diagonal line the lightest neutralino is heavier than the sleptons,
 and only the {direct} decays are allowed. Below the line the {indirect}
 decays  generally dominate, but the branching ratio of the {direct} (dashed
 lines) to  {indirect} (full lines) decays depends on the magnitude of the
 coupling $\lambda_{ijk}$. The two choices of $\lambda_{122}$ and
 $\lambda_{133}$ correspond to the best and worst case exclusions for the
 {indirect} decays, respectively.
 Fig. a) shows the selectron limit in the gaugino region for $\mu=-200\gevcc$.
 Fig. b) and c) show the mass limits on smuons and staus.}
\end{center}
\end{figure}

Limits are calculated for the two extreme cases of $100\%$ direct or $100\%$ indirect decay
modes. This generally corresponds to the two cases when the   slepton or the
neutralino is the LSP, respectively. If $\Gamma({\tilde l \ra l \nu}) \sim
\Gamma({\tilde l \ra l \chi})$, up to $50\%$ {\em mixed topologies} are also expected
for neutralino  LSPs in some regions of parameter space. In this case the exclusion region 
would lie in between the two extreme cases of direct or indirect decays, the
exact location of this region depending on the magnitude of the R-parity violating coupling. 
The limits for the direct and indirect decay modes are shown in
Fig.~\ref{sleps}.
The three choices of
couplings  for the direct decays  
correspond to the three possible decay topologies for right-handed sleptons,
which are listed in Table~{\ref{acop.topos}}. 
\begin{table}
\begin{center}
\begin{tabular}{|c||c|c|c|}
\hline 
\hline 
Topology & \multicolumn{3}{c|}{Coupling} \\ \hline 
& ${{\tilde e}_R}$ & ${{\tilde \mu}_R}$  & ${{\tilde \tau}_R}$  \\
\hline 
$ \onequarter ee \oplus \onehalf e\mu \oplus \onequarter \mu \mu$ & $\lambda_{121}$ &
$\lambda_{122}$ & $\lambda_{123}$ \\
$ \onequarter ee \oplus \onehalf e\tau \oplus \onequarter \tau \tau$ & $\lambda_{131}$ &
$\lambda_{132}$ & $\lambda_{133}$ \\
$ \onequarter \mu \mu \oplus \onehalf \mu\tau \oplus \onequarter \tau \tau$ & $\lambda_{231}$ &
$\lambda_{232}$ & $\lambda_{233}$ \\
\hline 
\hline 
\end{tabular}
\caption[.]{\em Acoplanar lepton topologies for right-handed sleptons, and their
    corresponding R-parity violating couplings.}
\label{acop.topos}
\end{center}
\end{table}
  In contrast to the limits on the
direct smuon and stau topologies, the direct selectron limits  show a strong
dependence on $M_{\chi}$ owing to the dependence of the cross section on
neutralino  {\em t}-channel interference\cite{bartl}.
Note that the interference is mostly 
destructive when $M_{\chi} \gsim M_{{\tilde l}_R}$.
The two choices $\lambda_{122}, \lambda_{133}$ for
the indirect decay modes correspond to the best- and worst-case exclusions,
respectively.

For the indirect decay modes, the limits on the sleptons for the
most conservative choice of coupling ($\lambda_{133}$), and for 
$M_{\chi} > 23\gevcc$ (the neutralino limit derived in
Section~\ref{chargino.limit}), are: $M_{{\tilde e}_R}>64\gevcc$ (gaugino region,
  $\mu=-200\gevcc$, $\tan\beta = 2$), 
  $M_{{\tilde \mu}_R}>62\gevcc$, $M_{{\tilde \tau}_R}>56\gevcc$.

\subsection{Sneutrinos}
 A sneutrino  can decay either directly to a pair of leptons, or indirectly to a
 neutrino and a  neutralino. The decays to charginos are
 kinematically inaccessible for most of the
 sneutrino mass range considered here (see also Section~\ref{chargino.limit}).
 The three types of topologies from the pair-production 
 of sneutrinos are again classified as the {\em direct}, {\em indirect} and {\em
   mixed topologies}. 

\begin{figure}
\begin{center}
\epsfig{figure=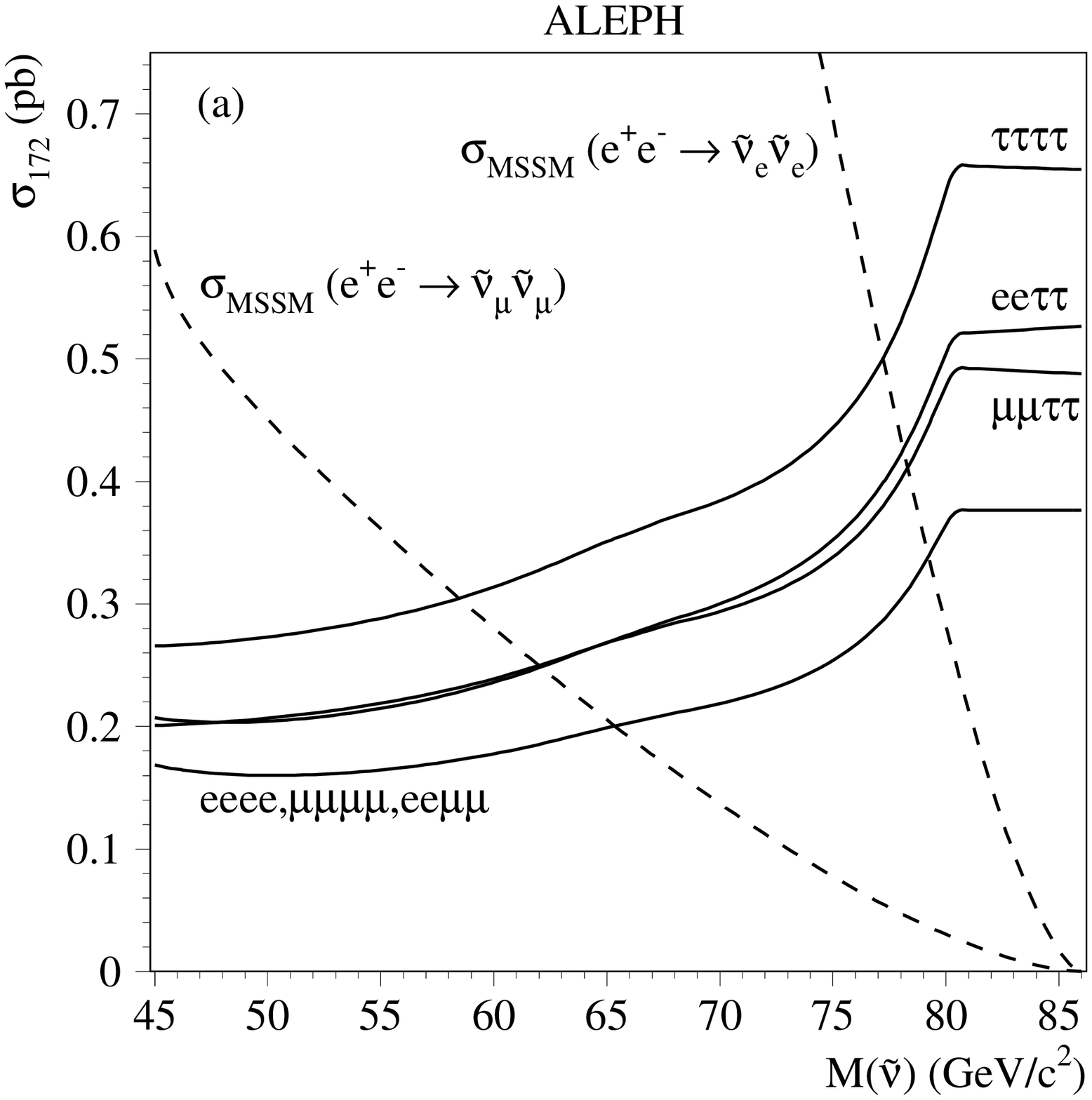,width=0.5\textwidth}\hfill
\makebox[\textwidth]{
\epsfig{figure=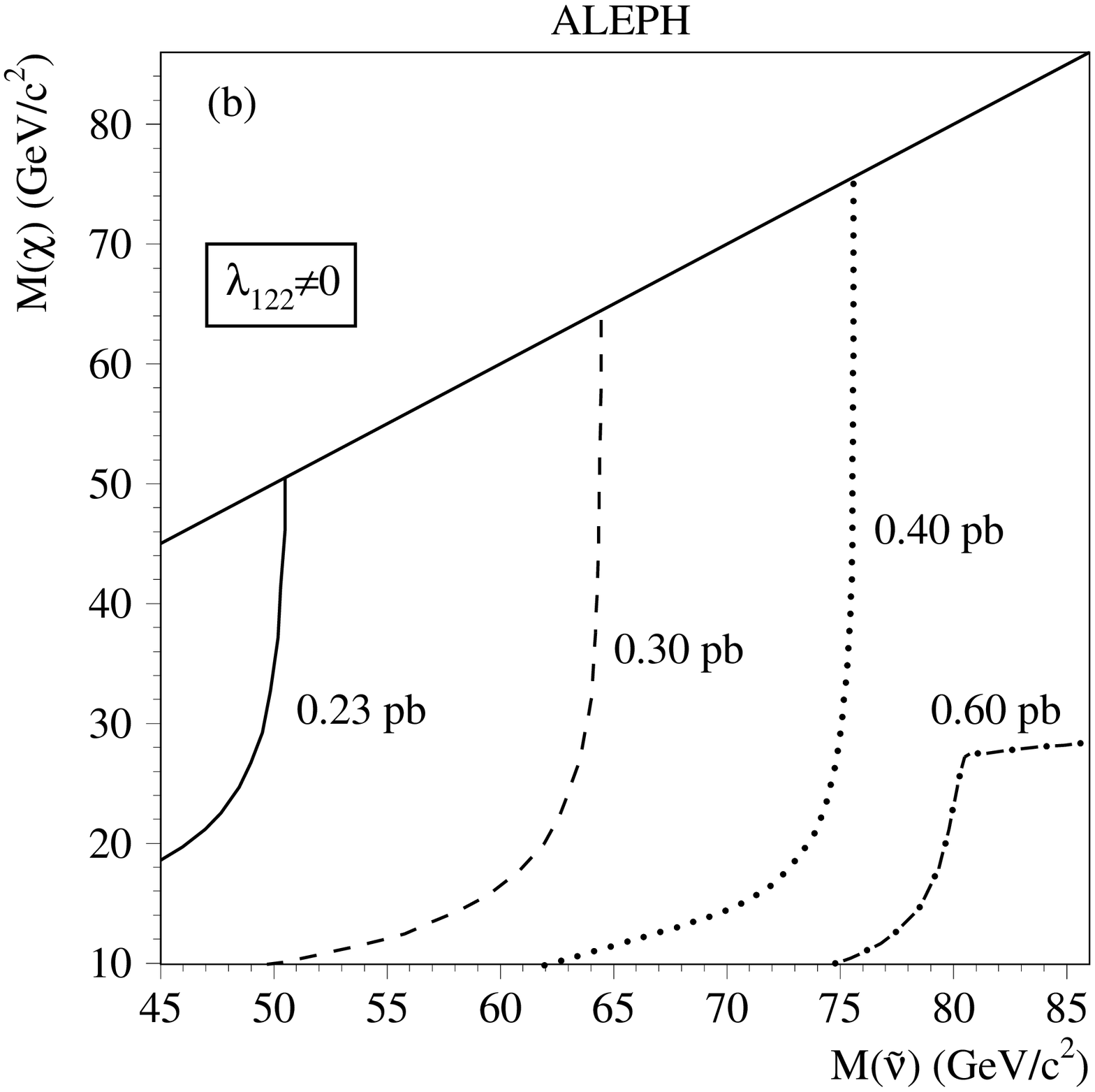,width=0.5\textwidth}
\epsfig{figure=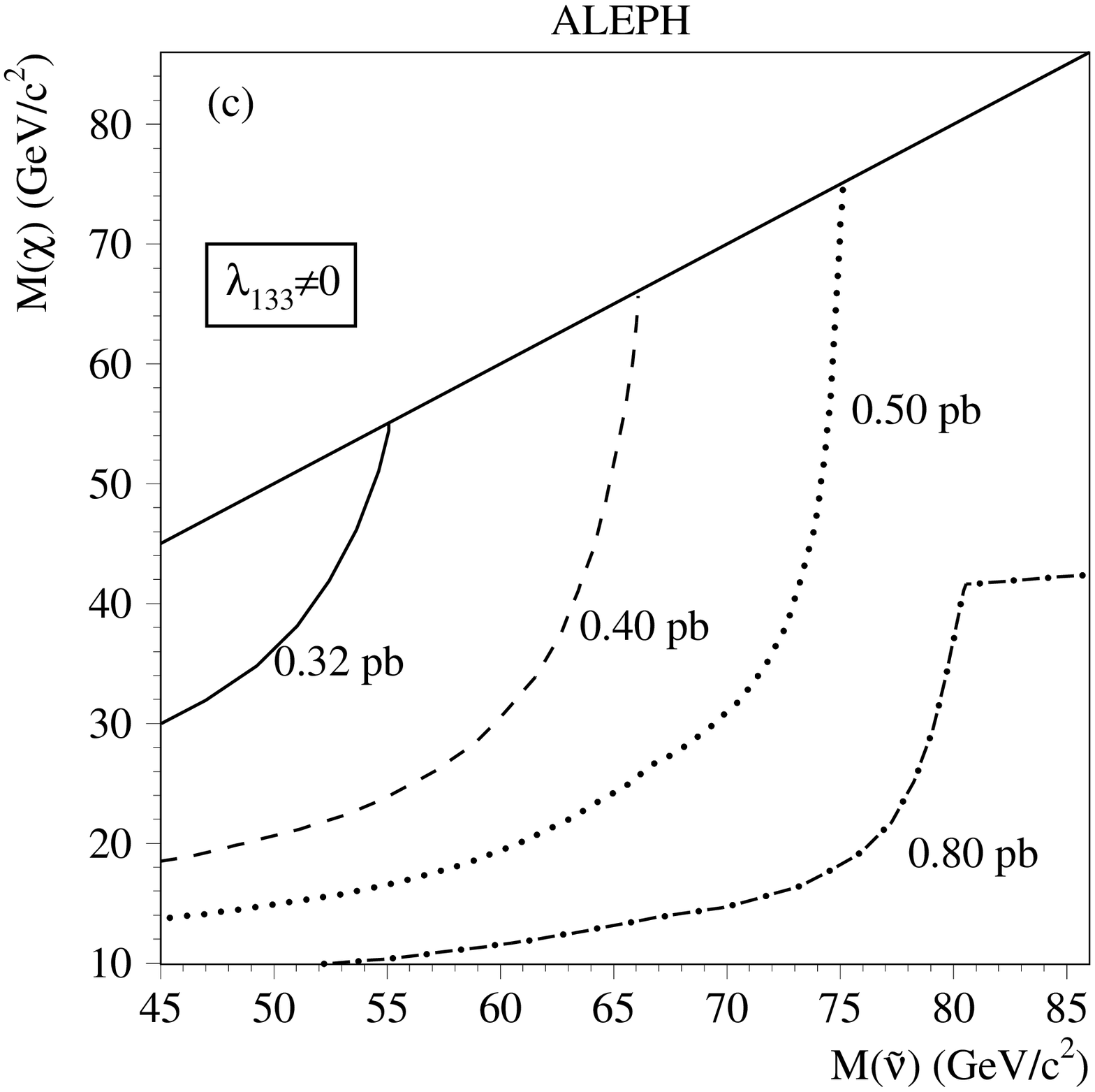,width=0.5\textwidth}}
\caption{\label{excl.xsec.snus}\em The 95\% C.L. sneutrino exclusion
  cross sections scaled to $\sqrt{s}=172\gev$.  For the purpose of these plots
  a $\beta^3/s$ cross section dependence, valid for  
 scalar pair-production in the {\em s}-channel,  was assumed.
Fig. a) shows $\sigma_{\rm excluded}^{172}$ for the direct decays of
 sneutrinos. Superimposed are
the MSSM cross sections at $\tan\beta = 2$ for electron-sneutrino production
($M_2=100\gevcc,\mu=-200\gevcc$) and muon-sneutrino production. Fig. b) and c) show
contours of $\sigma_{\rm excluded}^{172}$ for the indirect decays 
in the ($M_{\chi},M_{\snu}$) plane for the best-case exclusion
($\lambda_{122}$) and  the worst-case exclusion ($\lambda_{133}$).}
\end{center}
\end{figure}

  For the {\em direct topology} the ``Four Lepton'' selection is used.
  The efficiency of pair-produced sneutrinos decaying 
  into the final states $eeee$, $ee\mu\mu$,
  $ee\tau\tau$, $\mu\mu\mu\mu$, $\mu\mu\tau\tau$, $\tau\tau\tau\tau$  are
  calculated as a  function of the sneutrino mass. 
  The different final states  correspond to different choices of
  the generation indices $i,j,k$. The exclusion cross sections scaled to $\sqrt{s}=172\gev$ are
  derived combining the data samples from the various energies,
  and the result is shown in Fig.~\ref{excl.xsec.snus}a. Note that 
  sneutrinos have {\it  only one} direct decay mode  for
  a given choice of the  generation indices $i,j,k$. 

  For the {\em indirect topologies}, which consist of four leptons and four
  neutrinos, an inclusive combination  of the
  ``Four Leptons plus $\emiss$'' and the ``Four Lepton'' selection
  is used, the latter one increasing the selection  efficiencies in the
  region of small $\Delta M = M_{\tilde \nu} - M_{\chi}$. 
  The efficiencies for the sneutrino signal (c.f.\
  Table~{\ref{effics}}) primarily depend on the neutralino mass, and the
  lowest efficiencies are found for small $M_\chi$. 
  The $95\%$ C.L. exclusion cross sections  scaled to
  $\sqrt{s}=172\gev$  are shown in Fig.~\ref{excl.xsec.snus}b,c for the 
   best- and worst-case couplings $\lambda_{122}$ and $\lambda_{133}$. 

  As in the slepton case, pairs of sneutrinos can produce up to $50\%$ {\em
  mixed topologies}. The efficiencies for the mixed topologies, which are
  efficiently selected by the ``Six Leptons plus $\emiss$'' selection in
  combination with the ``Four Leptons plus $\emiss$'' and the ``Four Lepton''
  selections,  are generally
  higher than the efficiencies for indirect topologies, especially for
  low neutralino masses.

\begin{figure}
\begin{center}
\makebox[\textwidth]{
\epsfig{figure=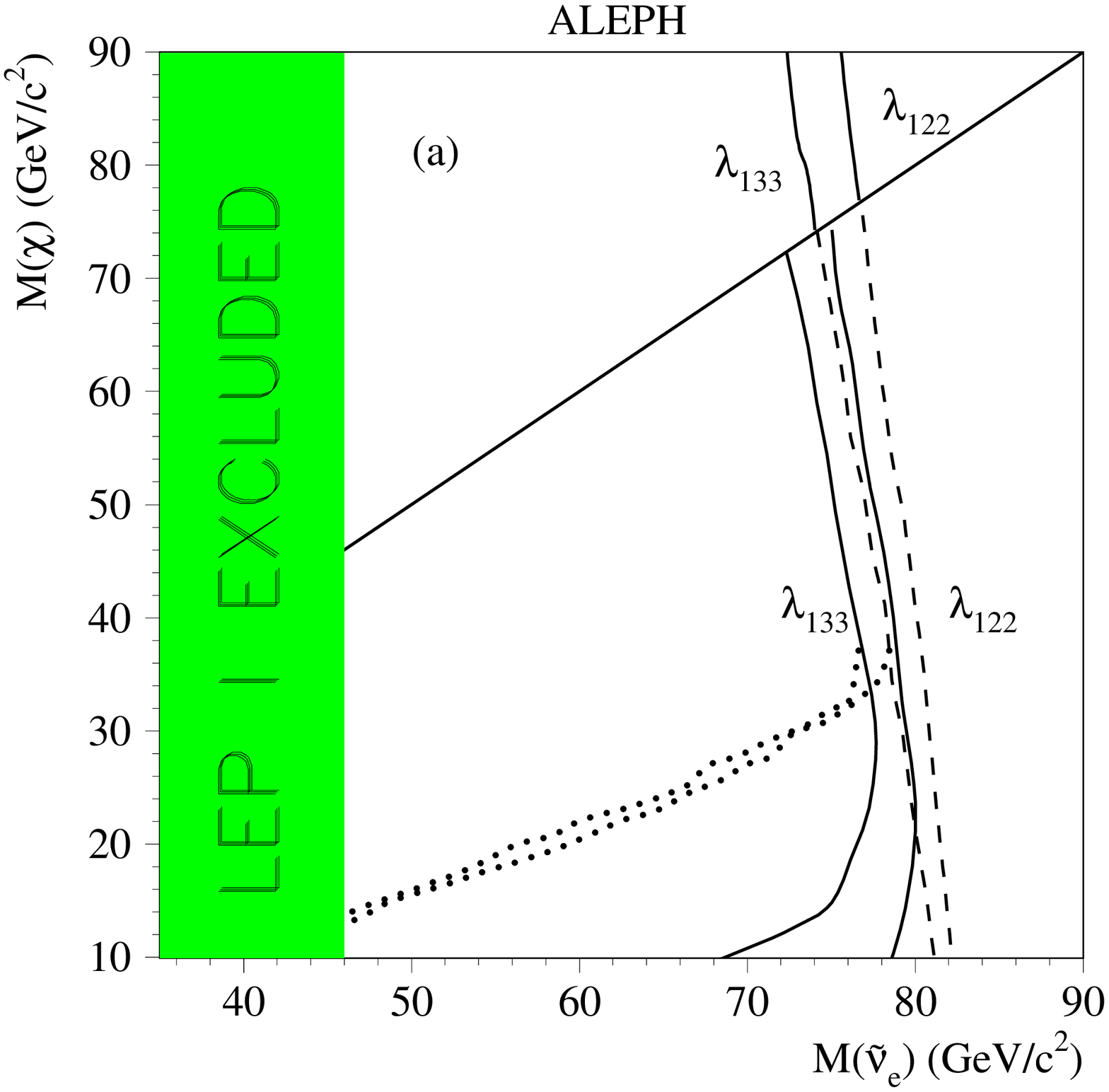,width=0.5\textwidth}\hfill
\epsfig{figure=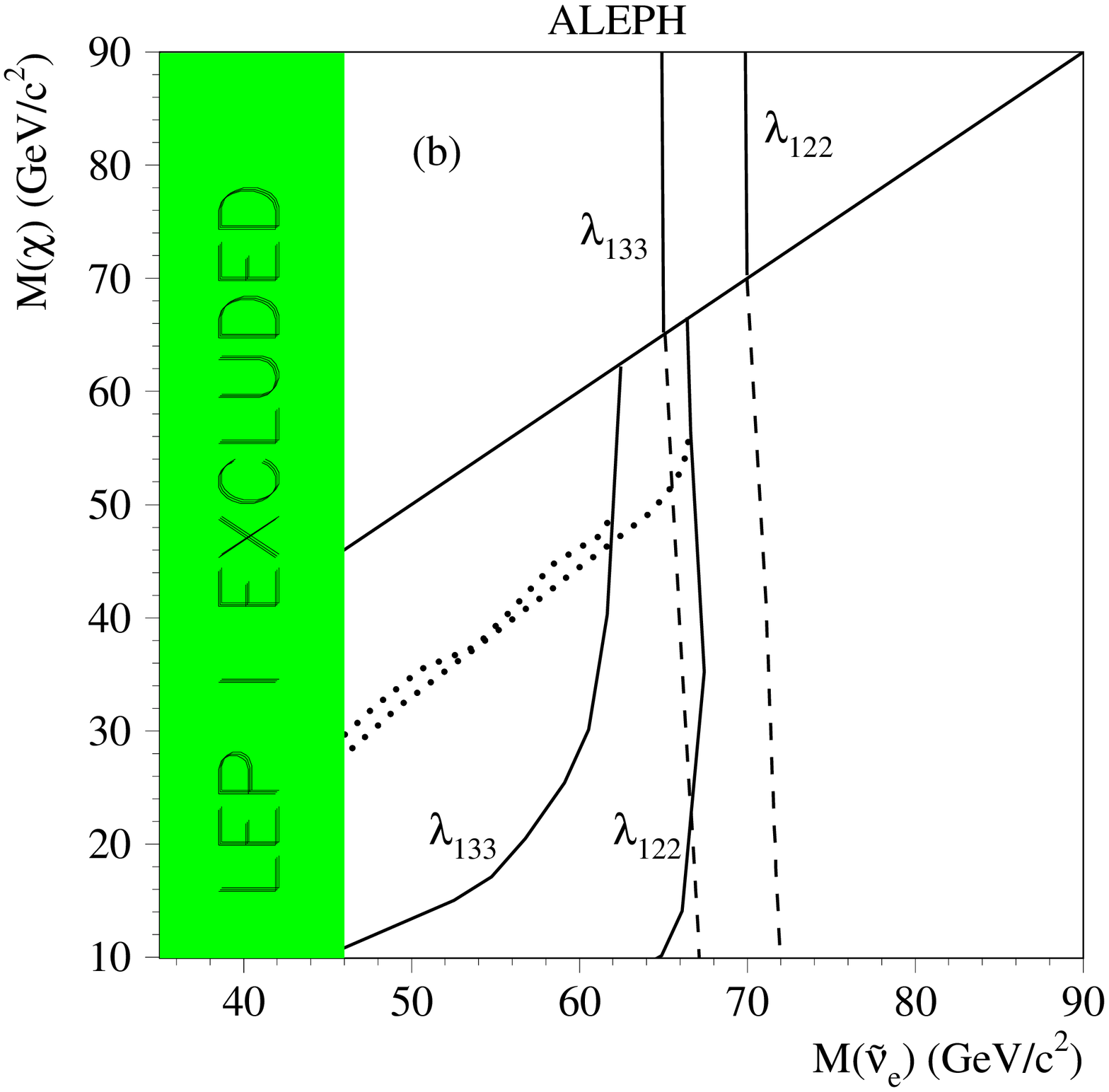,width=0.5\textwidth}}
\makebox[\textwidth]{
\epsfig{figure=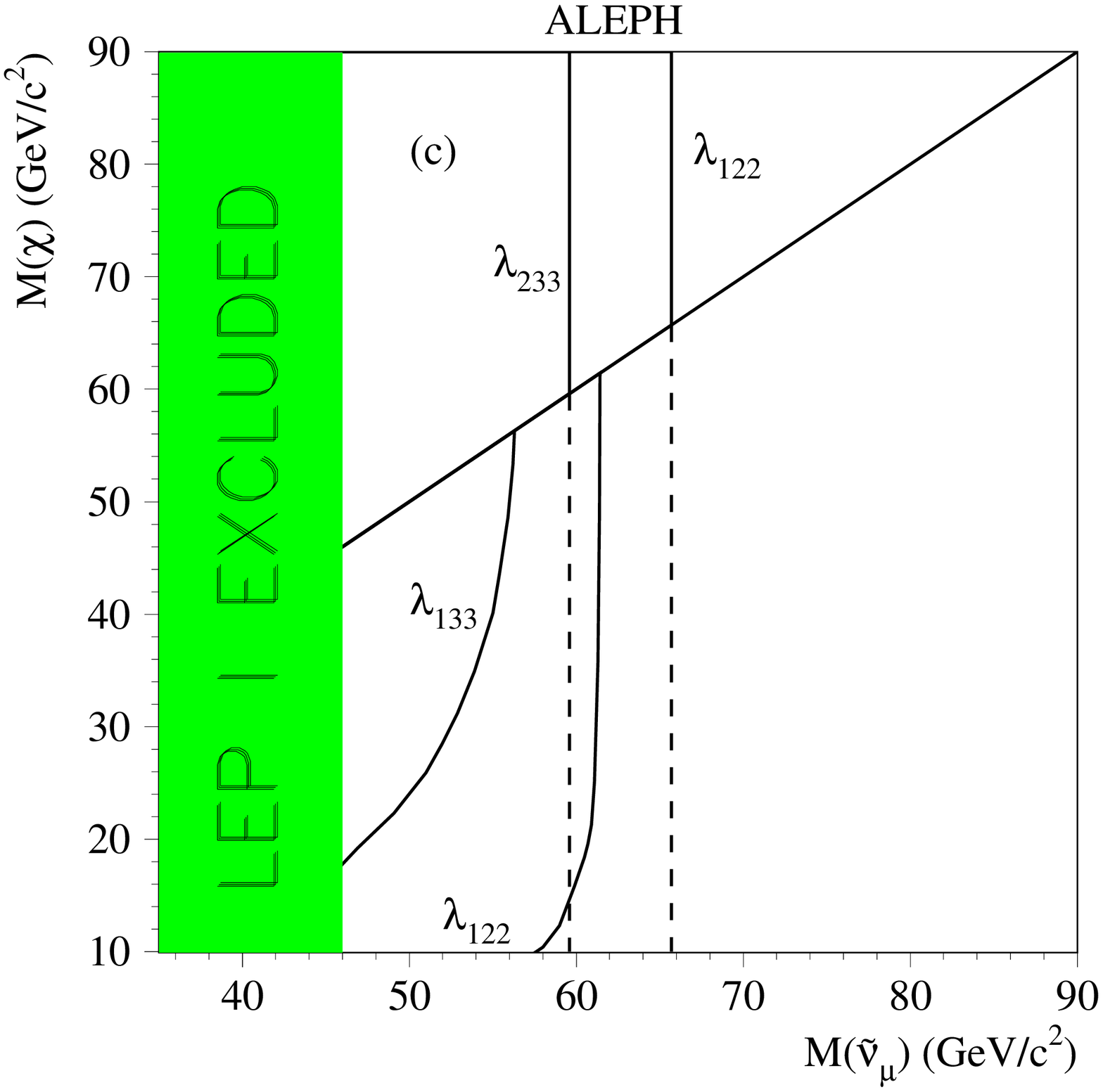,width=0.5\textwidth}\hfill
\epsfig{figure=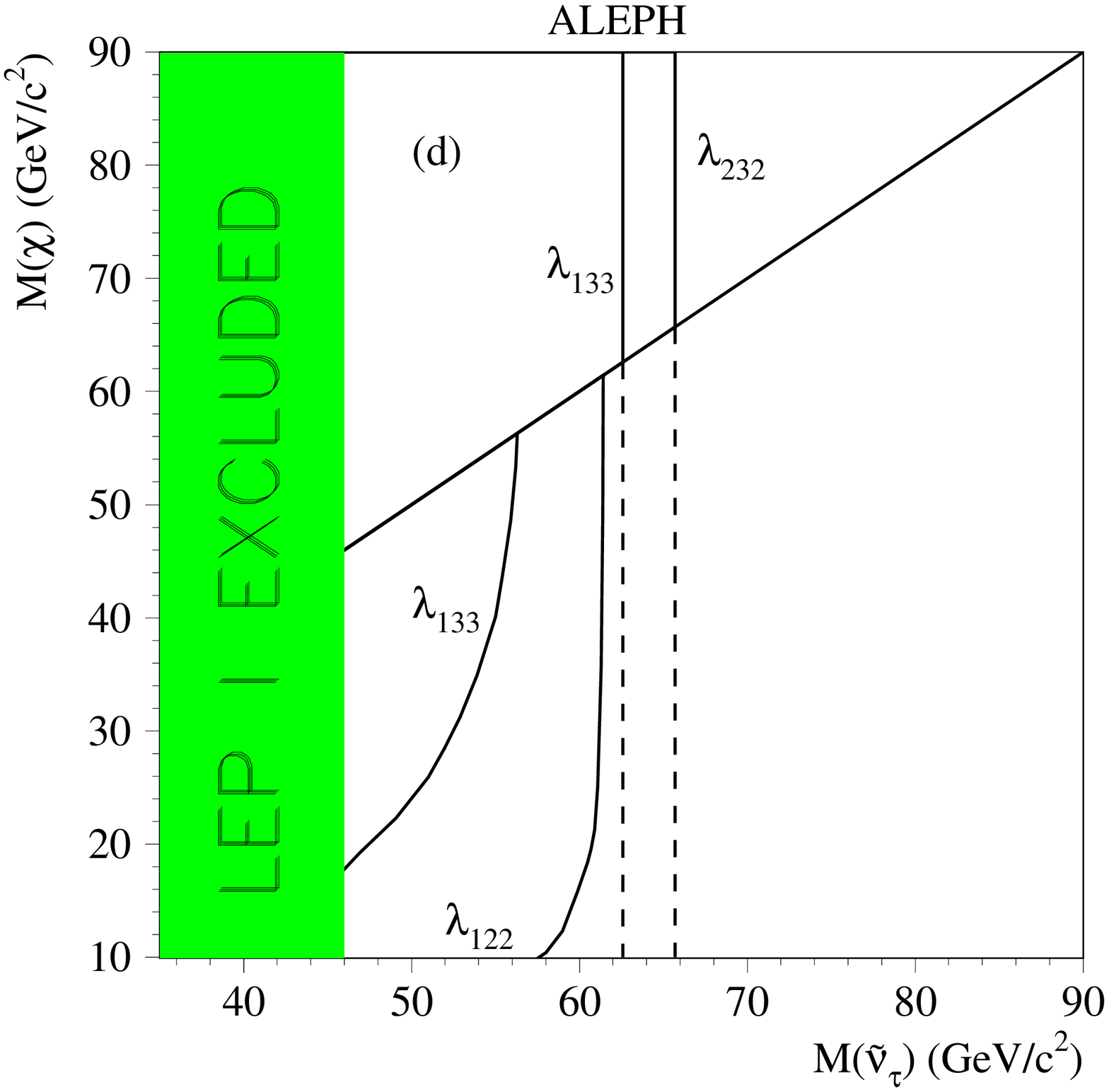,width=0.5\textwidth}}
\caption{\label{snus}\em The $95\%$ C.L. limits in the
  $(m_{\chi},m_{\tilde{\nu}})$ plane at $\tan\beta=2$. 
 Above the diagonal line the lightest neutralino is heavier than the
 sneutrinos, and only the {direct} decays are allowed.
 Below the line the {indirect}
 decays generally dominate, but the branching ratio of the {direct} (dashed
 lines) to {indirect} (full lines) decays depends on the magnitude of the
 coupling $\lambda_{ijk}$. The two
 choices of $\lambda_{122}$ and $\lambda_{133}$ correspond to the best and
 worst case exclusions for the {indirect} decays.
 Fig. a) and b) show the electron-sneutrino limit in the gaugino region
 ($\mu=-200\gevc$) and the higgsino region $(M_2=400\gevcc)$, respectively, 
 assuming BR(${\tilde \nu}_e \ra \nu_e \chi$)=100$\%$ for the indirect decays
 (full lines), and conservatively assuming zero efficiency for the
 cascade decays (dotted lines). Fig. c) and d) show the mass limits on
 muon- and tau-sneutrinos.}
\end{center}
\end{figure}

  Interpreting these results within the MSSM, the $95\%$ C.L. exclusion regions
  are derived in the  ($M_\chi,M_{\tilde \nu}$) plane, and are shown in
  Fig.~\ref{snus}.
  Exclusions for the two extreme cases of $100\%$ direct
  or $100\%$ indirect decay modes (which generally correspond to the two cases
  of sneutrino and neutralino  LSPs, respectively) are shown, 
  while the exclusion regions
  for the case when
  $\Gamma({\tilde \nu \ra l^+ l^-}) \sim \Gamma({\tilde \nu \ra \nu \chi})$
  (resulting in a substantial fraction of
  mixed topologies) would lie in between those two extreme cases. 
  The two choices of couplings for the direct and the indirect 
  topologies correspond to final states with a maximum number of muons or taus,
  resulting in  best- and worst-case exclusion limits, respectively. 

  For electron-sneutrinos, {\em t}-channel chargino exchange can
  enhance the cross section\cite{barger.et.al}, and this effect is shown by considering a 
  typical point in the gaugino region ($\mu=-200, \tan\beta=2$) and in the
  higgsino   region ($M_2=400\gevcc, \tan\beta=2, \mu<0$), assuming BR$({\tilde
  \nu} \ra \nu \chi$)=100\%. For $M_{\chi}<20$-$40\gevcc$, sneutrinos can 
  cascade to the chargino,  indicated by the dotted lines in Fig.~\ref{snus}a
  and b, which conservatively assume zero efficiencies for the 
  cascade decays. However, the cascade regions
  are  already excluded by the chargino and neutralino limits of
  Section~\ref{chargino.limit}, and are therefore not
  considered further.
  The sneutrino mass limits for the indirect decay modes, the
  most conservative choice of coupling ($\lambda_{133}$), and for 
 $M_{\chi} > 23\gevcc$ are: $M_{{\tilde \nu}_e}>72\gevcc$ (gaugino region,
 $\mu=-200\gevcc$, $\tan\beta = 2$), $M_{{\tilde \nu}_e}>58\gevcc$ (higgsino region,
 $M_2=400\gevcc$, $\tan\beta = 2, \mu<0$), 
  $M_{{\tilde \nu}_\mu}, M_{{\tilde \nu}_\tau} >49\gevcc$.

\subsection{Squarks}
\begin{figure}
\begin{center}
\makebox[\textwidth]{
\epsfig{figure=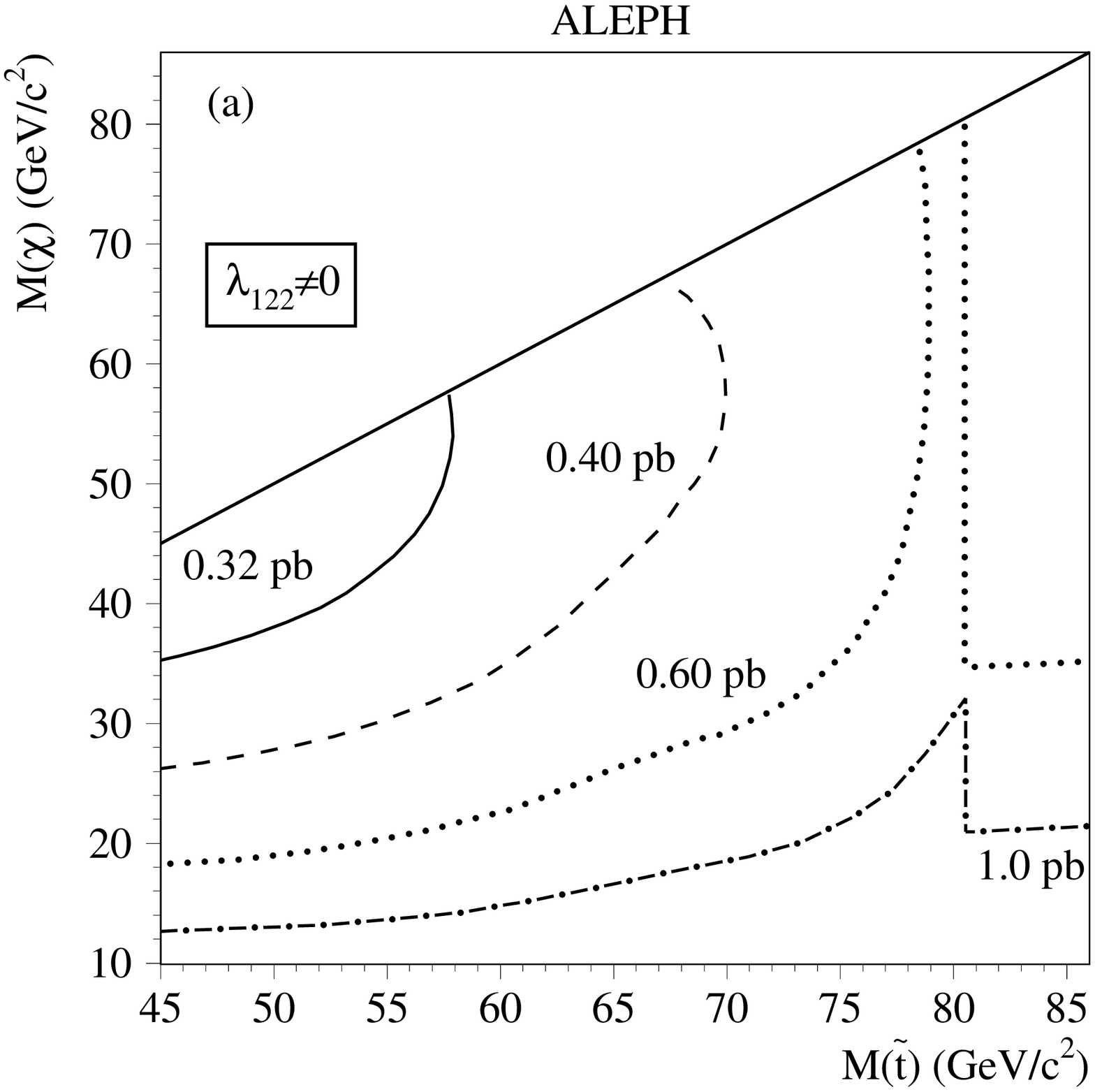,width=0.5\textwidth}\hfill
\epsfig{figure=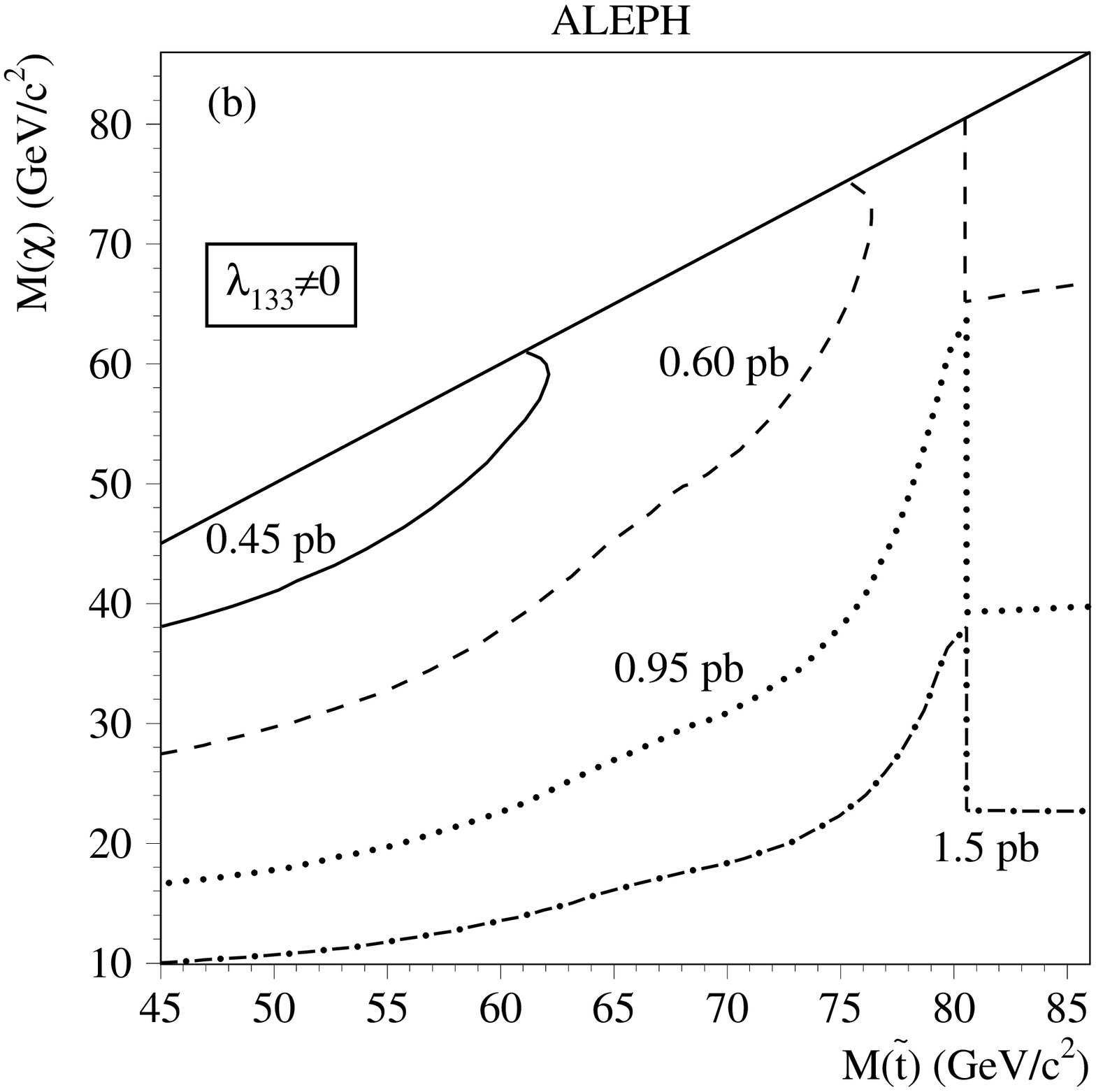,width=0.5\textwidth}}
\caption{\label{stop.excluded.xsecs}\em The $95\%$ C.L.  stop exclusion
  cross sections at $\sqrt{s}=172\gev$.  Fig. a) and b) show
contours of $\sigma_{\rm excluded}^{172}$ in the ($M_{\chi},M_{\tilde t}$) 
plane for the best-case exclusion ($\lambda_{122}$) 
and the worst-case exclusion ($\lambda_{133}$).}
\end{center}
\end{figure}

The stop and  sbottom cannot decay {\em directly} via the
purely leptonic $LL{\bar E}$ couplings, but they can decay {\em indirectly}
to the lightest neutralino, producing topologies with four leptons and
two jets plus a small amount of missing energy. 
Using the ``Leptons and Hadrons'' selection, efficiencies
(c.f.\ Table~\ref{effics}) for the stop and sbottom
signal are calculated as a function of the squark and neutralino masses and
for the three energies. Conservatively, sbottoms have been assumed to
hadronise before their decay throughout parameter space, as selection
efficiencies for this case are smaller compared to hadronisation after
the decay.
The excluded cross sections 
are shown in Fig.~\ref{stop.excluded.xsecs} for the two couplings 
$\lambda_{122}$ and  $\lambda_{133}$.

\begin{figure}
\begin{center}
\makebox[\textwidth]{
\epsfig{figure=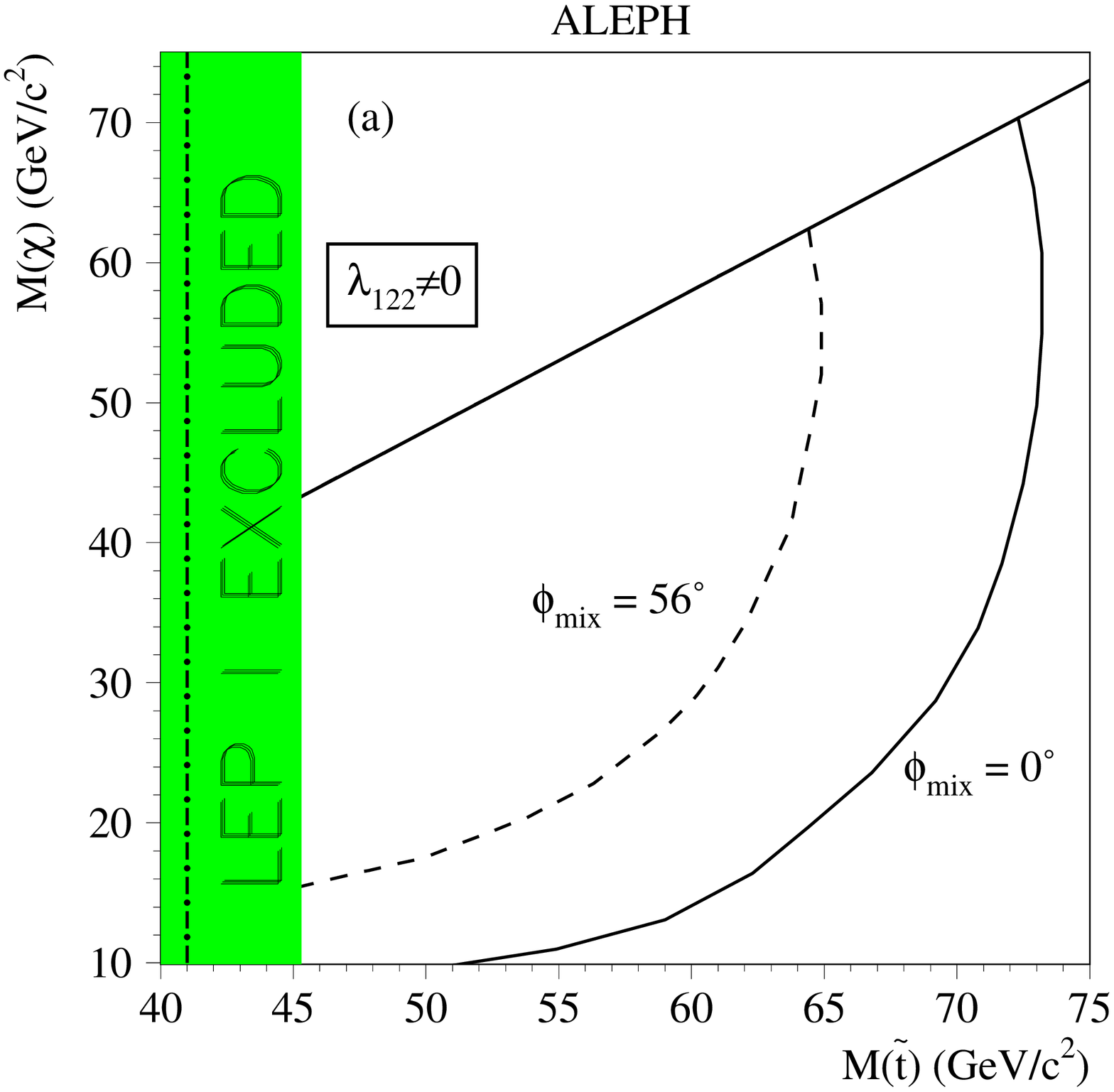,width=0.5\textwidth}\hfill
\epsfig{figure=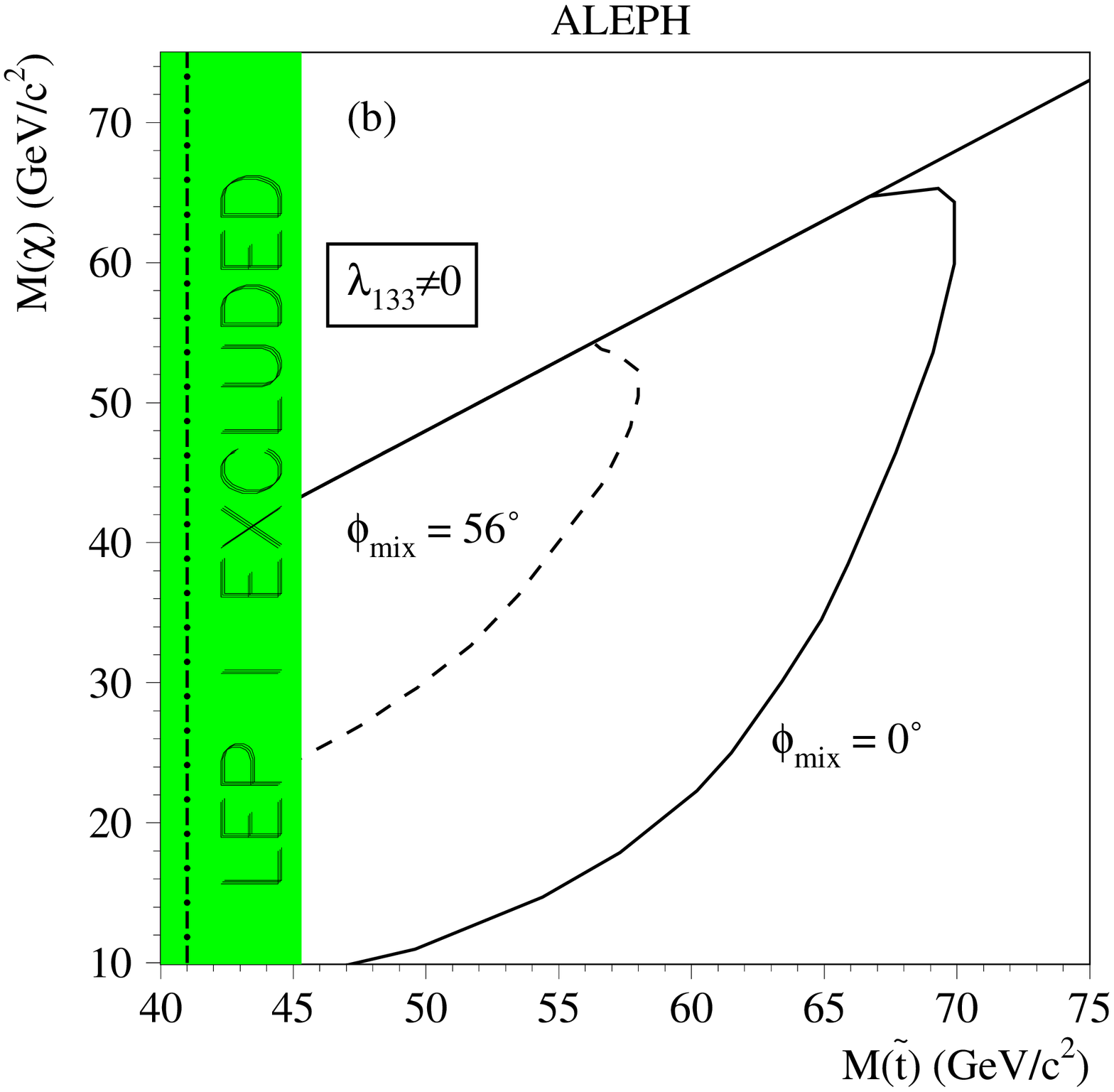,width=0.5\textwidth}}
\makebox[\textwidth]{
\epsfig{figure=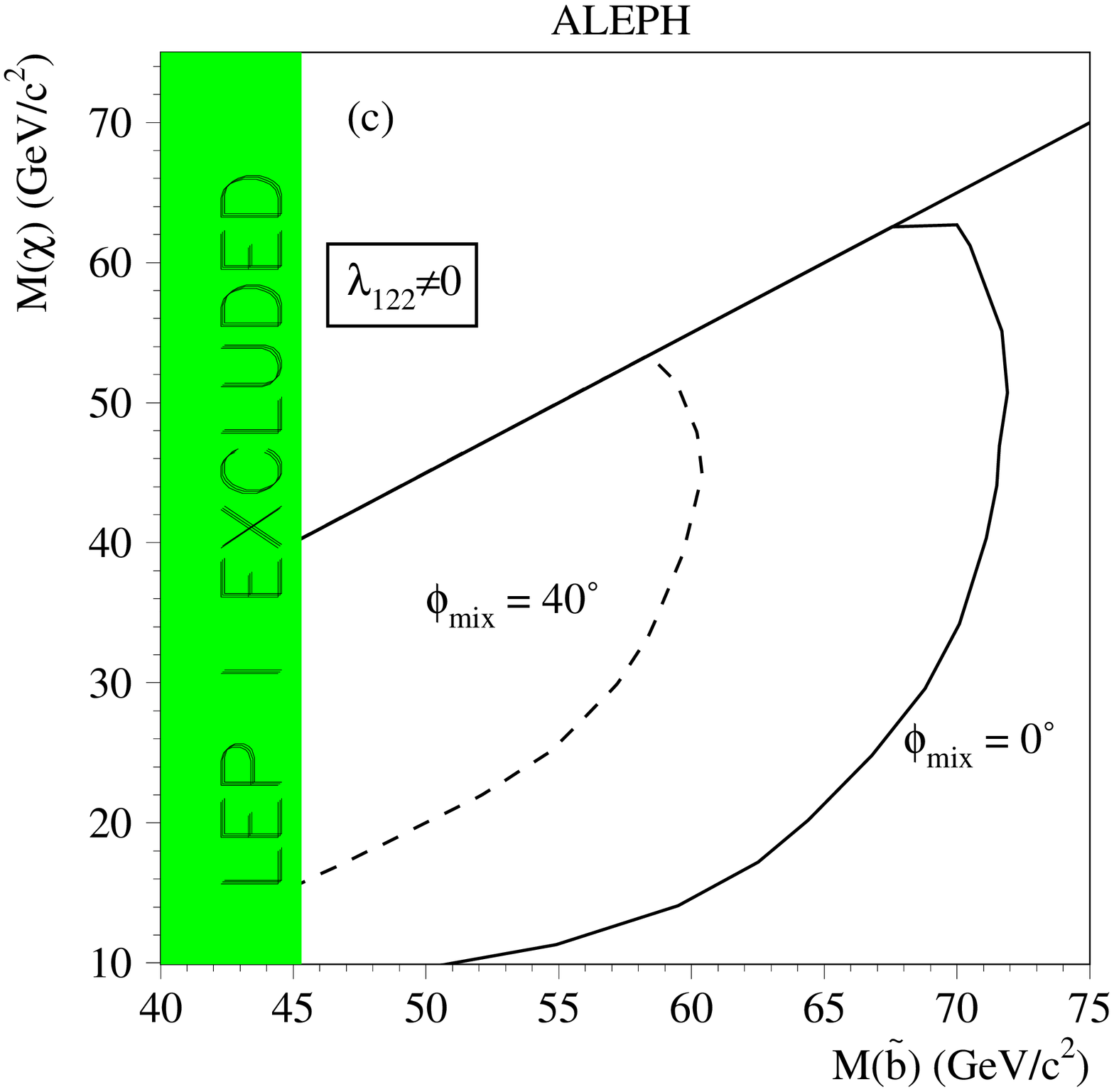,width=0.5\textwidth}\hfill
\epsfig{figure=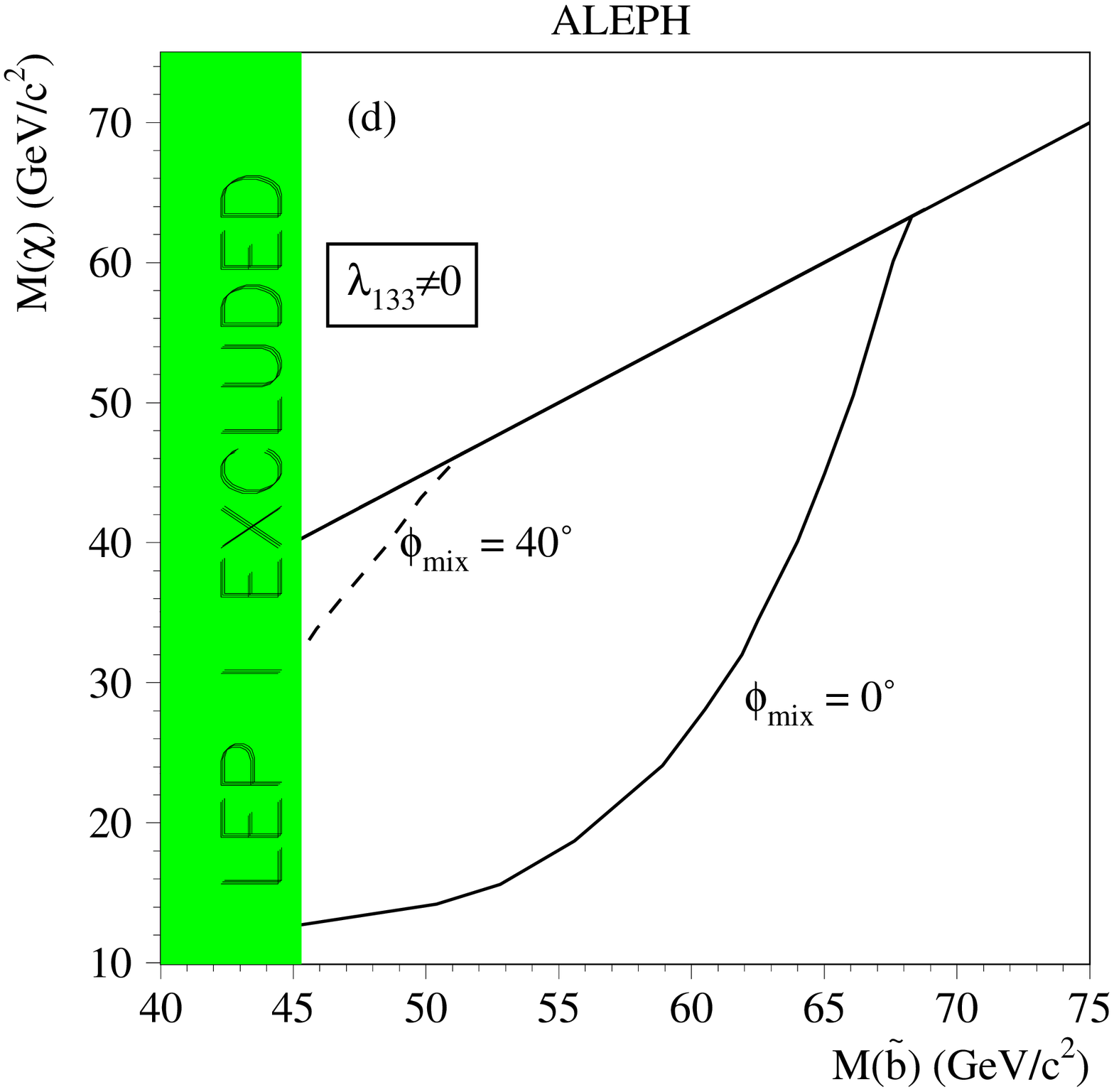,width=0.5\textwidth}}
\caption{\label{stops.excluded}\em The $95\%$ C.L. limits on the stop  and sbottom in the
    $(m_{\chi},m_{\tilde{q}})$    plane.
    The two  choices of $\lambda_{122}$ and $\lambda_{133}$ correspond to the best and
   worst case exclusions, respectively. The mass limits are shown for minimal squark
    mixing ($\phi_{mix}=0^\circ$), and for $\phi_{mix}=56^\circ, 40^\circ$ for stops
    and sbottoms, respectively. The LEP~1 limit for $\phi_{mix}=0^\circ$ (and
    $\phi_{mix}=56^\circ$ for stops -- dashed-dotted lines) is also  shown.}
\end{center}
\end{figure}

The limits in the ($M_\chi,M_{{\tilde q}}$) plane obtained within the MSSM
are shown in Fig.~\ref{stops.excluded} for the two choices of couplings
$\lambda_{122}, \lambda_{133}$, corresponding to the best- and worst-case
exclusions, respectively.
For stops, the results for the two mixing angles
$\phi_{mix} = 0^\circ, 56^\circ$ correspond to a maximal and
minimal ${\tilde t}_1$-Z coupling. 
The limits for the most conservative coupling ($\lambda_{133}$)
and $M_{\chi} > 23\gevcc$ are:
$M_{{\tilde t}_L}>60\gevcc$ and $M_{{\tilde  b}_L}>58\gevcc$
($\phi_{mix} = 0^\circ$), and $M_{{\tilde t}_1}>44\gevcc$
($\phi_{mix} = 56^\circ$).

\section{Conclusions}\label{conclusions}
A number of search analyses have been developed to select R-parity violating
SUSY topologies from the pair-production of sparticles.
It was assumed that the 
LSP has a negligible lifetime, and that only the  $LL{\bar E}$ couplings are
non-zero. Limits were derived under the assumption that only one coupling
$\lambda_{ijk}$ is non-zero, although the search analyses cover topologies 
which would be produced by the simultaneous presence of more than one coupling.
Particular emphasis was placed on making no assumption on the nature
of the LSP. The search analyses for the various topologies find no evidence for
R-parity violating Supersymmetry in the data collected at
$\sqrt{s}=$130--172\gev, and limits have been set within the framework of the
MSSM.

The decay modes of charginos and  heavier neutralinos
 were  classified according to topology into
 {\it indirect decay modes} to the lightest neutralino (which generally
 corresponds to neutralino LSPs), 
  and into {\it direct decay modes} to three leptons (which generally
 corresponds to slepton or sneutrino LSPs).
 At low values of $\tan\beta$, charginos are excluded up 
 to $M_{\chi^+}>85\gevcc$ for the indirect decays, and up to
$M_{\chi^+}>80\gevcc$ for the direct decays.
 For large $\tan\beta$,
the chargino limit drops to $M_{\chi^+}>78\gevcc$ and $M_{\chi^+}>73\gevcc$,
respectively.  The weakest  mass bound  on the lightest neutralino is found at $\tan\beta=1$,
where  $M_{\chi}>25\gevcc$ for the indirect (chargino) decays, and
$M_{\chi}>23\gevcc$ for the direct decays. The mass bound is much stronger at
large values of $\tan{\beta}$, where $M_{\chi}>47\gevcc$ and $M_{\chi}>45\gevcc$
for the two chargino decay modes at $\tan{\beta}>15$.
The limits for charginos and neutralinos hold for any choice
of the generation indices $i,j,k$ of the coupling $\lambda_{ijk}$, and
neutralino, slepton and sneutrino LSPs.

The mass limits for the sfermions are highly dependent on the choice of
the indices $i,j,k$ and the nature of the LSP, mainly owing to the much smaller
production cross section of scalars compared to the fermionic cross sections. 
For the indirect decay modes (where the sfermions decay to the lightest
neutralino) and the most conservative choice of coupling,
the mass limits at $\tan{\beta}=2$ are:
\begin{itemize}
\item{$M_{{\tilde e}_R}>64\gevcc$ (gaugino region),}
\item{$M_{{\tilde \mu}_R}>62\gevcc$,}
\item{$M_{{\tilde \tau}_R}>56\gevcc$,}
\item{$M_{{\tilde \nu}_e}>72\gevcc$ (gaugino region),}
\item{$M_{{\tilde \nu}_\mu}, M_{{\tilde \nu}_\tau} >49\gevcc$,}
\item{$M_{{\tilde t}_L}>60\gevcc$,}
\item{$M_{{\tilde  b}_L}>58\gevcc$.}
\end{itemize}
These mass limits considerably improve upon existing limits. 

\section{Acknowledgements}
It is a pleasure to congratulate our colleagues from the accelerator divisions
for the successful operation of LEP at high energy. 
We would like to express our gratitude to the engineers and 
support people at our home institutes without whose dedicated help
this work would not have been possible. 
Those of us from non-member states wish to thank CERN for its hospitality
and support.

\end{document}